%% file: main.tex
\documentclass[sigconf, nonacm]{acmart}

\usepackage{booktabs}
\usepackage{colortbl}
\usepackage{float}
\usepackage{listings}
\usepackage{soul}
\usepackage{multirow}
\usepackage[compact]{titlesec}

\usepackage{lipsum}

\usepackage{pythonhighlight}
\usepackage{listings}

\usepackage{todonotes}

\usepackage{graphicx}
\usepackage{booktabs}
\usepackage{verbatim}
\usepackage[noend]{algpseudocode}
\usepackage{algorithm}
\usepackage{amsmath}
\usepackage{paralist}
\usepackage{amsthm}
\usepackage{pifont}

\usepackage{booktabs}

\theoremstyle{remark}

\theoremstyle{definition}

\theoremstyle{remark}

\theoremstyle{remark}
\newtheorem{invariant}{\textbf{Invariant}}

\newenvironment{packeditemize}{
\begin{list}{$\bullet$}{
\setlength{\itemsep}{1.5pt}
\setlength{\labelwidth}{8pt}
\setlength{\leftmargin}{10pt}
\setlength{\labelsep}{3pt}
\setlength{\listparindent}{\parindent}
\setlength{\parsep}{1.5pt}
\setlength{\parskip}{1.5pt}
\setlength{\topsep}{1.5pt}}}{\end{list}}

\newcommand\vldbavailabilityurl{https://github.com/Froot-NetSys/promsketch}

\newcommand{\sysname}{PromSketch}
\newcommand{\mycaption}[3]{{\caption{\label{#1}{\bf \small #2. } {\em #3}}}}

\newcommand{\change}[1]{{\color{black}{#1}}}

\begin{document}
\title{Approximation-First Timeseries Monitoring Query At Scale}

\author{Zeying Zhu}
\affiliation{%
  \institution{University of Maryland}
}
\email{zeyingz@umd.edu}

\author{Jonathan Chamberlain}
\affiliation{%
  \institution{Boston University}
}
\email{jdchambo@bu.edu}

\author{Kenny Wu}
\affiliation{%
  \institution{University of Maryland}
}
\email{kwu588@terpmail.umd.edu}

\author{David Starobinski}
\affiliation{%
  \institution{Boston University}
}
\email{staro@bu.edu}

\author{Zaoxing Liu}
\affiliation{%
  \institution{University of Maryland}
}
\email{zaoxing@umd.edu}

\input{abstract}

\maketitle

\noindent\textbf{\em Note:} {\em This is the preprint version of the paper accepted to VLDB 2025. }

\input{Introduction}

\input{Background}

\input{Overview}

\input{Design}
\input{Implementation}

\input{Evaluation}

\input{RelatedWork}

\input{Conclustions}



\input{main.bbl}

\end{document}

%% file: abstract.tex
\begin{abstract}
Timeseries monitoring systems such as Prometheus play a crucial role in gaining observability of the underlying system components. These systems collect timeseries metrics from various system components and perform monitoring queries over periodic window-based aggregations (i.e., rule queries). However, despite wide adoption, the operational costs and query latency of rule queries remain high.
In this paper, we identify major bottlenecks associated with repeated data scans and query computations concerning window overlaps in rule queries, and present \sysname{}, an approximation-first query framework as intermediate caches for monitoring systems. It enables low operational costs and query latency, by combining approximate window-based query frameworks and sketch-based precomputation. 
\sysname{} is implemented as a standalone module that can be integrated into Prometheus and VictoriaMetrics, covering 70\% of Prometheus' aggregation over time queries. Our evaluation shows that \sysname{} achieves up to a \change{two orders of magnitude} reduction in query latency over Prometheus and VictoriaMetrics, while lowering operational dollar costs of query processing by two orders of magnitude compared to Prometheus and by at least $4\times$ compared to VictoriaMetrics with at most 5\% average errors across statistics.  
The source code has been made available at \url{\vldbavailabilityurl}.
\end{abstract}

%% file: Introduction.tex
\section{Introduction} \label{sec:intro}
Cloud-native timeseries monitoring systems such as Prometheus~\cite{promethues-soundcloud}, VictoriaMetrics~\cite{VictoriaMetrics}, and Grafana Mimir~\cite{mimir} are widely used as the cloud telemetry platform, where various metrics such as sensor readings~\cite{madden2003design}, IP network traffic information~\cite{5gmetric, snmp, cloudflare, junos, cranor2003gigascope}, and cluster CPU and memory utilization~\cite{kubernetesmetric, abraham2013scuba} are stored and monitored.  
Under the hood, such a monitoring system often consists of a timeseries database as the back-end and a dynamic query engine as the front-end, allowing users to perform various statistical queries over different time ranges to support downstream applications such as anomaly detection~\cite{vmanomaly, datadog-anomaly-detection}, attack detection~\cite{moosa2023detection, priovolos2021using, pope2021container}, and data visualization~\cite{grafana-dashboards}. 
Among all queries supported for these applications, {\em rule} queries~\cite{promethuesfunction, victoriarollup, awesomealerts} are often set up to periodically compute aggregated statistics over time ranges (i.e., repeated {\em time range queries}) and alert users if abnormal conditions are met (e.g., quantiles, top-K, cardinality). For instance, timeseries network flow data (e.g., source/destination IPs, ports, and protocols) can be monitored in the range of seconds to aggregate distinct source IPs targeting a specific host over a recent time window, indicating a potential Distributed Denial of Service (DDoS) attack~\cite{antonakakis2017understanding,liu2021jaqen}.

While Prometheus and its variants have been a de facto standard open-source tool to handle rule queries, they struggle with non-trivial operational costs and high query latency in practice. 
\change{In our evaluation}, an AWS Prometheus service running 10 rule queries every minute and monitoring just on {\em a single rack} would approximately take \$11,520 for query processing and \$9,256 for data ingestion per month (\S\ref{sec:bottlenecks}). Performing a quantile query over 100K-sample windows and 10K timeseries takes 15 min on a commodity server in our testbed. Our profiling reveals two major bottlenecks in rule queries that lead to high monitoring cost and query latency: 1) {\em repeated data scans} from storage and 2) {\em repeated query computations}, based on the observation that a single rule can perform time range queries over consecutive overlapping windows and different rules may also query the same overlapping windows. For example, a rule with 10-minute windows and 1-min evaluation intervals, or queries over different time windows (e.g., 2-, 5- and 10-minute), repeatedly access the overlapping portions of data among windows but Prometheus computes them separately.

While several existing efforts aim to address the bottlenecks of Prometheus, they fall short in one or more of the dimensions in operational cost, query latency, and query accuracy. Exact monitoring systems that optimize Prometheus (e.g., VictoriaMetrics~\cite{VictoriaMetrics}) can reduce query latency through better storage engine designs and data caching for lower data retrieval time, and applying parallel query computation for lower query evaluation time. However, they do not reduce operational costs as they do not address the repeated data scanning and computational bottlenecks. While one can consider applying pre-computation approaches similarly to those optimizing SQL queries~\cite{ho1997range,  shen2023lindorm}, they tend to support a fixed time window and limited statistics such as sum and max, with small cost and performance improvements.

Alternatively, approximate analytics (e.g., sampling- and sketch-based) offer a promising approach to trade off estimation accuracy for further lower operational costs and query latency~\cite{manousis2022enabling} of complex queries. \change{Applications often require near real-time analytics and tolerate approximate but highly accurate results, such as datacenter alerts~\cite{google-alerting, lim2020approximate, agrawal2017low}, network measurements~\cite{estan2002new, wu2023microscopesketch}, and more~\cite{koren2009collaborative, chang2017streaming, yang2020joltik}. } However, practical issues of low accuracy and low query generality remain. Sampling-based approaches (e.g., \cite{thanos-downsampling, lu2010simple}) can provide an estimation for any queries but suffer from worse and unpredictable accuracy for complex statistics such as quantiles and entropy. Sketch-based analytics (e.g., \cite{masson2019ddsketch, karnin2016optimal, manousis2022enabling, acharya1999aqua, charikar2002finding}) and  sliding window sketches (e.g.,\cite{ben2019succinct, ben2016heavy, chabchoub2010sliding, arasu2004approximate}) can provide strong accuracy guarantees over querying statistics of a fixed window but are limited to answering certain queries. 

In this paper, we revisit the promises of approximate analytics to improve operational efficiency and performance in timeseries monitoring systems. We present {\bf\sysname{}}, an approximate query cache that improves operational cost and query latency by up to two orders of magnitude while preserving high accuracies (e.g., >$95\%$). In contrast to  Prometheus' independent query handling, \sysname{} is a framework that is able to ``sketch and cache'' a wide range of recent windows and statistics in the fast storage (e.g., main memory) to mitigate the bottlenecks of repeated data scans and query computations from overlapping windows (as in Fig.~\ref{fig:prometheus_diagram}).

\sysname{} is built on the combination of two key ideas. First, \change{\sysname{} caches a range of {\em intermediate results} rather than caching raw data or final query results integrated in today's timeseries monitoring systems~\cite{VictoriaMetrics}}. This is a practical choice because (1) a raw data cache does not help reduce repeated query computations, and the memory usage can be prohibitively large; and (2) a query result cache misses the opportunities to optimize drill-down queries that are not predefined. Thus, we adopt an extended sliding window model based on the Exponential Histogram~\cite{datar2002maintaining, ivkin2019know} to maintain a list of intermediate results (called buckets) covering consecutive intervals of sizes varying exponentially in a timeseries. At query time, we can linearly merge these buckets to obtain the final results on any {\em sub-window} of the large window.
We consider this intermediate result cache as a balance between caching raw data and final results.

\begin{figure}[t]
\centering
\includegraphics[width=0.99\linewidth]{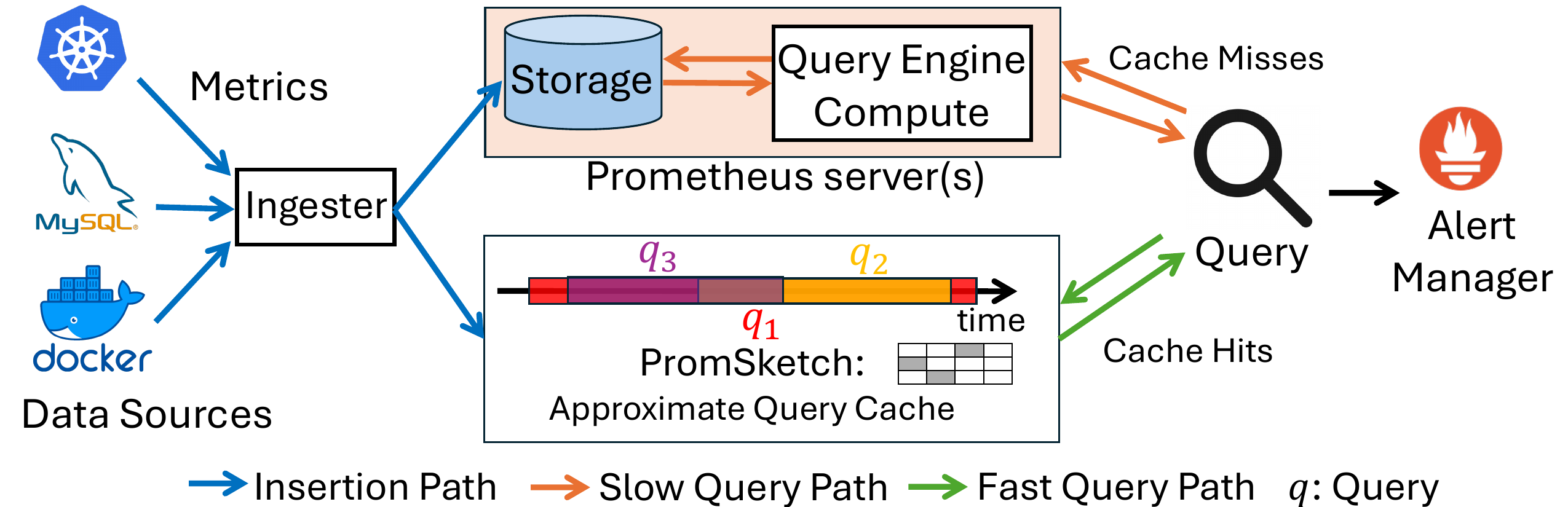}
\vspace{-2mm}
\mycaption{fig:prometheus_diagram}{A Prometheus monitoring ecosystem with \sysname{}}{\footnotesize } 
\vspace{-2mm}
\end{figure}

Second, we provably combine the extended sliding window model with popular linear sketches to support various query functions. There are a potentially large number of concurrent timeseries and query functions that need to be monitored (e.g., quantiles, entropy, and $L_2$ norm of CPU and memory usages from various Kubernates~\cite{kubernetes} nodes). We want to cache as many timeseries as possible given a memory budget. While  exact data structures can be used to store intermediate results, they cannot scale to a large number of timeseries. To optimize memory usage, we extend the Exponential
Histogram model with KLL sketch~\cite{karnin2016optimal} and universal sketching~\cite{braverman2010zero, liu2016one}, and prove their memory-accuracy efficiency both theoretically and empirically, with system optimizations to reduce system runtime and operational costs.

We implement \sysname{} as a Go package in 5K lines of code that is compatible with Prometheus and VictoriaMetrics, two popular open-source timeseries monitoring systems, extending PromQL and covering 70\% of Prometheus' aggregation over time functions. \sysname{} is also portable to other Prometheus-like systems such as ~\cite{mimir, thanos}. 
Our extensive experiments show that:  (1) \sysname{} offers robust accuracy (mean error $\leq$5\%) while reducing operational costs of query processing by 400$\times$ compared to Prometheus and at least 4$\times$ compared to VictoriaMetrics and (2) it reduces end-to-end query latency by up to \change{two orders of magnitude} over Prometheus and VictoriaMetrics. \sysname{}'s precomputation overhead is moderate as 1.3$\times$ to 3$\times$ of  non-precomputed/cached Prometheus.  \change{In summary, we make the following contributions. 
\begin{packeditemize}
\item We systematically analyze the rule queries in the popular timeseries monitoring systems and identify the bottlenecks and cost consequences of repeated data scans and overlapped query computations on the cloud. (\S\ref{sec:background})

\item To mitigate the bottlenecks, to the best of our knowledge, \sysname{} is the first work to (1) introduce an end-to-end approximate intermediate caching design for various time ranges and statistics in timeseries monitoring, (2) propose the combination of Exponential Histogram and different types of sketches (e.g., KLL and Universal Sketching) to support various time windows and query statistics, and (3) analytically prove the guarantees of such constructions. (\S\ref{sec:design})
\item We provide ready-to-plugin \sysname{} to both single-machine and distributed systems with cloud-native architecture (\S\ref{sec:impl}), and show its benefits on real-world and synthetic datasets over baseline systems (\S\ref{sec:eval}).
\end{packeditemize} }

%% file: Background.tex
\section{Background and Motivation}\label{sec:background}
In this section, we introduce background of timeseries monitoring systems, present motivating scenarios, and discuss the limitations of existing monitoring systems and new design opportunities.

\begin{figure}[t]
\centering
\includegraphics[width=0.8\linewidth]{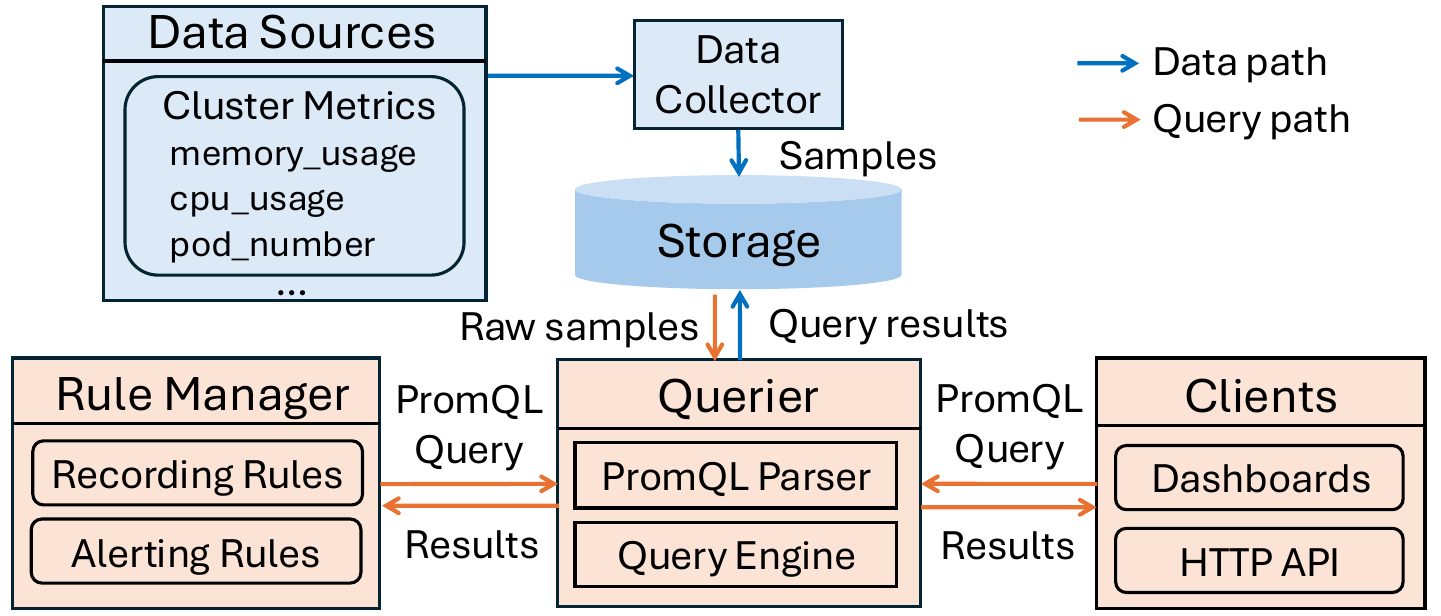}
\vspace{-2mm}
\mycaption{fig:monitoring_architecture}{The architecture of a  timeseries monitoring system}{\footnotesize }
\vspace{-2mm}
\end{figure}
 
\subsection{Timeseries Monitoring Systems}

A monitoring system usually collects, stores, and queries timeseries data from various sources. Fig.~\ref{fig:monitoring_architecture} shows a typical timeseries monitor architecture. {\em Data collectors} scrape metrics and send them to storage. The {\em storage engine} appends new data to the timeseries without modifying previous data. Users can issue queries in PromQL~\cite{promql} via various clients, including rule queries for periodic monitoring and alerting. The {\em query engine} retrieves data samples from storage and computes results based on query expressions. The rule query results can be stored for reuse.

\noindent\textbf{Data Model.} Timeseries data are streams of timestamped values belonging to the same data source and the same set of labeled tags. They span over two dimensions: (1) the time dimension, which consists of data samples each associated with a timestamp belonging to one timeseries; and (2) the label dimension, which consists of samples from many different data sources and label tags at a given timestamp. A data sample can be represented by $\rho = (l, t, v)$, where $l=(d_1, d_2, ..., d_m)$ contains $m$ label dimensions, $t$ is the timestamp, and $v$ is the data value, either a 64-bit floating point (e.g., CPU usage) or string (e.g., IP address).  
An example timeseries of \texttt{cpu\_usage} metric recording the CPU usage of each node and each core can be represented as \texttt{cpu\_usage\{node\_id=``node0'',cpu\_id=``0''\}},  specified with labels of node ID and CPU core ID.

\begin{figure}[t]
\centering
\includegraphics[width=0.85\linewidth]{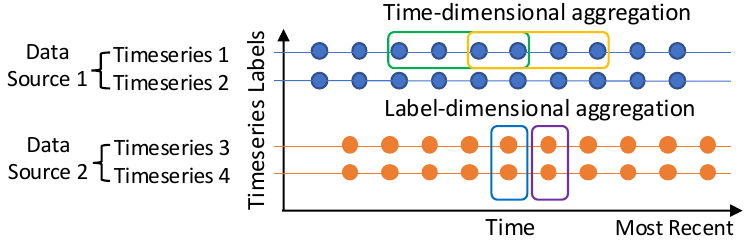}
\vspace{-3mm}
\mycaption{fig:query_types}{Timeseries data  and different types of queries \change{in Prometheus}}{\footnotesize Each row is a timeseries identified by its labels. 
}
\vspace{-3mm}
\end{figure}

\noindent\textbf{Queries. } Monitoring systems like Prometheus offer various queries for \change{downstream applications}, especially alerting and recording rules~\cite{prometheus-rules}. Users can define {\em rules} to automatically execute periodic monitoring queries and track alerts~\cite{instant-query-grafana}. 
A rule query mainly consists of three parts as below: 1) a rule type, either recording rule (which stores results for future use), or alerting rule (which sends alerts based on conditions); 2) a rule evaluation interval $T_{eval}$, defining the evaluation frequency; and 3) a query expression, with an optional alert condition for triggering alerts in alerting rules. 
\begin{python}
  rule:
  - type: record|alert
  - evaluation_interval: T_eval
  - expr: PromQL expression, [alert condition]
\end{python}

Formally, a query expression can be defined as 
$Q_R= \{q(\rho), \rho: t_{cur} - T_{q} \leq t \leq t_{cur} \bigwedge d_{i_1} = x_{i_1} \bigwedge \dots \bigwedge d_{i_m} = x_{i_m}\}$ , with query function $q$ on a set of timeseries data samples for $T_{q}$ query window looking back from current time $t_{cur}$, and $d_{i_1}, \dots, d_{i_m}$ a subset of the label dimensions to condition on. The query function $q$ can be an aggregation over the time dimension or label dimensions at a given timestamp. Examples of timeseries data samples and rule queries are shown in Fig.~\ref{fig:query_types}. 
Aggregation queries, such as quantiles, Top-K, entropy, and cardinality, are important for understanding statistics without focusing on single values in timeseries and are more efficient than querying raw data samples. Summarizing the data over time with aggregation query functions and periodic rule queries, essentially, it forms a sliding window model. In this paper, we aim to optimize the time-dimensional aggregation queries.

\subsection{Motivating Scenarios}

\noindent\textbf{Network Flow Monitoring. } DDoS attacks can occur when attackers use TCP SYN floods to exhaust bandwidth or server resources via a botnet~\cite{antonakakis2017understanding, marzano2018evolution}. Victims can detect ongoing attacks by monitoring the volume and entropy of SYN packets from multiple source IPs targeting a single destination~\cite{1378371}. 
For example, operators can track DDoS indicators for a virtual machine using alerting rules with a 5-second evaluation interval as follows~\cite{liu2021jaqen}. Detection queries can identify the start and the end of a DDoS attack by monitoring flow changes and comparing metrics against alert thresholds, requiring frequent queries across various time windows (e.g., 10s, 5s
) due to the uncertainty of optimal window sizes. In this case, each target server may receive millions of packets per second~\cite{ec2-pps}, requiring the time windows of being over 100K to 1 million data samples.

\begin{python}
 rules:
 - evaluation_interval: 5s
 - type: alert
 - expr: entropy_over_time(src_ip{vm="instance1"}[10s]) 
         > entropy_threshold 
 - expr: distinct_over_time(src_ip{vm="instance1"}[10s])
         > volume_threshold
 - expr: distinct_over_time(src_ip{vm="instance1"}[5s]) 
         > volume_threshold
\end{python}

\noindent\textbf{Cloud Resource Scaling.} Cloud-native platforms autoscale resources, such as pods, to reduce costs~\cite{google-scaling} based on aggregated statistical queries (e.g., averages, quantiles) over time windows for metrics like CPU, memory, and pod counts from monitoring tools~\cite{prom-autoscale}. For instance, recording rules can query each container's 0.95-quantiles for memory and CPU usage, and average pod counts over the past 5 minutes, storing the results for quick retrieval by the cloud resource scheduler and downstream applications as below. Standard Google Cloud clusters can have up to 256 pods per node and up to 100 nodes per cluster~\cite{gke-quotas}. Thus, cluster-level monitoring  can easily result in 100K or more timeseries to query on.

\begin{python}
rules:
- evaluation_interval: 1m
- type: record
- expr: quantile_over_time(0.95, 
         container_memory{dimension="used"}[5m])
- expr: quantile_over_time(0.95, 
         container_cpu{dimension="used"}[5m])
- expr: avg_over_time(pod_number[5m])
\end{python}

\change{In summary, rule query use cases involve monitoring queries that repeatedly query the same metrics over time with varying window sizes, various statistical functions, and the ability to handle large data volumes, while being sensitive to query latency, providing critical observability to anomaly detection~\cite{turnbull2018monitoring, mart2020observability, priovolos2021using, huang2017gray}, security checking~\cite{sides2015yo, antonakakis2017understanding}, and cloud performance monitoring~\cite{david2021kubernetes, turnbull2018monitoring}. }

\subsection{Operational Cost and Bottleneck Analysis} \label{sec:bottlenecks}

We start by comparing the operational costs of two representative systems in Table~\ref{table:dollar_cost}, monitoring a 1000-node Kubernetes cluster with each node having 1000 metrics, storing 268 billion data samples per month. 10 concurrent rule queries run every minute and each query processes 8 billion samples. 
Cost estimates follow AWS Prometheus Pricing~\cite{aws-price}, which charges by storage and samples processed, and a typical cloud billing model used by VictoriaMetrics~\cite{ec2-billing, vm-price}, which charges based on resource usage such as memory, vCPUs. We defer detailed analysis in \S\ref{sec:eval}. Query processing comprises 55\% of the total costs in Prometheus and 95\% of in VictoriaMetrics.

We analyze bottlenecks in Prometheus and VictoriaMetrics to identify sources of high query costs. Using Golang pprof and the testbed in \S~\ref{sec:eval}, we profile recording rule queries with extended time windows. For example, we test with a 10,000-second query window, 100ms sample interval, and 1s evaluation interval, benchmarking the \texttt{quantile\_over\_time(0.99,metric[10000s])} query. Table~\ref{table:prometheus_cpu_profiling} shows the CPU profiling results and the top two bottlenecks.

\begin{table}[t]
\centering
\mycaption{table:dollar_cost}{Comparison of operational cost between several systems}{\footnotesize ``PS-PM'' and ``PS-VM'' refer to Prometheus- and VictoriaMetrics-based integrations of \sysname{}.}
\vspace{-3mm}
\resizebox{1\columnwidth}{!}{
\begin{tabular}{@{}l|l|l|l|l@{}}
\toprule
\textbf{Costs} & \textbf{AWS Prometheus~\cite{aws-price}}      & \textbf{PS-PM}  & \textbf{VictoriaMetrics~\cite{vm-price}}  & \textbf{PS-VM}  \\
\midrule
\textbf{Storage} & \$70, \$0.03/GB-Month  & \$70 & \$348 & \$348 \\ 
\textbf{Data Ingestion} &  \$9186, $\sim$\$0.35/10M samples &\$9186 & incl.  & incl. \\ 
\textbf{Query Processing} &  \$11520, \$0.1/B samples  & \$28.6  & $\geq \$7443$ & $\leq \$1833$ \\
\midrule
\textbf{Total Costs}& \$20776  & \$9284.6  & $\geq \$7791$ & $\leq \$2181$  \\
\bottomrule
\end{tabular}
}
\end{table}

\begin{table}[t]
\centering
\mycaption{table:prometheus_cpu_profiling}{CPU hotspots of evaluating a quantile rule query in Prometheus and VictoriaMetrics}{}
\vspace{-3mm}
\resizebox{0.98\columnwidth}{!}{
\begin{tabular}{@{}c|c|c|c@{}}
\toprule
\multirow{ 2}{*}{\textbf{Func/Call Stack}}  & \multicolumn{2}{c|}{\textbf{CPU Time}}  & \multirow{ 2}{*}{\textbf{Description}}\\
\cmidrule(rl){2-3}
  &\textbf{Prometheus} & \textbf{VictoriaMetrics}      \\
\midrule
\textbf{Data Scanning} & {\bf 41\%}  & {\bf 80.2\%}   & Fetch data from storage\\
\textbf{Query computation} & {\bf 27.6\%} & {\bf 11.7\%}   & Aggregation queries in rule\\
\textbf{Go Garbage Collector}  & 24.7\% & 4.3\%   & Golang garbage collector\\
\textbf{mcall} & 4.5\% & 0.8\%  & Golang runtime scheduling \\
\bottomrule
\end{tabular}
}
\end{table}

\noindent\textbf{Bottleneck 1: { Repeated data scans from storage.}} Data scanning from storage accounts for over 40\% of CPU time, marking the primary bottleneck. This is due to repeated scans of data, even when query windows overlap in rule queries or when there are concurrent queries from multiple users.

\noindent\textbf{Bottleneck 2: { Repeated query computations. }} The second major bottleneck is quantile query calculation. In both VictoriaMetrics and Prometheus, periodic rule queries are computed independently rather than as sliding window queries. They re-execute the entire query computation for overlapping portions, without leveraging intermediate results from previous overlapping windows.

\subsection{Prior Work and Limitations}

\noindent\textbf{Exact monitoring systems.} Prior work reduces timeseries query latency and costs via three categories. The first enhances storage engines with better indexing (e.g., InfluxDB~\cite{InfluxDB}, VictoriaMetrics~\cite{VictoriaMetrics}), optimized storage schemas (e.g., Heracles~\cite{wang2021heracles}), and improved compression techniques (e.g., Gorilla~\cite{pelkonen2015gorilla}).  These methods reduce storage costs and retrieval latency but don't address computational bottlenecks from repeated data scans and overlapping windows.

The second improves query performance by utilizing parallel query processing, query sharding, and precomputation. Parallel processing (e.g., VictoriaMetrics~\cite{vm-parallel-query}) distributes the query computation across CPU cores, splitting tasks by time series. Query sharding (e.g., Mimir~\cite{mimir}, Thanos~\cite{thanos}) reduces memory usage by partitioning the query by time range or timeseries and processing each partition sequentially. While both methods can reduce query latency through hardware parallelism or reduced memory impact from Go garbage collection, parallel processing does not lower overall query costs, and query sharding, while reducing memory costs, still maintains redundant computational overheads across queries.
Precomputation, e.g., LindormTSDB~\cite{shen2023lindorm},  computes predefined statistics for fixed time intervals during data ingestion.
While this reduces operational costs and query latency by reducing query redundancy from window overlapping, they support only basic statistics (e.g., sum, max) and set fixed intervals in advance.

The third employs key-value caches (e.g., fastcache~\cite{fastcache} in VictoriaMetrics, Memcached~\cite{memcached} or Redis~\cite{redis.io} in Grafana Mimir~\cite{mimir, mimir-redis}) to accelerate queries, including metadata-cache, index-cache, chunk-cache, and result-cache~\cite{mimir-redis}. Metadata and index caches accelerate timeseries searches by mapping metrics to database indexes but don't remove repeated data retrieval or computational overhead. Chunk-cache stores data in memory, reducing disk retrieval time but is limited by memory capacity and doesn't address repeated computations. Result caches store query results, but frequent changes in query statistics and time ranges limit cache reuse.

\noindent\textbf{Approximate Query Processing (AQP).}  In monitoring systems, approximate results are often sufficient for downstream applications~\cite{liu2016one, ivkin2019know, liu2021jaqen, papadopoulos2016peas}, offering the chance to trade off minor accuracy for lower query latency and operational costs~\cite{manousis2022enabling}, using sampling or data summarization for time-window queries.

{\em Sampling} for aggregation queries has been widely explored in Approximate Query Processing (AQP) by pre-processing data samples for query-time use~\cite{park2018verdictdb, peng2018aqp++, lu2010simple, agarwal2013blinkdb}. Monitoring systems like Thanos~\cite{thanos-downsampling} apply downsampling to reduce data retrieval and computation costs. While sampling-based frameworks offer broad applicability across various statistics and support the sliding window model~\cite{lu2010simple}, their accuracy guarantees weaken for complex statistics (e.g., quantiles~\cite{karnin2016optimal}) and suffer from larger errors when zooming into small sub-windows with a low fixed sampling rate, due to limited sample availability.

{\em Sketch}-based analytics offer bounded accuracy-memory trade-offs in sub-linear space~\cite{greenwald2001space, masson2019ddsketch, karnin2016optimal, liu2016one}, creating compact summaries during ingestion and estimating statistics with provable error bounds. 
Sliding window sketches are often designed for specific query types, maintaining summaries for the entire window, such as sliding sum~\cite{ben2019succinct}, 0-1 counting~\cite{datar2002maintaining}, heavy hitter detection~\cite{ben2016heavy, basat2018memento}, distinct counting~\cite{chabchoub2010sliding}, and sliding quantiles~\cite{arasu2004approximate}. While implementing each individually supports diverse queries, it introduces per-statistic effort and lacks sub-window query support within the recent window, leading to additional maintenance overhead.  Recent approaches~\cite{ivkin2019know} extend fixed sliding window frameworks~\cite{datar2002maintaining, braverman2007smooth} to support arbitrary sub-windows and accommodate various sketch types as subroutines, making them well-suited for periodic rule queries with varying window sizes and statistical requirements.

\noindent\textbf{Summary and Opportunities. } Existing solutions fall short in the tradeoffs among operational costs, query latency, and accuracy.  
Our analysis reveals a key optimization opportunity in removing query redundancy due to overlapping windows. Since periodic rule queries often share overlaps, caching appears as an effective approach to reduce redundant data scans and query computations. However, caching all raw data samples is not a scalable choice and does not reduce the computational costs from window overlaps. Moreover, caching some final results is an ad-hoc choice to only optimize a few predefined queries. Thus, caching intermediate results that are precomputed and flexible enough to query a wide range of windows becomes a well-informed choice.

%% file: Overview.tex
\section{\textmd{\sysname{}}: System Overview} \label{sec:overview}

\noindent\textbf{\sysname{} Architecture.} We illustrate the system components in Fig.~\ref{fig:workflow}. \sysname{} maintains an in-memory approximate cache. Data samples are ingested into both backend storage and the cache by the data ingester. \sysname{} precomputes intermediate results for the most recent windows of timeseries selected by rule queries. When a rule query is issued, the querier first checks the cache for the required time range and statistics. If found, the query retrieves estimated results from \sysname{} with reduced latency; otherwise, it falls back to the original TSDB query engine to scan raw data and compute exact statistics. The final query results, whether from \sysname{} or the exact engine, are then returned to users.

\noindent\textbf{Challenges and Key Ideas.} To realize the vision of \sysname{}, we address several key design challenges:
 
\noindent\textbf{Challenge 1: Caching many recent query windows and results. } Query statistics and functions are various in real use cases. Caching all samples ensures generality but is memory-intensive, while caching only final results limits optimization for unforeseen drill-down queries. A raw data cache also fails to reduce redundancy from overlapping query windows.

\noindent\textbf{Key Idea.} We extend the window-based approximate query framework (e.g., Exponential Histogram~\cite{datar2002maintaining, ivkin2019know}) to be sub-window-capable as a flexible intermediate query cache along the time dimension for each timeseries. The cache stores intermediate results for many sub-windows within a large recent time window, allowing reuse for overlapping portions of query windows (e.g., one can query 5-, 10-, 15-min windows within a cached 30-min window without recalculating from scratch). It can support multiple and arbitrary sub-windows and different query functions through using different internal data structures to maintain intermediate results.

\noindent\textbf{Challenge 2: Caching a large number of timeseries. } Number of active timeseries that need to be monitored can be large~\cite{shen2023lindorm}, requiring caching as many timeseries as possible within a memory budget.  While exact data structures in window-based frameworks to store the intermediate results offers high accuracy, it needs a large amount of memory as exact query processing and cannot scale to a large number of timeseries. 
    
\noindent\textbf{Key Idea.} To reduce memory usage, we integrate approximate methods, such as sketches and sampling, to work as compact and low-latency intermediate data summarizations in the framework. For instance, we propose proper combinations of Exponential Histogram and KLL, and formally establish space-error bounds.

\noindent\textbf{Challenge 3: Efficient caching of various query statistics. } Users often query different statistics over the same timeseries, such as distinct counting and entropy of source IPs for DDoS attack detection. To support as many query statistics as existing exact monitoring systems, a strawman solution is to cache each statistic with a separate sketch instance for each timeseries. However, this approach introduces per-statistic efforts and large memory costs.
    
\noindent\textbf{Key Idea.} To avoid per-statistic effort, recent advances in universal sketching~\cite{braverman2010zero, liu2016one}  allows a single sketch instance to support multiple target query functions, such as $L_0$, $L_1$, $L_2$ norms and entropy, instead of requiring a separate sketch for each function. We combine universal sketching with EH to support multiple statistics simultaneously~\cite{ivkin2019know}, and proposes a novel optimization that combines exact maps and universal sketching as EH buckets, reducing memory footprint while improving accuracy.

\begin{figure}[t]
\centering
\includegraphics[width=0.96\linewidth]{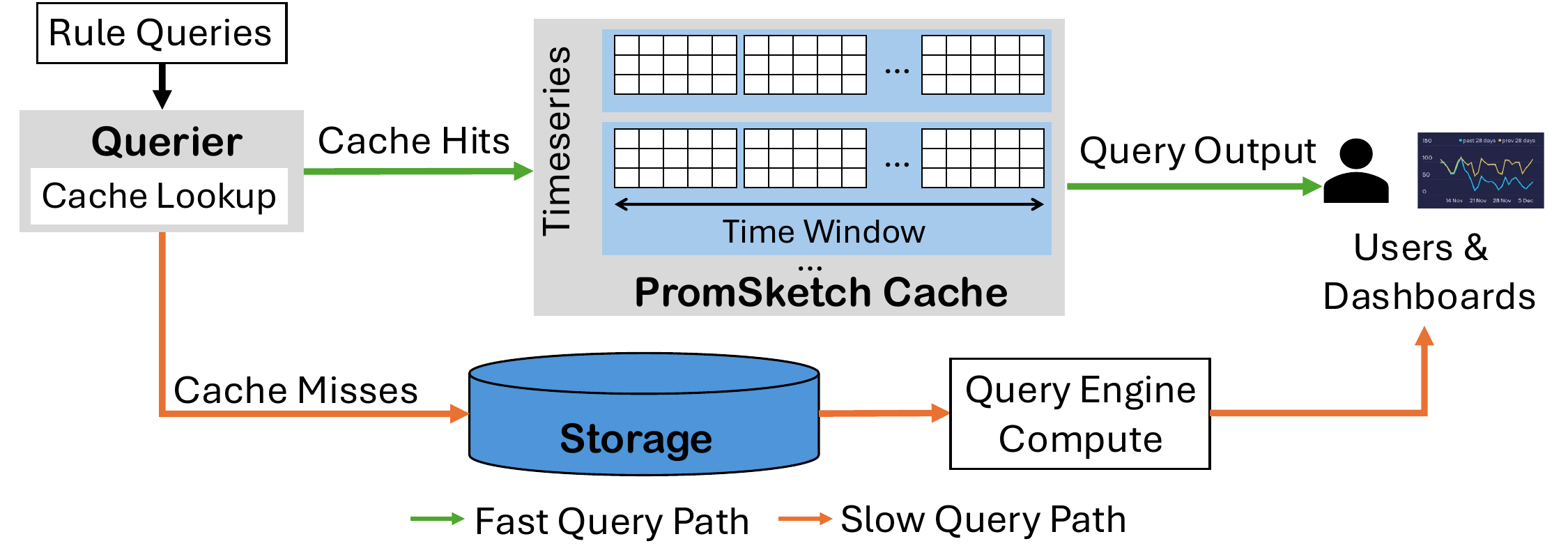}
\vspace{-3mm}
\mycaption{fig:workflow}{\sysname{} Architecture}{\footnotesize  
} 
\end{figure}

%% file: Design.tex
\section{\sysname{} Detailed Design} \label{sec:design}
We introduce sub-window query frameworks as \sysname{} cache, present its algorithmic building blocks, with provable accuracy-space bounds and detailed system design.

\subsection{Window-based Frameworks as a Cache} \label{sec:sliding_window}

Periodic time interval aggregation queries, such as Alerting rules and Recording rules, are essentially {\em sliding window} queries along the time. 
These queries maintain statistics of the most recent $T$ time window $W=(t-T,t)$. Users can also query any statistics over a sub-window $(t_1, t_2) \subseteq W$ for zoom-in diagnosis of applications such as anomaly localization. To cache as many query windows as possible within limited memory budgets, approximate window-based frameworks that maintain sliding windows and sub-window structures are viable options. Currently, there are two general approximate window-based frameworks providing $o(N)$ memory with good estimations for a recent window $W$ of $N$ items: Exponential Histogram (EH)~\cite{datar2002maintaining} and Smooth Histogram (SH)~\cite{braverman2007smooth}. Intuitively, Exponential Histogram maintains non-overlapping buckets whose bucket sizes are exponentially growing when buckets are older; Smooth Histogram maintains overlapped buckets that covering time ranges with different start time points, as Fig.~\ref{fig:eh_and_sh} shows.

\begin{figure}[t]
\centering
\includegraphics[width=0.95\linewidth]{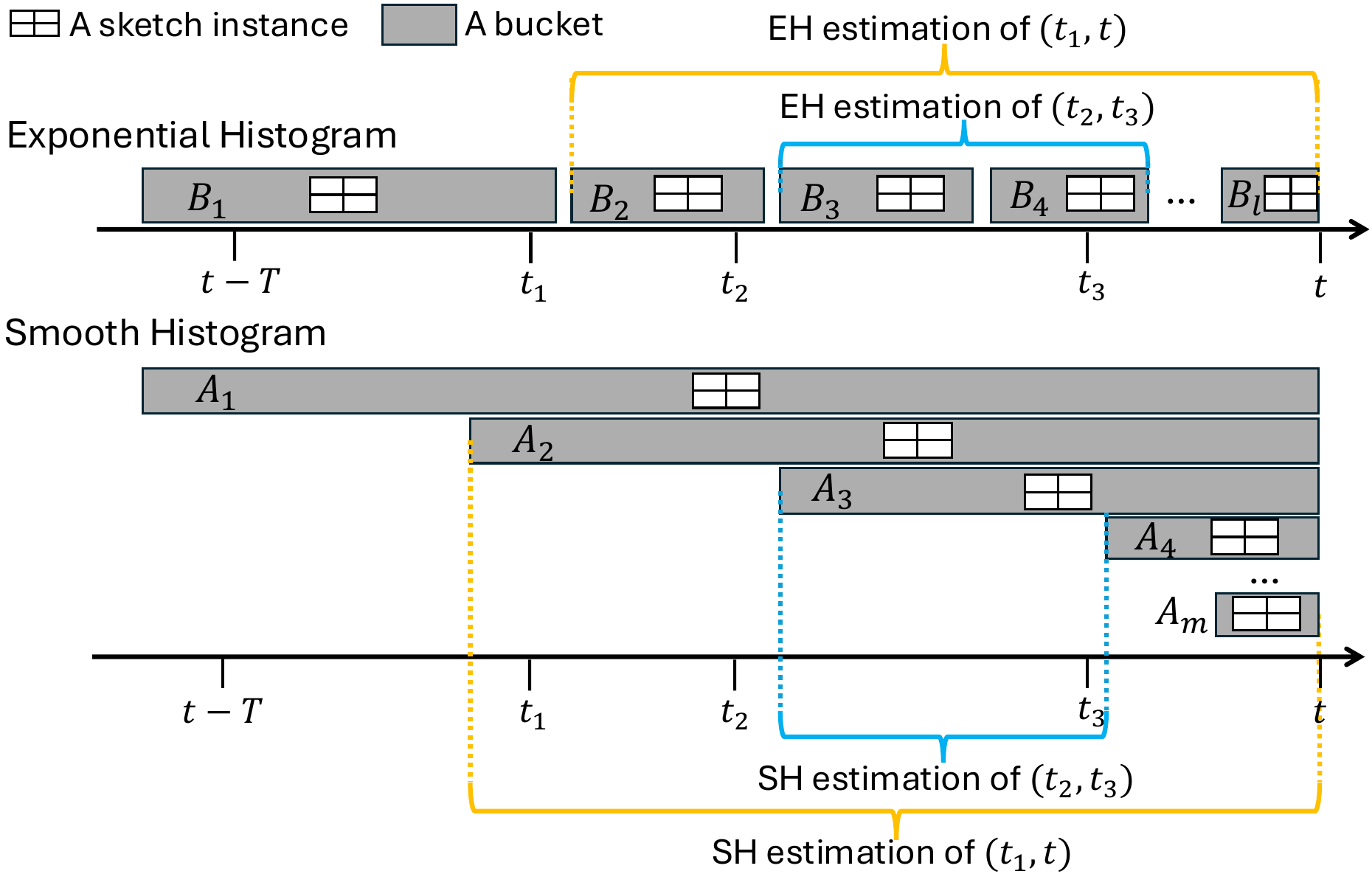}
\vspace{-2mm}
\mycaption{fig:eh_and_sh}{Exponential Histogram (EH)~\cite{datar2002maintaining} and Smooth Histogram (SH)~\cite{braverman2007smooth} structures and window-based queries}{\footnotesize $(t_1, t)$ and $(t_2, t_3)$ are sub-window queries within the most recent $T$ time window and current time $t$. 
}
\end{figure}

\noindent\textbf{Exponential Histograms}~\cite{datar2002maintaining} suggests to break the most recent window $W=(t-T, t)$ into a sequence of $l$ non-overlapping intervals (buckets) $B_1, B_2, \dots, B_l$. Window $W$ is covered by $\bigcup_{i=1}^l B_i$, and contains all $B_i$ except $B_1$. Then, if a target function $f$ admits a composable sketch, maintaining such a sketch on each bucket can provide us with an estimator for $f$ on a window $W'=\bigcup_{i=2}^l B_i$. $f(W)$ is sandwiched between $f(W')$ and $f(B_1 \cup W')$. Therefore, a careful choice of each bucket endpoints provides control over the difference between $f(W)$ and $f(W')$. When the window slides, new buckets are introduced, expired buckets are deleted, and buckets in between are merged. The EH approach admits non-negative, polynomially bounded functions $f$ which in turn enable a composable sketch and are weakly additive, i.e., $\exists C_f \geq 1$, such that $\forall S_1, S_2$:
\setlength{\belowdisplayskip}{3pt} \setlength{\belowdisplayshortskip}{3pt}
\setlength{\abovedisplayskip}{3pt} \setlength{\abovedisplayshortskip}{3pt}
\begin{equation}\label{eq:eh}
    f(S_1) + f(S_2) \leq f(S_1 \cup S_2) \leq C_f (f(S_1) + f(S_2)).
\end{equation}

We show the intuition of querying a sub-window by the following example: $q=(t_2, t_3)$ as depicted in Fig.~\ref{fig:eh_and_sh}. In the example, $q$ is sandwiched between $ B_2 \cup B_3 \cup B_4$ and $B_3$, where $f(\bigcup_{j=2}^l) = (1\pm \varepsilon) f(t_2, t)$ and $f(\bigcup_{j=4}^l) = (1\pm \varepsilon) f(t_3, t)$. Intuitively, one can expect that $f(t_2, t_3)$ can be approximated by $f(B_3 \cup B_4)$ with an additive error of $\pm \varepsilon f(t_2, t)$\change{, related to the suffix $(t_2, t)$}.

\noindent\textbf{Smooth Histograms}~\cite{braverman2007smooth} buckets $A_1, \dots, A_m$ overlap. An example sub-window query $q=(t_2, t_3)$ is sandwiched between $f(A_2)$ and $f(A_4)$ and can be approximated by $f(A_3 - A_4)$ with SH buckets, if the sketches preserve approximation upon subtraction. 

We choose Exponential Histogram (EH) over Smooth Histogram (SH) in \sysname{} for two main reasons.  \change{First, SH requires subtractive properties between sketches while EH requires only additive mergeability, which most sketches support, allowing us to analyze error bounds for more window/sketch combinations.}
Second, for potentially large-scale data ingestion, EH can offer better system performance. \change{Specifically, when inserting an item, EH only needs inserting into the newest (and smallest) bucket, with an amortized $O(1)$ insertion cost~\cite{datar2002maintaining}, while SH inserts the item into every active bucket, resulting in an $O(\log N)$ insertion cost~\cite{braverman2007smooth}.} Additionally, EH typically has smaller buckets than those of SH because they represent non-overlapping sub-windows, and thus require smaller inner data structure allocations.

\subsection{Algorithmic Building Blocks}\label{sec:building_blocks}
\change{Next, we introduce algorithmic building blocks of \sysname{}, novelly combining EH window framework and configurable sketches with provable error guarantees. \sysname{} supports 70\% of queries in Prometheus as in Table~\ref{table:supported_queries}.}

\begin{table}[t]
\centering
\mycaption{table:supported_queries}{Example Prometheus aggregation-over-time queries supported by \sysname{}}{\footnotesize \sysname{} can support 70\% existing aggregation over time queries in Prometheus and introduces capabilities for currently unsupported queries. \texttt{count\_over\_time} can be supported by multiple algorithms. }
\vspace{-3mm}
\resizebox{1\columnwidth}{!}{
\begin{tabular}{@{}l|lll@{}}
\toprule
{EHKLL (Algo~\ref{alg:eh_quantile})} & \texttt{quantile\_over\_time} & \texttt{min\_over\_time} & \texttt{max\_over\_time} \\
\midrule
{EHUniv (Algo~\ref{alg:eh_univ})} & \texttt{count\_over\_time} & \texttt{entropy\_over\_time} & \texttt{l2\_over\_time} \\
& \texttt{distinct\_over\_time} & \change{\texttt{topk\_over\_time}}\\
\midrule
{Uniform Sampling} & \texttt{count\_over\_time} &\texttt{sum\_over\_time} &  \texttt{avg\_over\_time} \\ 
 & \texttt{stddev\_over\_time} & \texttt{stdvar\_over\_time} & \\
\bottomrule
\end{tabular}
}
\end{table}

\begin{algorithm}[t]
\begin{algorithmic}[1]
 \State \textbf{Input}: EH item count error $\epsilon_{EH}$, KLL rank error $\epsilon_{KLL}$, confidence level $\delta$, time window range $T$  
 
\Function{Update}{$t$, item}
\State Maintain EH with $k_{EH}=O(\frac{1}{\epsilon_{EH}})$ based on Invariant~\ref{invar:eh_count} and Invariant~\ref{invar:eh_bucket}.
\State Each bucket $B_i$ maintains a KLL sketch $KLL_i$ with $\epsilon_{KLL}$.
\EndFunction
\Function{Query}{$t_1, t_2, \phi$}
\State Find $B_i=(b^0, b^1)$ and $B_j=(b^2, b^3)$ s.t.: $t_1\in B_i$ and $t_2 \in B_j$
\State Compute the merged sketch $KLL_{merge}=\bigcup_{i+1 \leq r \leq j} KLL_r$
\State \Return $\phi$-quantile: $x_q=KLL_{merge}.query(\phi)$
\EndFunction
\end{algorithmic}
\caption{\label{alg:eh_quantile} EHKLL: Quantiles Based on EH}
\end{algorithm}

\subsubsection{EH+KLL for Quantiles (EHKLL).}\label{sec:ehkll}  Quantile-based rule queries such as \texttt{quantile\_over\_time}, \texttt{min\_over\_time}, \texttt{max\_over\_time}, consist of querying data samples over a time range and a $\varphi \in [0,1]$, representing $\varphi$-quantile (e.g., \texttt{min} and \texttt{max} correspond to the 0-quantile and 1-quantile). \change{We present a novel  construction for arbitrary sub-window quantiles, using an EH with each bucket as a KLL sketch ~\cite{karnin2016optimal} to maintain quantiles, as shown in  Alg.~\ref{alg:eh_quantile} (EHKLL). The user specifies the KLL rank error $\epsilon_{KLL}$, EH error $\epsilon_{EH}$, confidence level $\delta$, and size (in time range $T$ or data count $N$) of the most recent window.   When inserting a data sample, it is added into the latest bucket $B_l$'s KLL sketch. If needed, buckets merge~\cite{datar2002maintaining} to maintain EH invariants based on the number of items in each EH bucket. For a query with time range $T=[t_1, t_2]$, we identify two buckets that contain $t_1$ and $t_2$, merge the KLL sketches between the two buckets based on EH to construct a merged sketch. }

\noindent\textbf{Feasible Quantile Sketches with EH. } \change{We integrate KLL sketch~\cite{karnin2016optimal} as the quantile estimator because of feasibility to aggregate the estimation errors from both EH window framework and each bucket's sketch consistently. At a high level, we can choose quantile sketches between two types -- one provides rank error guarantees, and the other provides relative error guarantees. } A rank error  $\epsilon_{rank}$ approximate quantile sketch receives items $x_1, x_2, \dots, x_n$, and allows one to approximate the rank of any query item up to additive error $\epsilon_{rank} n$ with probability at least $1-\delta$.  The rank of a query $x$ is the number of items in the queried window such that $x_i \leq x$.  Given a $\varphi$-quantile query $x_{\varphi}$, a relative error $\epsilon_{rel}$ approximate quantile sketch outputs $\Tilde{x}_{\varphi}$ such that $|\Tilde{x}_{\varphi} - x_{\varphi}| \leq \epsilon_{rel} x_{\varphi}$.  
Since an EH maintains buckets with rank error guarantees based on Invariant~\ref{invar:eh_count} and Invariant~\ref{invar:eh_bucket} as below (\cite{datar2002maintaining}) and KLL is a representative sketch with rank errors, 
we explore the novel combination of EH+KLL and analyze the aggregated rank error bounds. 
\setlength{\topsep}{0pt} 
\setlength{\partopsep}{0pt} 
\begin{invariant} \label{invar:eh_count}
Define $k_{EH}=\frac{1}{\epsilon_{EH}}$ and assume $\frac{k_{EH}}{2}$ is an integer; otherwise, we replace $\frac{k_{EH}}{2}$ by $\lceil \frac{k_{EH}}{2} \rceil$. At all times, the bucket sizes $C_1, \dots, C_l$ satisfy $\frac{C_j}{2(1 + \sum_{i=j+1}^{l}C_i)} \leq \frac{1}{k_{EH}}$, for all $1 \leq j \leq l$.
\end{invariant}
 
\begin{invariant} \label{invar:eh_bucket}
 At all times, the bucket sizes are nondecresing, i.e., $C_1 \leq \dots \leq C_{l-1} \leq C_l$. Further, the bucket sizes are constrained to the following: $\{1, 2, 4, \dots, 2^{l'}\}$ for some $l' \leq l$ and $l' \leq \log \frac{2N}{k_{EH}} + 1$. For every bucket size other than the size of the last bucket, there are at most $\frac{k_{EH}}{2} + 1$ and at least $\frac{k_{EH}}{2}$ buckets of that size.
\end{invariant}

\noindent\textbf{EHKLL Error Guarantee. } We prove the error bound as follows. First considering queries over the entire sliding window, the oldest bucket with size $C_1$ that we discard for quantile estimation can contribute at most $C_1$ rank difference from the accurate answer. The newest bucket $C_l$ exactly aligns with the sliding window query boundary and introduces no errors. Therefore, based on Invariant~\ref{invar:eh_count}~\cite{datar2002maintaining}, the rank error caused by window framework EH is at most $\frac{2}{k_{EH}}$. Assuming a KLL sketch has $\epsilon_{KLL}$ rank error of the quantile estimation after taking all $N$ items, given its mergeability, the final estimated rank error is $\epsilon_{EHKLL} \leq 2\epsilon_{EH} + \epsilon_{KLL}$.
Alg.~\ref{alg:eh_quantile} can be used to query all quantiles including $min$ ($\phi = 0$) and $max$ ($\phi = 1$). Based on \cite{karnin2016optimal}, a KLL sketch needs $O(\frac{1}{\epsilon_{KLL}}\log^2\log(1/\delta\epsilon_{KLL}))$  bits of memory for estimations of all quantile. There are on the order of $O(\frac{1}{\epsilon_{EH}} \log N)$ EH buckets if we maintain EH buckets based on the number of samples inserted to each bucket. Therefore, the total memory needed by Alg.~\ref{alg:eh_quantile} is $O(\frac{1}{\epsilon_{KLL}}\log^2\log(1/\delta\epsilon_{KLL}) \cdot \frac{1}{\epsilon_{EH}} \log N)$, which gives at most $2\epsilon_{EH} + \epsilon_{KLL}$ normalized rank error for sliding window queries with $N$ samples in the most recent window.

Next, we extend the error guarantee for arbitrary sub-window queries with EHKLL. The sub-window query is answered by merging buckets $i+1$ to $j$ in Alg.~\ref{alg:eh_quantile}. Similarly, the bucket $B_i$ discarded in calculation contributes at most $C_i$ rank difference; the bucket $B_j$ included in the calculation contributes at most $C_j$ rank difference. In total, the rank difference from EH sub-window query is at most $C_i - C_j$. If we provide rank error bound against $(t_1, t_2)$, the rank error from EHKLL is $\epsilon_{EHKLL} \leq \frac{C_i - C_j}{N_{t_1} - N_{t_2}} + \epsilon_{KLL} \leq \frac{N_{t_1}}{N_{t_1}-N_{t_2}} \frac{C_i-C_j}{(1+\sum_{r=i+1}^{l}C_r)} + \epsilon_{KLL} \leq 2{\epsilon_{EH}}\frac{N_{t_1}}{N_{t_1}-N_{t_2}} + \epsilon_{KLL}$, where $N_{t_i}$ refers to the number of samples between time $t_i$ to current time $t$. If we provide rank error bound against $(t_1, t)$, similarly, the rank error from EHKLL is $\epsilon_{EHKLL} \leq \frac{C_i-C_j}{(1+\sum_{l=i+1}^{m}C_l)} + \epsilon_{KLL} \leq 2\epsilon_{EH} + \epsilon_{KLL}$.

\subsubsection{Universal sketching with EH (EHUniv).} \label{sec:ehuniv}  Next, we support estimations of complex statistics that are not fully supported by current systems, such as $L_2$ norm, distinct counting, and entropy. These are often used together and queried on the same timeseries, but naive solutions require separate caching per statistic. For example, in DDoS attack detection, one would need a distinct counting sketch for cardinality and another for entropy, doubling memory usage. \change{Therefore, we leverage a universal sketch as a subroutine in EH to answer multiple statistics with one sketch, as shown in Alg.~\ref{alg:eh_univ} (EHUniv).  The update process inserts the item into the newest EH bucket while maintaining EH buckets according to EH invariants.
If EH invariants are violated for a pair of buckets $(B_{j-1}, B_{j})$, they are merged into a new bucket $B_{j-1}'$, whose universal sketch combines those of $B_{j-1}$ and $B_j$. The update process has an amortized merge time $O(1)$ and a worst-case merge time of $O(k_{EH} \log N)$, where $N$ is the item number in the most recent window. During queries, after finding the two buckets that contains the query window start time $t_1$ and end time $t_2$, it merges all buckets in between, including the last bucket but excluding the first bucket. Finally, the supported statistic can be answered by Recursive GSum algorithm~\cite{braverman2016streaming} and the merged universal sketch}.

\begin{algorithm}[t]
\begin{algorithmic}[1]
 \State \textbf{Input}: $L_2$ error target $\epsilon$, confidence level $\delta$, time window $T$ 
\Function{Update}{$t$, item}
\State Maintain EH for $L_2^2$  with $k_{EH} = O(1/\epsilon^2)$ based on Invariant~\ref{invar:eh_3} and Invariant~\ref{invar:eh_4}.
\State On each bucket $A_i$ maintain a universal sketch per bucket with error target $\epsilon$.
\EndFunction
\Function{Query}{$t_1, t_2$}
\State Find $B_i=(b^0, b^1)$ and $B_j=(b^2, b^3)$ s.t.: $t_1\in B_i$ and $t_2 \in B_j$
\State Compute the merged sketch $Univ_{merge}=\bigcup_{i+1 \leq r \leq j} Univ_r$
\State Query $gsum$ from $Univ_{merge}$ based on Recursive GSum algorithm~\cite{braverman2016streaming} (also Alg. 4 in \cite{ivkin2019know}) \label{line:ehuniv_merge}
\State \Return $gsum$
\EndFunction
\end{algorithmic}
\caption{\label{alg:eh_univ} EHUniv: GSum Based on EH}
\end{algorithm}

\noindent\textbf{EHUniv Benefits.} \change{The benefits of integrating universal sketch is the ability to maintain a single sketch for querying multiple statistics rather than creating separate sketches for each, and its natural mergeability. These statistical functions can be summarized as {\em GSum}, allowing a single \emph{universal} sketch instance to maintain multiple statistics~\cite{liu2016one, braverman2016streaming}.} The {\em GSum} function is defined as $G = \sum_{i=1}^m g(f_i)$, where $f_i$ is the frequency of data sample $\text{data}_i$ and $g: \mathbf{N} \rightarrow \mathbf{N}$ is a function.  \change{The class of {\em GSum} functions covers many practical monitoring functions, including $L_0$ (distinct counting), $L_1$ norms (\texttt{count\_over\_time}), $L_2$ norm, entropy\footnote{Entropy is measuring the diversity/uncertainty of the data in a timeseries, defined as $H \equiv - \sum_{i=1}^m \frac{f_i}{n} \log (\frac{f_i}{n})$~\cite{lall2006data,liu2016one}, where $n$ is the total item count in a window, $m$ is the number of distinct items.}, and TopK-frequent item finding}.
We provide the basics of universal sketching here and refer to \cite{braverman2016streaming,liu2016one} for more details. 
Theorem 2 in \cite{braverman2016streaming} states that if $g(x)$ grows slower than $x^2$, drops no faster than sub-polynomially, and has predictable local variability, then there is an algorithm that outputs an $\epsilon$-approximation to $G$, using sub-polynomial space and only one pass over
the data. For a universe with $M$ different items in a data stream, a universal sketch maintains $\log M$ parallel copies of a ``$L_2$-heavy hitter'' (L2-HH), e.g., using a Count Sketch~\cite{charikar2002finding} as the L2-HH subroutine. Then, leveraging a Recursive GSum algorithm~\cite{datar2002maintaining} (\change{also described in Alg. 4 of \cite{ivkin2019know}, we omit the details here}), a universal sketch estimates the statistical function $g$ by recursively computing $g$ on founded heavy hitters in $\log M$ layers of Count Sketches.  
\change{Following \cite{ivkin2019know}}, EH buckets can be maintained based on $L_2$ norm and thus can support L2-HH routines: \cite{ivkin2019know} improves EH~\cite{datar2002maintaining}'s results on $L_2$ that by maintaining $k_{EH}=O(\frac{1}{\epsilon^2})$ and $C_f=2$, EH can provide $\epsilon$-approximation for $L_2$ on the sliding window by maintaining Invariant~\ref{invar:eh_3} and \ref{invar:eh_4}, where $f(B_j) = L_2^2(B_j), j=1, \dots, l$.

\begingroup
\setlength{\abovedisplayskip}{1pt}
\setlength{\belowdisplayskip}{1pt}
\begin{invariant} \label{invar:eh_3}
$f(B_j) \leq \frac{C_f}{k_{EH}} \sum_{i=j+1}^l f(B_i) $.
\end{invariant}
\begin{invariant} \label{invar:eh_4}
$f(B_{j-1})+f(B_j) > \frac{1}{k_{EH}} \sum_{i=j+1}^l f(B_i)$.
\end{invariant}
\endgroup

\noindent\textbf{EHUniv Error Guarantee.} According to Theorem 7 in \cite{datar2002maintaining}, the EH approach required $O(k_{EH} s(\epsilon, \delta) \log N)$ bits of memory, where $s(\epsilon, \delta)$ is the amount of memory needed for a sketch to get a $(1+\epsilon)$-approximation with a failure probability of at most $\delta$. Theorem 3.5 in \cite{ivkin2019know} shows that a Count-Sketch-based $L_2$ heavy hitter algorithm based on EH with $k_{EH} = O(\frac{1}{\epsilon^2})$ can solve $(\epsilon, L_2)$-heavy hitters problem in the Sliding Window and Sub-Window query using $O(\epsilon^{-4}\log^3 N \log \delta^{-1})$ memory bits. As shown in \cite{ivkin2019know}, Recursive Sketch with $(g, \epsilon)$-heavy hitter algorithm which finds all $i$ such
that $g(f_i(t_1,t_2)) \geq \epsilon G(t_1,t)$ will return $\hat{G}(t_1,t_2) = G(t_1,t_2) \pm \epsilon G(t_1,t)$ and errors with probability at most 0.3, and with $O(\log M)$ space overhead. Therefore, using EH and buckets of universal sketches with Count Sketches for L2-HH subroutines, Alg.~\ref{alg:eh_univ} estimates GSum statistics using $O(\epsilon^{-4}\log^3 N \log M \log \delta^{-1})$ bits of memory.

\smallskip\noindent\textbf{\change{EHUniv Optimizations. }} Straightforward EHUniv implementation can incur large memory usage, as universal sketches in EH need to be configured with the same parameters and memory for mergability among buckets, where each sketch cannot be too small to guarantee good accuracy for a bucket. However, newer EH buckets maintained in the window are usually very small-sized (e.g., sizes $1, 2, 4, \dots$). To optimize EHUniv memory usage and runtime, we \change{propose to use} exact item frequency maps for smaller buckets (when sizes are below \change{the sketch memory}) and universal sketches for larger buckets.  The hybrid sketch/map construction reduces memory footprint and per-item update time, and improves accuracy because maps provide deterministic results for small buckets. When a map size exceeds the threshold, the map is converted into a universal sketch. Querying an interval among active buckets may access maps or sketches. If the time range includes only maps, we merge selected maps in the time range to calculate item frequencies and statistics. If it includes only sketches, we merge them and apply Recursive GSum to answer the query. When both maps and sketches are present, we merge the maps into one item frequency map, update the universal sketch with these frequencies, and combine all sketch buckets with the updated sketch to answer it.

\subsection{\change{Single-Machine \sysname{}}} \label{sec:single_machine}

\change{\sysname{}, as an intermediate result cache, can be applied to both single-node and distributed monitoring systems. We first introduce the end-to-end design for integrating it into a single-node system with the Promethues architecture.}

\noindent\textbf{\change{\sysname{} data ingester.}} \change{Data ingester inserts collected timeseries data samples into both the backend TSDB and corresponding \sysname{} cache instances in parallel.}

\noindent\textbf{\change{Rule manager.}} \change{Rule manager issues rule queries. When it initiates a rule query, it signals the query engine that a query is periodic and eligible for caching. The query engine then creates a \sysname{} instance. If rule configurations are updated and certain rules are removed, the corresponding \sysname{} instances are remoevd.}

\noindent\textbf{\change{\sysname{} query engine.}} \change{The query engine registers a \sysname{} cache instance when it first executes a rule query, with timeseries ID (or name), statistical function, and query window size. If multiple rule queries {\bf\em share the same timeseries and statistical function but have different window sizes}, the \sysname{} cache expands its window range to the largest query window for best possible caching. 
When evaluating an \texttt{aggregation\_over\_time} query, the query engine first checks whether the timeseries has been precomputed by \sysname{}. If available, it computes results using \sysname{}; otherwise, it retrieves raw data samples from the cache or storage to perform exact query computation. \sysname{} supports evaluating multiple timeseries sequentially (e.g., integrating with Prometheus) or in parallel across multiple cores (e.g., integrating with VictoriaMetrics).

\sysname{} is designed to be compatible with PromQL-like query languages, including those used by Prometheus, VictoriaMetrics, and more~\cite{thanos, mimir}, with \texttt{aggregation\_over\_time} functions.
To support the \sysname{} cache with the query engine, we extend the query parser to include an option for utilizing the \sysname{} cache at the entry point of the query's Abstract Syntax Tree, which is widely used to parse \texttt{aggregation\_over\_time} functions in PromQL. This preserves the original query syntax and allows outer functions to process results from the \sysname{} cache. 
For queries that first aggregate by timeseries label and then by time (e.g., {avg\_over\_time(max(metric)[10s])}), we initiate a \sysname{} instance with the inner aggregated timeseries (e.g., max(metric)) as input. 
Similarly, for queries that join timeseries before applying time-based aggregation, the joined timeseries samples are inserted into a \sysname{} instance.
In this work, we focus on optimizing aggregation over time functions, leaving optimizations for label-dimension aggregation to future work.}

\noindent\textbf{\sysname{} cache considerations.} \sysname{} uses a hash table for timeseries indexing as Prometheus for prototyping. For each timeseries, insertions and queries are performed concurrently. Inserting to a \sysname{} instance may require reconstruction of its EH buckets. Therefore, we add a Read-Write lock between query and insertion threads for each \sysname{} instance, allowing multiple concurrent reads to the buckets while permitting only one insertion at a time. \change{\sysname{} cache is maintained dynamically: If some rules are removed by the users, \sysname{} will remove cache instances that are no longer needed. Optionally, \sysname{} can also integrate VictoriaMetrics' timeseries index cache for accelerated sketch instance look-up. Moreover, \sysname{} has several reliability and data ordering considerations: (1) When a running \sysname{} fails, the in-memory cache can be rebuilt with old data from storage, with another \sysname{} instance accepting new data with current timestamps. In this case, queries are answered by merging two instances.  (2) \sysname{} has the same out-of-order data model as VictoriaMetrics~\cite{vm-backfilling} and Prometheus~\cite{prometheus-ood}, where only the data samples with current timestamp ranges are accepted and out-of-order/duplicated samples should be rejected. In practice, \sysname{} cache is placed after the deduplication and reordering component in VictoriaMetrics~\cite{vm-dedup}.}

\subsection{\change{Extension to Distributed \sysname}} \label{sec:distributed}
\change{To demonstrate a distributed system design, we integrate \sysname{} into a cluster-based VictoriaMetrics-like architecture~\cite{cluster-vm}), where all {\em ingester}, {\em cache}, {\em query}, and {\em storage} components can be running as container nodes as microservices. This cloud-native design brings several benefits --- The timeseries set is dynamically partitioned using consistent hashing~\cite{karger1997consistent} for possible scaling up and down, with each \sysname{} node independently maintaining its own data shard without sharing data with other nodes. This is different from VictoriaMetrics' design of tightly coupling the query and caching in same nodes. Ingester and query nodes access timeseries based on the partitioning. With this design, the number of \sysname{} cache and the number storage/query nodes can be adjusted accordingly based on the load. Moreover, distributed \sysname{}  also allows replicas of cache nodes for fault tolerance in case of node failure as well as leveraging the auto-scaling mechanism provided by Kubernates~\cite{cluster-vm}. }

%% file: Implementation.tex
\section{Implementation} \label{sec:impl}
We implement \sysname{} as a plugin with 5K lines of Go code and integrate it into Prometheus (release-2.52) and VictoriaMetrics (release-v1.102.0). \sysname{} is also compatible with other Prometheus-like systems, such as Grafana Mimir~\cite{mimir}. For integration, for example, users can apply a $\sim$30-line patch to Prometheus.

\noindent\textbf{Algorithm implementation optimizations: } To further improve \sysname{}'s system performance, we implement two optimizations following \cite{yang2020joltik} for universal sketches in Alg.~\ref{alg:eh_univ}:
\noindent (1) \textit{One layer update: } We update only the lowest sampled layer per insertion, reducing the layers updated to one per insertion. 
\noindent (2) \textit{Pyramid memory: } We use larger Count Sketches for upper layers and smaller ones for lower layers, while preserving high accuracy, given that the layered-sampling in universal sketching reduces the data size reaching the lower layers.

\noindent\textbf{Extending to more statistics:}  We implement approximate caching for additional statistics using sliding window uniform sampling~\cite{lu2010simple} for statistics like average, sum, standard deviation and variance.
This algorithm maintains a sample set $S$ of the most recent sliding window with time range $T$. New items are added with a probability $p\in(0, 1)$, and outdated samples are discarded. For a query over $(t_1, t_2)$, it answers queries with  samples within the range from $S$.

%% file: Evaluation.tex
\begin{figure*}[t]
    \centering
     \begin{minipage}[t]{.25\linewidth}
    \includegraphics[width=1\linewidth]{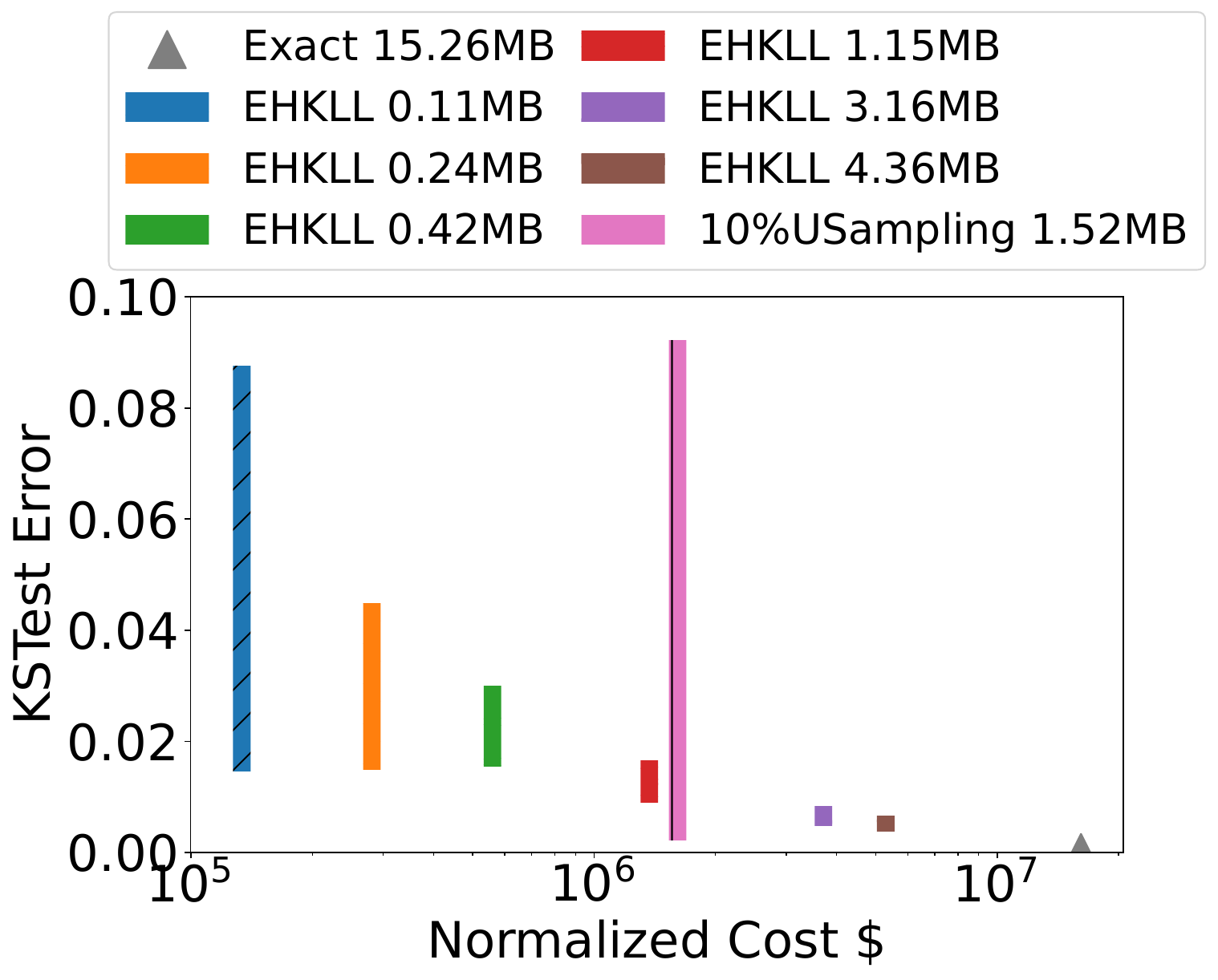}
    	\centering
    	\vspace{-.2in}
    	
    	{\footnotesize(a) Quantile\_over\_time, Google}
    \end{minipage}
    \begin{minipage}[t]{.45\linewidth}
     \includegraphics[width=1\linewidth]{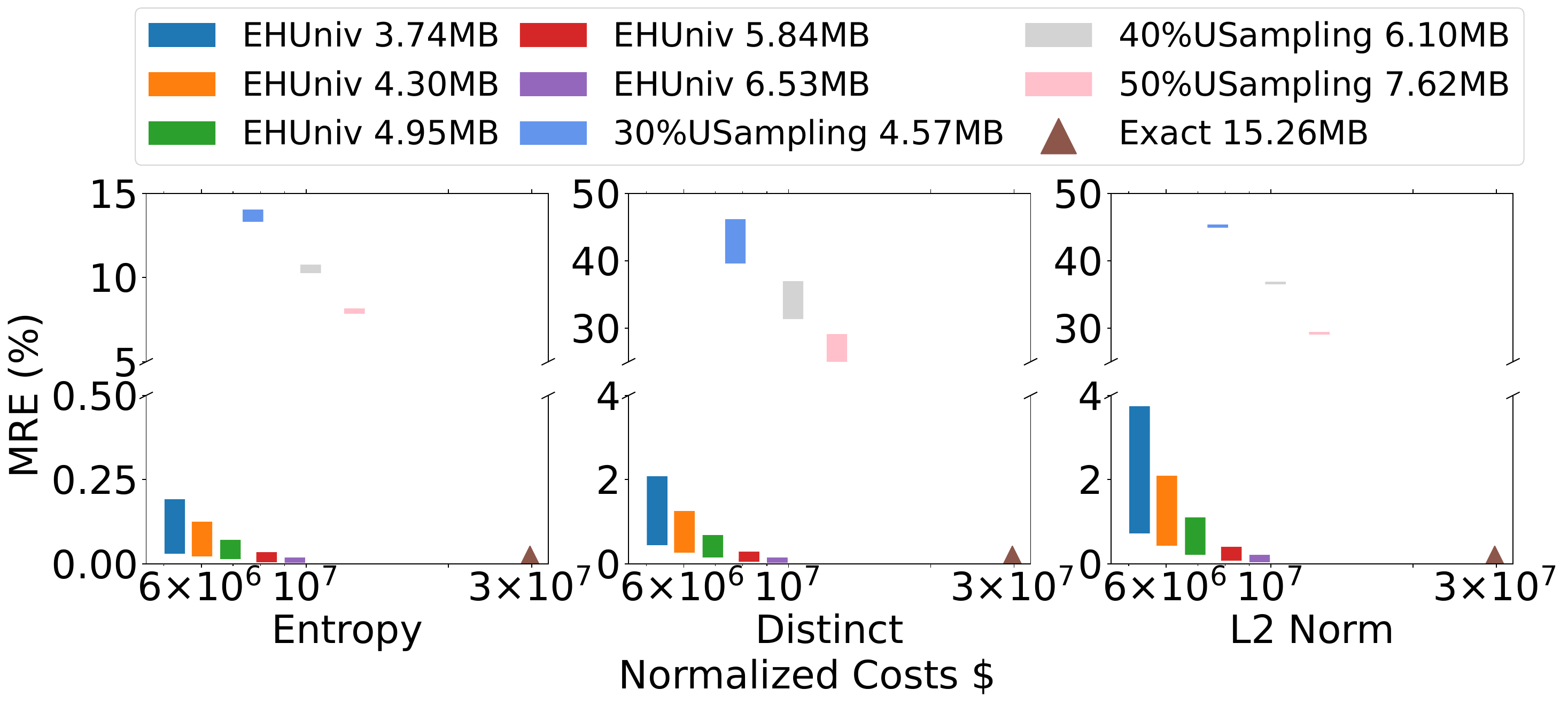}
    	\centering
            \vspace{-.2in}
    	
    	{\footnotesize (b) GSum-statistics, CAIDA2018}
    \end{minipage}
    \begin{minipage}[t]{.28\linewidth}
     \includegraphics[width=1\linewidth]{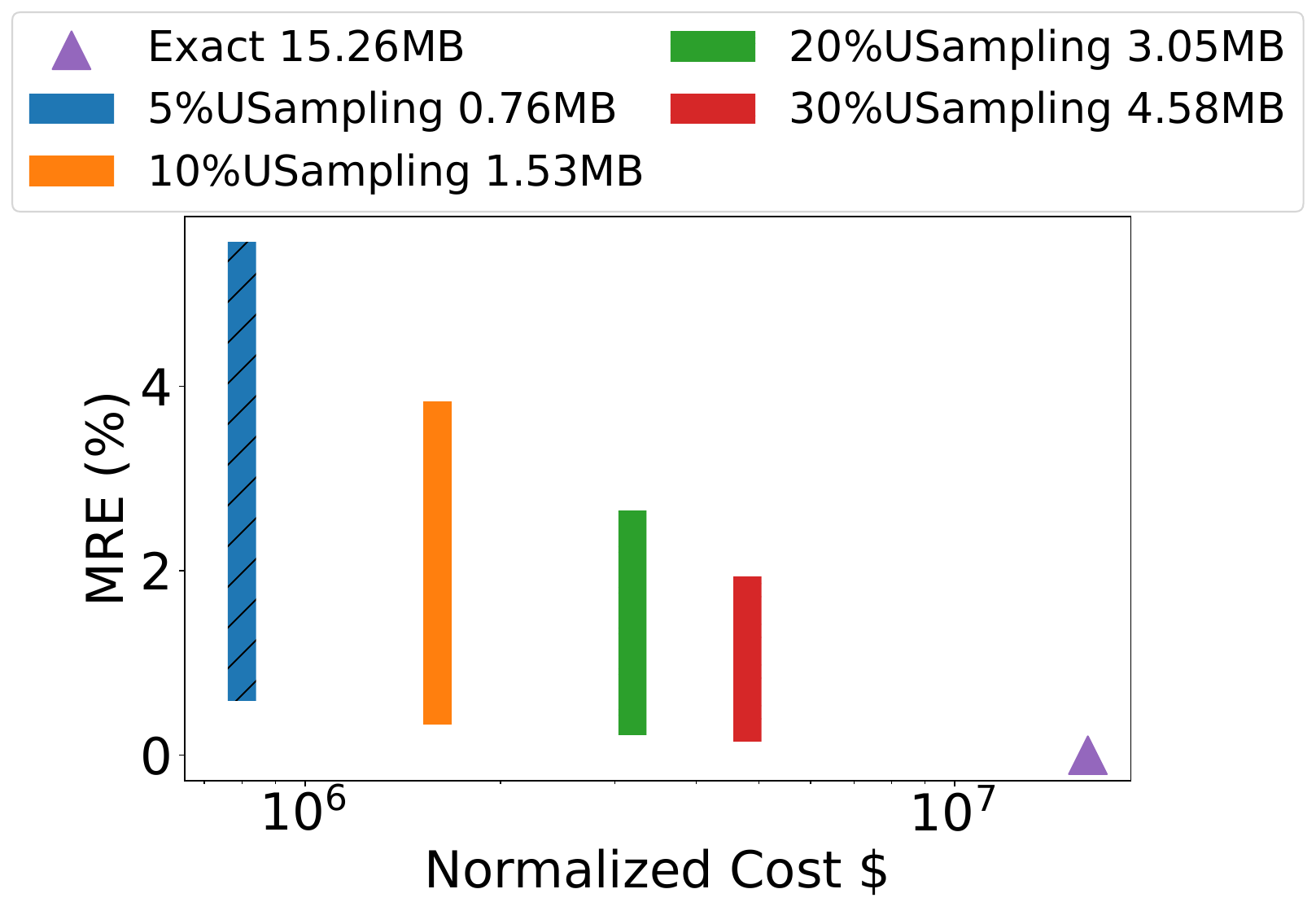}
    	\centering
    	\vspace{-.2in}
    	
    	{\footnotesize(c) Avg\_over\_time, Google}
    \end{minipage}
    \vspace{-4mm}
    \mycaption{fig:cost_accuracy}{\change{Query cost-accuracy analysis of multiple zoomed-in sub-window queries}}{\footnotesize Normalized costs refer to compute (microseconds) and memory usage (Bytes) costs per set of queries (10K-, 100K, 1M-sample sub-windows) in total. Each bar shows the error region of multiple sub-window queries given a statistic. 
    } 
\end{figure*}

\section{Evaluation}\label{sec:eval} We evaluate \sysname{}'s end-to-end performance and provide a sensitivity analysis and demonstrate that:
\begin{packeditemize}
\item \sysname{} offers $\leq 5\%$ mean errors across statistics \change{at 5$\times$ to 75$\times$ less system costs} (compute and memory) of exact query engines. Thus, the operational costs are reduced by 400$\times$ and 4$\times$ compared to Prometheus and VictoriaMetrics, respectively.
\item \sysname{} offers up to \change{two order of magnitude} smaller query latency than Prometheus and  VictoriaMetrics.
\item \change{\sysname{} maintains up to 8$\times$ faster ingestion than alternative fixed sliding window designs when querying multiple metrics and time windows, representing a moderate 1.3$\times$ to 3$\times$ slowdown compared to non-cached Prometheus depending on the number of timeseries and \sysname{} algorithm choices.}
\end{packeditemize}

\noindent\textbf{Testbed: } Our single-machine experiments run on an Ubuntu 20.04 system with a 32-core Intel Xeon Gold 6142 (2.6GHz), 384GB DRAM, and a 1TB SATA HDD. \change{The cluster experiments use three servers, each with a 24-core AMD 7402P (2.8GHz), 128GB DRAM, 1.6TB NVMe SSDs, and 100Gbps NICs.}

\noindent\textbf{Datasets: } We use two synthetic and two real-world traces.  (1) {\em Synthetic traces:} We generate Zipf-distributed (obeying $Pr(k) = (1+k)^{-1.01}, k \in N$, \textsl{Zipf}) \change{10M} data samples at configurable intervals \change{and 10M uniform distribution data (\textsl{Uniform})}, with values in $[0, 100000]$, for each timeseries. \change{We create a dynamic dataset (\textsl{Dynamic}) that transitions between Zipf, uniform, and normal distributions ($\mu = 50000, \sigma=10000$), generating 1M data points per distribution in a continuous cycle. }  (2) {\em Real-world traces:} We use Electronic Power dataset (\textsl{Power})~\cite{misc_individual_household_electric_power_consumption_235}, and \change{Google Cluster data v3}~\cite{google-cluster-trace} (\textsl{Google}), where we use start\_time as time and average\_usage.memory as memory\_usage. For EHUniv evaluation, we use CAIDA datasets~\cite{caida-dataset} of source IP addresses from a NYC Equinix backbone on 2018-12-20 (\textsl{CAIDA2018}) and 2019-01-17 (\textsl{CAIDA2019}). \sysname{} sensitivity analysis is conducted on the first \change{20M points}, and 2M points of \textsl{Power} due to length constraints. 

\noindent\textbf{Baselines and evaluation metrics:} We compare \sysname{} with the (1) \textsl{Prometheus} system. (2) \textsl{VictoriaMetrics} single-node version~\cite{vm-single-version}.  From the space of approximate analytics engines, we compare \sysname{} against (3) \textsl{Uniform Sampling}: We implement uniform sampling and insert the sampled data into an array-based caching layer for Prometheus, tested with varied sampling ratios. We evaluate rule query latency, insertion throughput, memory usage, accuracy, and operational costs. Accuracy is assessed by Mean Relative Error (MRE) against exact statistics. For quantile, min, and max estimations, we use the Kolmogorov-Smirnov test (KSTest)~\cite{berger2014kolmogorov} to compare CDF differences.

\begin{figure*}[t]
    \centering
    \begin{minipage}[t]{0.56\linewidth}
     \includegraphics[width=1\linewidth]{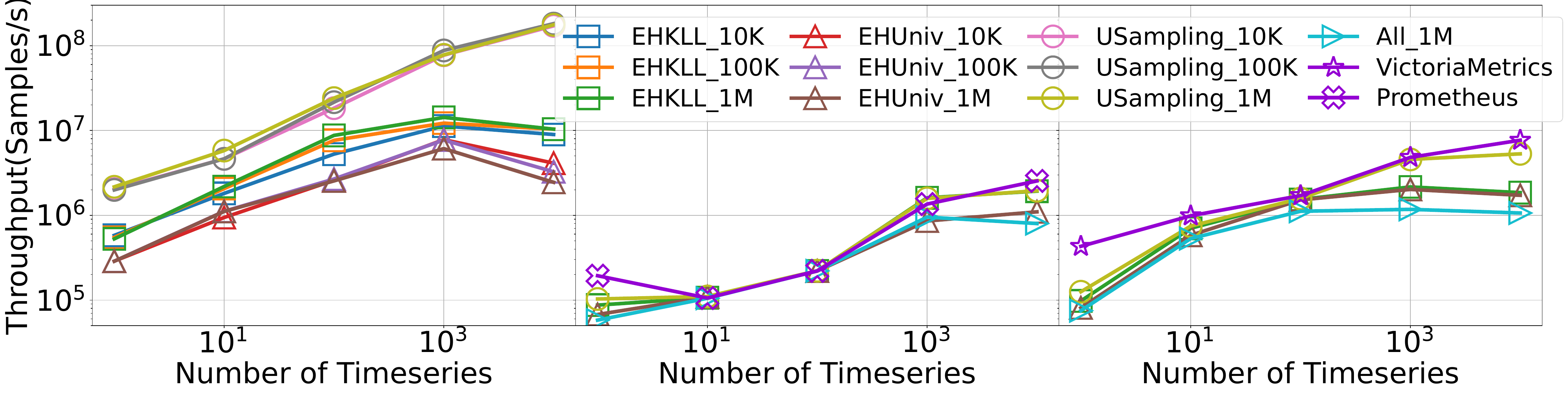}
    	\centering
    	\vspace{-.2in}
    	
    	{\footnotesize \hspace{8mm} (a) \sysname{}-only  \hspace{6mm}  (b) \sysname{}+Prometheus    \hspace{3mm}  (c) \sysname{}+VictoriaMetrics}
    \end{minipage}  
    \begin{minipage}[t]{.21\linewidth}
     \includegraphics[width=1\linewidth]{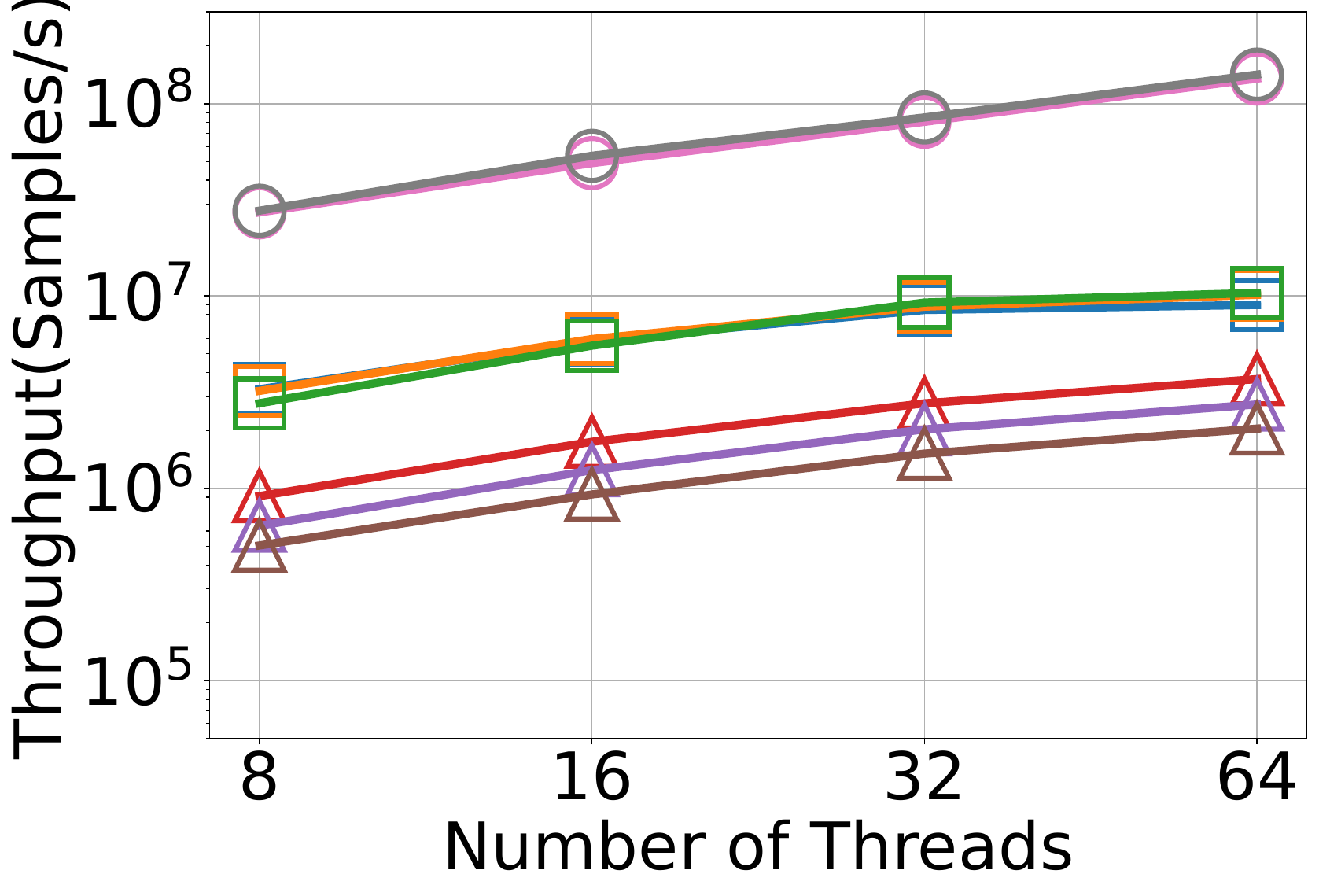}
    	\centering
    	\vspace{-.2in}
    	
    	{\footnotesize(d) \sysname{}-only Scalability}
    \end{minipage}
    \begin{minipage}[t]{.22\linewidth}
     \includegraphics[width=1\linewidth]{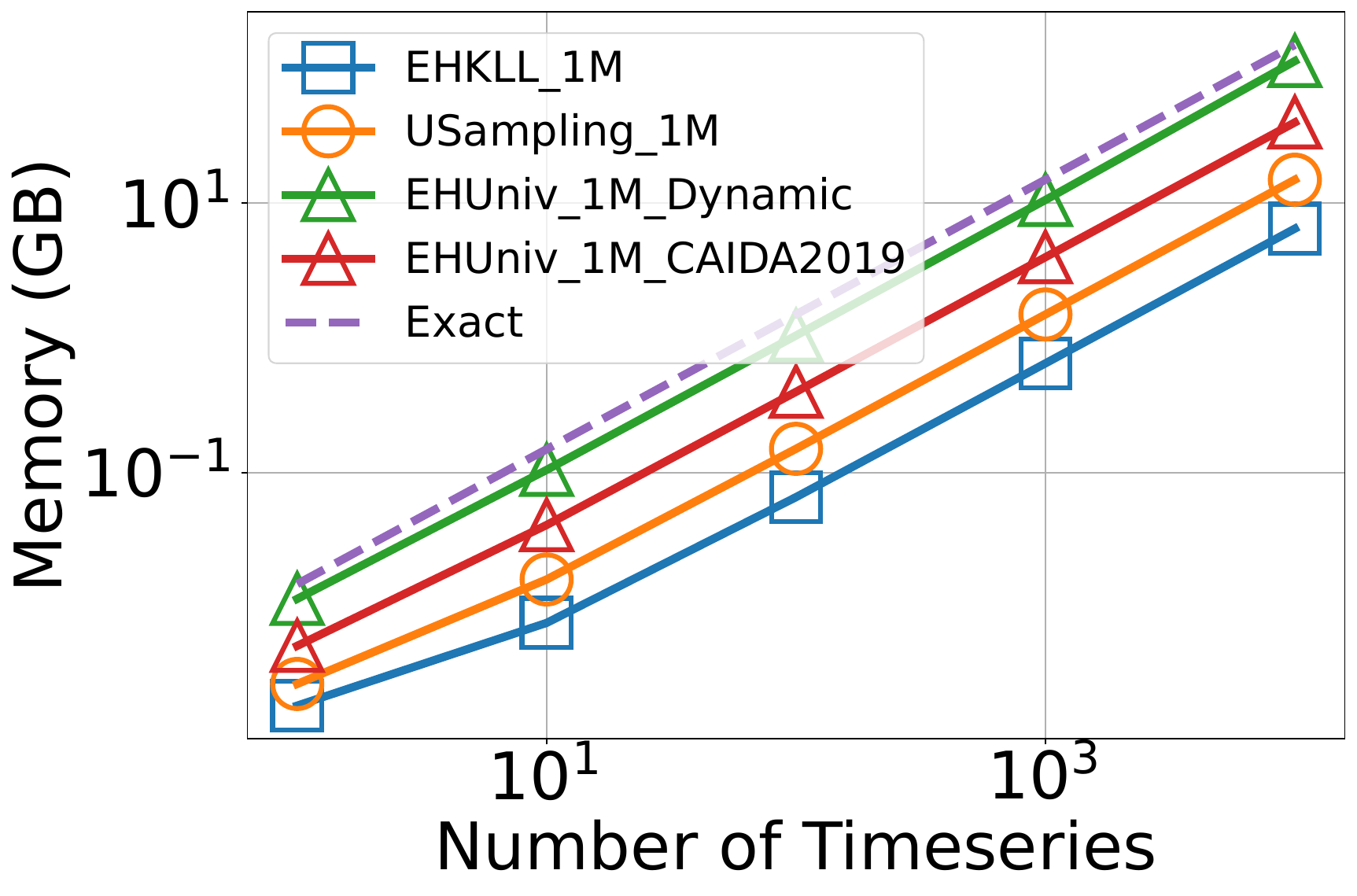}
    	\centering
    	\vspace{-.2in}
    	
    	{\footnotesize(e) \#Timeseries vs. Memory}
    \end{minipage}
    \vspace{-4mm}
    \mycaption{fig:insertion_throughput}{\change{Single-Machine data insertion throughput with different timeseries numbers and sliding window sizes on Dynamic dataset for (a) \sysname{}, (b) \sysname{} integrated to Prometheus, and (c) \sysname{} integrated to VictoriaMetrics; (d) insertion scalability with different thread numbers on Zipf dataset and 10K timeseries; and (e) memory usage with different timeseries numbers}}{\footnotesize \change{All\_1M refers to running all three algorithms together with 1M-sample window per series. In (e), we show memory consumption for the Dynamic dataset, with EHUniv also tested on CAIDA2019.} 
    } 
\end{figure*}

\subsection{End-to-End \sysname{} Performance}  \label{eval:production}

For evaluating query latency and insertion throughput, we set uniform sampling at a 10\% sampling rate; for EHKLL, we set KLL space limit parameter~\cite{karnin2016optimal} $k_{KLL}=256$ (where $k_{KLL}=(1/\epsilon_{KLL}) \sqrt{\log(1-\delta)}$), and EH error parameter $k_{EH}=50$, for 5\% KSTest errors; for EHUniv, we set $k_{EH}=20$ for 5\% relative errors.

\subsubsection{Operational Costs and Accuracy.}

\noindent\textbf{Compute and memory costs vs. accuracy.} Under a cloud billing model, users pay for resources such as memory and CPU cores. Fig.\ref{fig:cost_accuracy} shows the normalized operational costs for concurrent queries, including insertion, query compute, and memory usage (excluding storage and network). The costs are normalized by computation time and memory usage. 
Each figure shows the average errors of different sub-window sizes, with confidence levels shown as a region after five runs. Fig.~\ref{fig:cost_accuracy}(a) and (c) depict queries with sub-window sizes of 1M, 100K, and 10K samples, using a fixed 1M-sample sliding window. The zoom-in queries mimic anomaly detection: the user first queries a 1M window, then splits it into ten 100K sub-windows for further queries, and finally divides the last 100K sub-window into ten 10K sub-windows for finer granularity.
Fig.~\ref{fig:cost_accuracy}(b) shows the mean relative errors and confidence levels for \texttt{entropy}, \texttt{distinct} counting, and $L_2$ norm, with a 1M-sample sliding window and zoomed-in sub-windows of suffix length from 100K to 1M with an interval of 10K samples. Queries are issued every 100K samples.
For quantiles and GSum-statistics, \sysname{} offers better cost-accuracy tradeoffs than uniform sampling. \sysname{} maintains less than 5\% errors even for the smallest sub-windows (10K samples) while reducing costs by 75$\times$ for EHKLL, 10$\times$ for Sampling, and 5$\times$ for EHUniv compared to the exact baseline.

\noindent\textbf{Cloud operational cost estimations.} We compare operational costs of Prometheus, VictoriaMetrics, and \sysname{} integration in Table~\ref{table:dollar_cost}. \sysname{} respectively reduces the query processing cost by about 400$\times$ compared to AWS Prometheus Pricing and at least 4$\times$ compared to VictoriaMetrics, while not increasing the storage and data ingestion costs.  For a 1000-node Kubernetes cluster collecting 1000 metrics per node per second for a month, the total ingestion is 268B samples, requiring 1M samples/s. Assuming each metric has 20 labels with 100 unique values and averaging 30 bytes per label and 2 bytes per sample after compression~\cite{aws-price}, and 10 rule queries running 24/7, querying every minute with 8000 timeseries and 1M samples per series, the cost breakdown is as follows:
(1) {\bf AWS Prometheus pricing~\cite{aws-price}:} Data ingestion costs \$9,186, storing 2336GB of metrics and labels costs \$70/month, and query processing costs \$11,560. This is an estimate for 10 alerting rules, and we envision the cost to be at least several orders of magnitude more when it scales up. 
(2) {\bf VictoriaMetrics~\cite{vm-price}:} Using the typical cloud billing model~\cite{ec2-billing} and assuming each data sample has 64-bit floating point value and 64-bit timestamp associated after decompression, 10 queries concurrently require $10\times$ 8 Billion $\times$ 16B = 1192 GB memory, costing at least \$7,443/month for compute (using x2idn.24xlarge~\cite{ec2-billing} with 96 vCPUs and 1.5TB memory). Storage costs \$348/month, with no data ingestion charge~\cite{vm-price}. (3) {\bf \sysname{}-PM: } With Prometheus, \sysname{} only processes each sample once, costing \$28.6/month for query processing (\$0.1/Billion samples). 
(4) {\bf \sysname{}-VM:} With VictoriaMetrics, \sysname{} uses $\sim$3MB per timeseries for a 1M-sample window and 5\% error target, requiring $10\cdot 8000 \cdot 3\text{MB}=234$ GB, which can be handled by an m6g.16xlarge instance (64 vCPUs, 256GB) at \$1,833/month. Overlapping queries on the same timeseries will further reduce costs.

\begin{table}[t]
\centering
\mycaption{table:rule_latency}{\change{Total concurrent rule query latency on 10K-, 100K-, and 1M-sample windows}}{\footnotesize ``PM'' and ``VM'' stand for Prometheus and VictoriaMetrics, and ``PS-PM'' and ``PS-VM'' refer to corresponding \sysname{} integration. 
}
\vspace{-3mm}
\resizebox{1\columnwidth}{!}{
\begin{tabular}{@{}l|l|l|l|l|l@{}}
\toprule
\textbf{Statistics} & \textbf{Datasets} & \textbf{PM}  & \textbf{PS-PM ($\downarrow$)} & \textbf{VM}  & \textbf{PS-VM ($\downarrow$)}  \\
\toprule
\textbf{0.9-Quantile} & Zipf & 4155s  & 32.3s ($137\times$) & 63.0s & 3.1s  ($20\times$) \\ 
 & \change{Dynamic} & \change{5005s} & \change{27.7s ($181\times$)} & \change{96.1s} & \change{3.2s ($30\times$)} \\
\midrule
\textbf{Max}  & \change{Zipf} & 1102s & 31.5s ($35\times$) & 27.0s & 3.1s ($9\times$) \\
 & \change{Dynamic} & \change{1421s} & \change{25.1s ($57\times$)} & \change{29.3s} & \change{3.2s ($9\times$)} \\
\midrule
\change{\textbf{0.5-Quantile}} & \change{Zipf} & \change{12858s} & \change{64.9s ($198\times$)} & \change{207.5s} & \change{6.5s ($32\times$)} \\
\change{\textbf{+0.9-Quantile}} & \change{Dynamic} & \change{7537s} & \change{53.1s ($142\times$)} & \change{493.2s} & \change{6.5s ($76\times$)} \\
\midrule
\change{\textbf{0.5-Quantile+Max}}  & \change{Zipf} & \change{21494s} & \change{105.9s ($203\times$)} & \change{402.8s} & \change{12.9s ($31\times$)} \\
\change{\textbf{+0.9-Quantile+Min}} & \change{Dynamic} & \change{11383s} & \change{116.9s ($97\times$)} & \change{1014.4s} & \change{12.9s ($78\times$) } \\
\bottomrule
\textbf{Distinct}  & \change{Zipf} & 1779s  & 46.1s ($39\times$) & 29.4s  & 2.2s ($14\times$)   \\
 & \change{Dynamic} & \change{2593s} & \change{23s ($113\times$)} & \change{100.1s} & \change{1.9s ($53\times$)} \\
 & \change{CAIDA2019} & \change{1688s} & \change{28.8s ($59\times$)} & \change{31.5s} & \change{2.0s ($16\times$)} \\
\midrule
\textbf{Entropy}  & \change{Zipf} & 2042s & 53.6s ($38\times$) & 48.6 s  &  2.8s ($17\times$) \\
 & \change{Dynamic} & \change{7871s} & \change{34.0s ($231\times$)} & \change{179.9s} & \change{1.8s ($100\times$)}  \\
 & \change{CAIDA2019} & \change{2105s} & \change{33.0s ($64\times$)} & \change{44s} & \change{2.3s ($19\times$)} \\
\midrule
\textbf{$L_2$ Norm}  & \change{Zipf} &1940s &50.4s ($39\times$) & 42.26 s & 2.62s ($16\times$)  \\
 & \change{Dynamic} & \change{3562s} & \change{26.0s ($137\times$)} & \change{280.8s} & \change{1.77s ($158\times$)} \\
 & \change{CAIDA2019} & \change{1969s} & \change{30.7s ($64\times$)} & \change{42s} & \change{2.0s ($21\times$)} \\
\midrule
\change{\textbf{Distinct}} & \change{Zipf} & \change{7517s} & \change{112s ($67\times$)} & \change{415.48 s} & \change{5.2s ($81\times$)}\\
\change{\textbf{+Entropy}} & \change{Dynamic} & \change{13266s} & \change{73.0s ($182\times$)} & \change{1146s} & \change{9.9s ($116\times$)} \\
\change{\textbf{+$L_2$ Norm}} & \change{CAIDA2019} & \change{6866s} & \change{77.5s ($89\times$)} & \change{336s} & \change{6.5s ($52\times$)} \\
\bottomrule

\textbf{Average}  & \change{Zipf} & 1158s  & 9.4s ($123\times$) & 1.84s & 0.93s ($2\times$) \\ 
 & \change{Dynamic} & \change{1164.8s} & \change{9.4s ($124\times$)} & \change{2.44s} & \change{0.96s ($2.5\times$)} \\

\midrule
\change{\textbf{Average}}  & \change{Zipf} & \change{2492s} & \change{18.4s ($135\times$)} & \change{17.9s} & \change{1.1s ($16\times$)} \\ 
\textbf{\change{+Stddev}} & \change{Dynamic} & \change{2520s} & \change{20.5s ($123\times$)} & \change{23.3s} & \change{1.1s ($21\times$)} \\
  
\bottomrule
\change{\textbf{0.9-Quantile+Max}}  & \change{Zipf} & \change{8985s} & \change{180.2s ($50\times$)} & \change{349s} & \change{9.2s ($38\times$)} \\ 
\textbf{\change{+Average+Distinct}} & \change{Dynamic} & \change{13177s} & \change{88.6s ($154\times$)} & \change{590s} & \change{9.0s ($65\times$)} \\
 
\bottomrule
\end{tabular}
}
\end{table}

\subsubsection{Single-Machine Rule Query Latency.} \label{sec:single_rule_latency}
We evaluate \sysname{}'s end-to-end query latency in Prometheus and VictoriaMetrics across different statistics and query window sizes on 10K time series each with 100ms data insertion interval. \change{Table~\ref{table:rule_latency} reports total latencies for concurrent drill-down queries over the most recent $10^5$-, $10^4$-, and $10^3$-second time windows for each listed statistics.}  The total latency is averaged from 10 runs after a warm-up of $10^5$ seconds. Each \texttt{aggr\_over\_time} query aggregates statistics across all timeseries. \change{With Alg.~\ref{alg:eh_quantile}, \sysname{} improves quantile query latencies by up to $203\times$ over Prometheus, and $78\times$ over VictoriaMetrics. With Alg.~\ref{alg:eh_univ}, \sysname{} improves distinct counting, entropy, and $L_2$ norm query latencies by up to 231$\times$ compared to Prometheus and up to 158$\times$ compared to VictoriaMetrics. With 10\% uniform sampling, \sysname{} improves \texttt{average} and \texttt{stddev} query latencies by $135\times$ over Prometheus and $21\times$ over VictoriaMetrics.} The smaller improvement for \texttt{average} is due to VictoriaMetrics's data caching optimization for \texttt{avg}, while it does not optimize complex queries such as quantiles and distinct counting. The overall smaller improvements on VictoriaMetrics compared to Prometheus are attributed to VictoriaMetrics' more efficient storage engine.

\begin{table}[t]
\centering
\mycaption{table:cluster_rule_latency}{\change{VictoriaMetrics(w/ \sysname{}) cluster version total rule query latencies (seconds) with different nodes}}{\footnotesize \change{We show the speedup comparing VictoriaMetrics w/ \sysname{} and without.} }
\vspace{-3mm}
\resizebox{0.8\columnwidth}{!}{
\begin{tabular}{@{}ccccc@{}}
\toprule
\multirow{ 2}{*}{\change{\#Timeseries}}  & \multicolumn{2}{c}{\change{VictoriaMetrics}}   & \multicolumn{2}{c}{\change{w/\sysname{}}} \\
\cmidrule(rl){2-3} \cmidrule(rl){4-5}
  & \change{1-node}  & \change{3-nodes} & \change{1-node ($\downarrow$)}  & \change{3-nodes ($\downarrow$)}     \\
\midrule
\change{20K} &\change{74.49}  & \change{44.43} & \change{4.53 ($16\times$)} & \change{3.22 ($14\times$)} \\
\change{40K} &\change{141.62}  & \change{121.58} & \change{6.95 ($20\times$)} & \change{4.04 ($30\times$)} \\
\bottomrule
\end{tabular}
}
\end{table}

\begin{table}[t]
\centering
\mycaption{table:cluster_insert}{\change{VictoriaMetrics+\sysname{} cluster version insertion throughput (M/s) with different nodes and timeseries numbers}}{}
\vspace{-3mm}
\resizebox{0.6\columnwidth}{!}{
\begin{tabular}{@{}cccc@{}}
\toprule
{\change{\#Timeseries}}  & \change{1-node} & \change{2-nodes} & \change{3-nodes}     \\
\midrule
\change{20K}  & \change{1.33}  &\change{2.56} &\change{3.86} \\
\change{40K}   & \change{1.30} &\change{2.60}  &\change{3.79} \\
\change{80K}  & \change{1.30} &\change{2.57} &\change{3.79} \\
\bottomrule
\end{tabular}
}
\end{table}

\subsubsection{Single-Machine Insertion Throughput.} \label{sec:single_insert_throughput}
To evaluate the impact of \sysname{} precomputation on insertion throughput with the backend Prometheus and VictoriaMetrics TSDBs, we set the data timestamp interval as 100ms and measure the duration of inserting 60-hour data (2.16M samples) with varying timeseries numbers. Each approximation method is tested with 10K-, 100K- and 1M-sample window sizes. \change{Fig.~\ref{fig:insertion_throughput} shows the insertion throughput of \sysname{} and each approximation method running alongside a backend TSDB, with timeseries insertion evenly distributed across CPU cores. 
Fig.~\ref{fig:insertion_throughput} (a) shows the insertion throughput of \sysname{} alone without inserting into backend TSDB, can achieve 10M samples/s for EHKLL, 2M samples/s for EHUniv, and 100M samples/s for 10\% uniform sampling. We show the 1M-sample window size for each algorithm in integrated systems.
For Prometheus integration (Fig.~\ref{fig:insertion_throughput}(b)), with 10K number of timeseries, uniform sampling, EHKLL, EHuniv, and running all three algorithms together have $1.3\times$, $1.3\times$, $2.3\times$, and $3.1\times$ smaller insertion throughput compared to Prometheus. For integration with VictoriaMetrics (Fig.~\ref{fig:insertion_throughput}(c)), \sysname{} achieves over 1.7M samples/s insertion throughput with 100 to 10K timeseries. With VictoriaMetrics, uniform sampling, EHKLL, EHUniv, and running all three algorithms together has $1.4\times$, $4.1\times$, $4.5\times$, and $7.2\times$ smaller insertion throughput compared to VictoriaMetrics, respectively, with 10K number of timeseries. The discrepancy is attributed to VictoriaMetrics' faster storage backend.}
\change{To evaluate \sysname{} insertion scalability, we fix the timeseries number at 10K and vary the thread count in a single-machine setting.
We measure throughput across three algorithms using 10K-, 100K-, and 1M-sample time windows. Fig.~\ref{fig:insertion_throughput} (d) shows that each algorithm scales linearly as the number of threads increases.}

\subsubsection{\change{Distributed \sysname{} Performance.}}\label{sec:distributed_evaluation}
\change{We integrate \sysname{} into the VictoriaMetrics cluster version~\cite{cluster-vm}, distributing query, sketch, and storage nodes across 3 servers. 
 
\noindent\textbf{Rule Query Latency.} In latency experiments, data is collected every second per series through a single ingestion node, with a synthetic data generator issuing Zipf-distribution data of different timeseries. Table~\ref{table:cluster_rule_latency} shows the total query latencies for four statistics (0.9-quantile, max, average, and distinct) with 1K-, 10K-, and 100K-second query windows (12 concurrent queries).  With \sysname{} cache, total latency can reduce by up to 30$\times$ compared to VictoriaMetrics Cluster. Scaling from 1 to 3 nodes in \sysname{} reduces latency by 1.4$\times$ for 20K timeseries and 1.7$\times$ for 40K timeseries, representing a sub-linear speedup due to not all query nodes being used by the cluster scheduler.

\noindent\textbf{Insertion.} Table~\ref{table:cluster_insert} shows the throughput with different nodes and timeseries when inserting data to all three algorithm instances for each timeseries in VictoriaMetric + \sysname{}. With increasing number of server nodes, \sysname{}'s ingestion performance can scale linearly from 1.33M/s to 3.86M/s with a large number of concurrent timeseries. We observe that changing the number of monitored timeseries (from 20K to 80K) does not show a significant performance impact, indicating the feasibility of supporting larger-scale cloud infrastructure monitoring. }

\begin{figure}[t]
\centering
\begin{minipage}[t]{.485\linewidth}
     \includegraphics[width=1\linewidth]{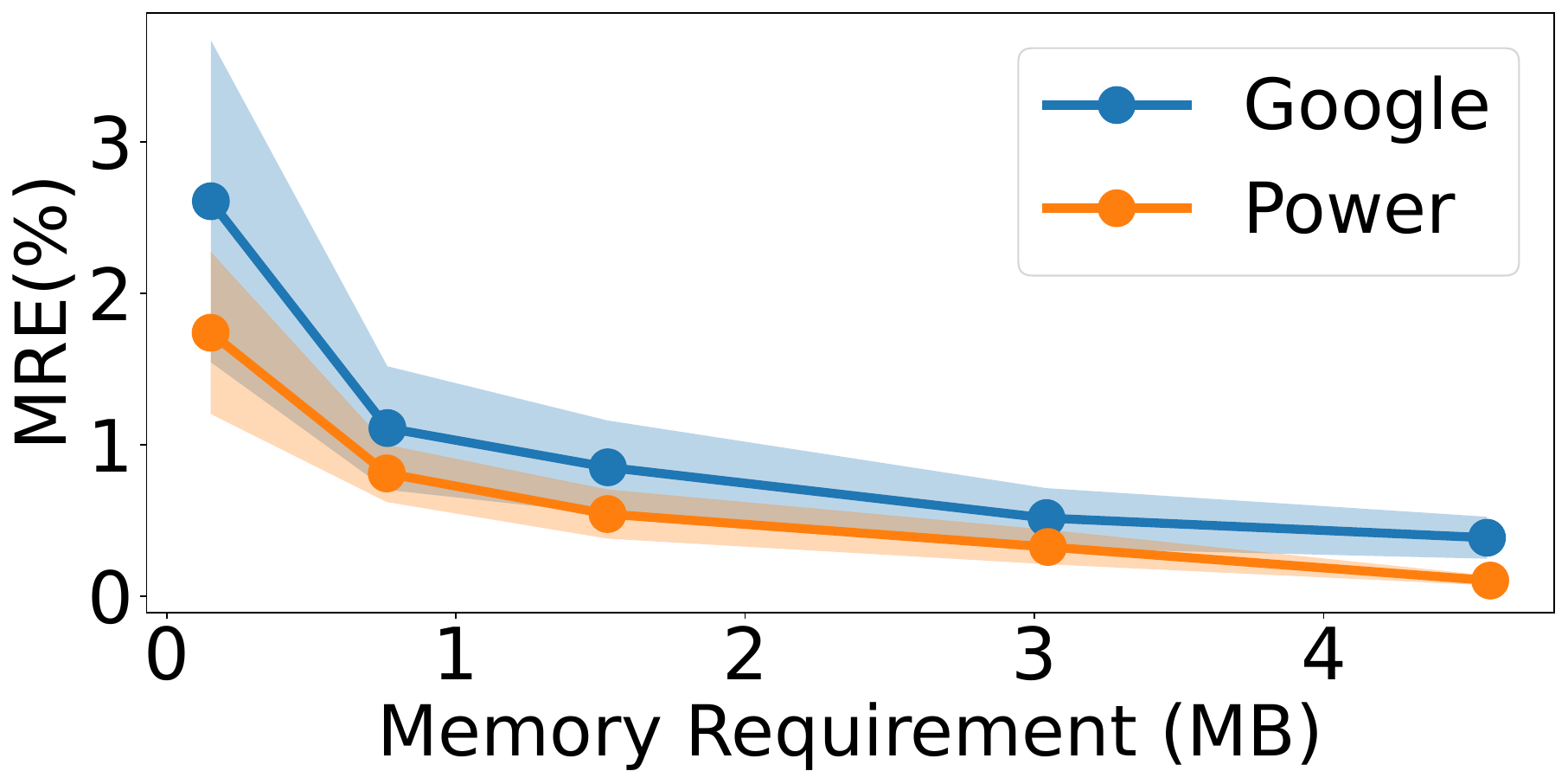}
    	\centering
    	\vspace{-.2in}
        
    	{\footnotesize(a) 300-600K Interval}
    \end{minipage}
    \begin{minipage}[t]{.49\linewidth}
     \includegraphics[width=1\linewidth]{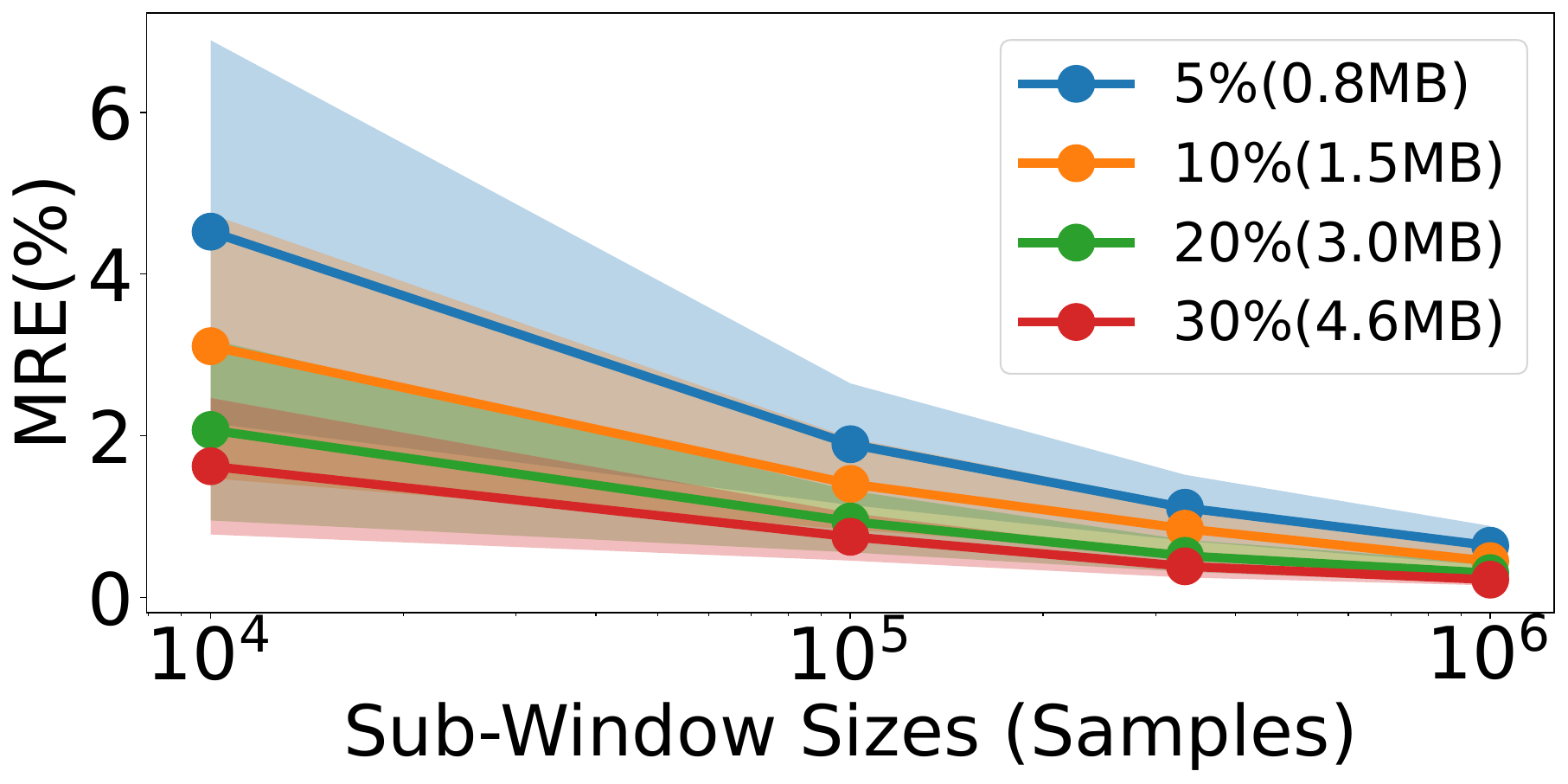}
    	\centering
    	\vspace{-.2in}
        
    	{\footnotesize(b) Multi Sub-windows, Google}
    \end{minipage}
    \vspace{-3mm}
\mycaption{fig:avg_error}{Uniform sampling's mean average estimation error for (a) memory-error in 300-600K intervals, and (b) various sub-windows}{\footnotesize \change{A legend label \texttt{x\%} refers to a configuration with \texttt{x\%} sampling rate.}}
\end{figure}

\begin{figure}[t]
\centering
\begin{minipage}[t]{.49\linewidth}
     \includegraphics[width=1\linewidth]{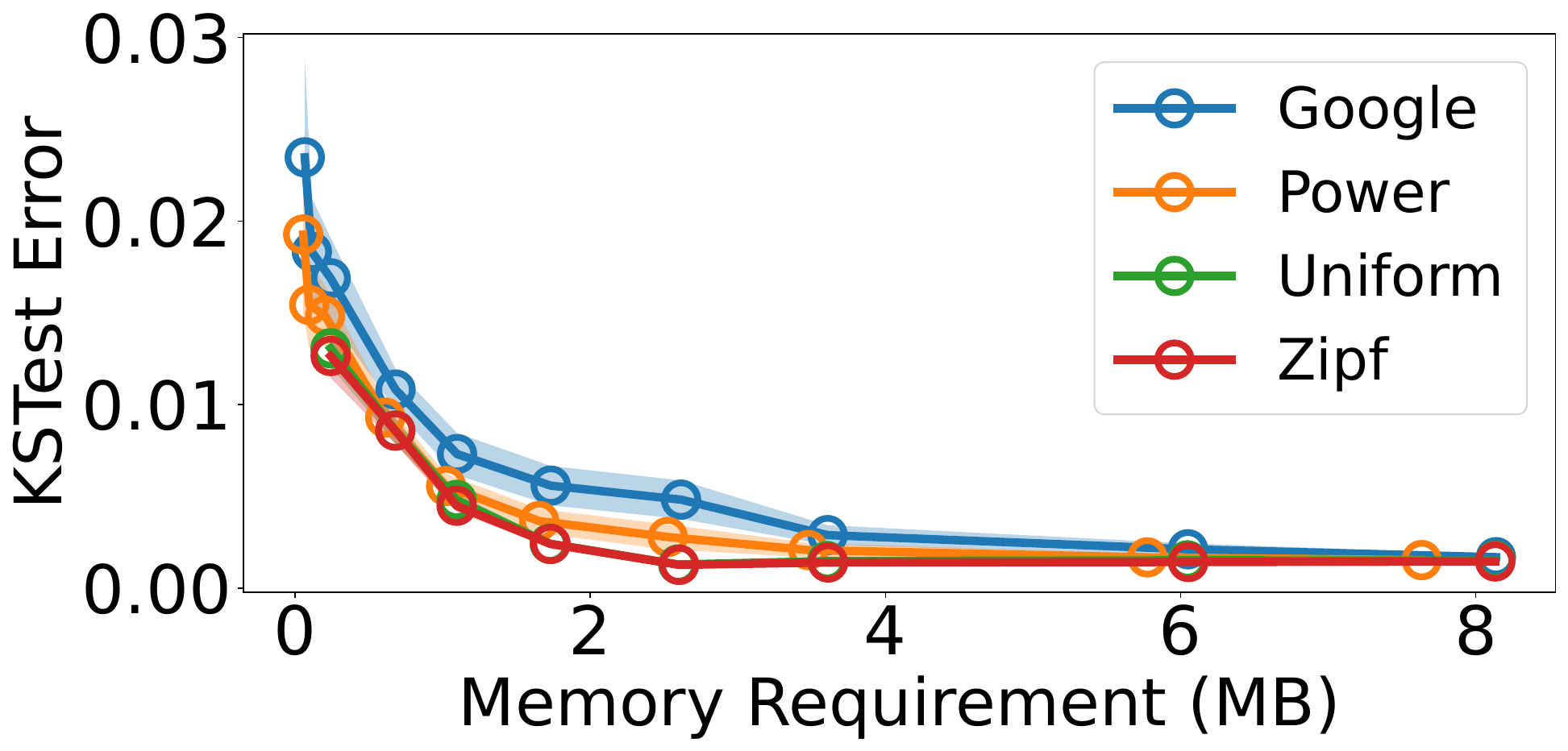}
    	\centering
    	\vspace{-.2in}
        
    	{\footnotesize(a) 300-600K Interval}
    \end{minipage}
    \begin{minipage}[t]{.49\linewidth}
     \includegraphics[width=1\linewidth]{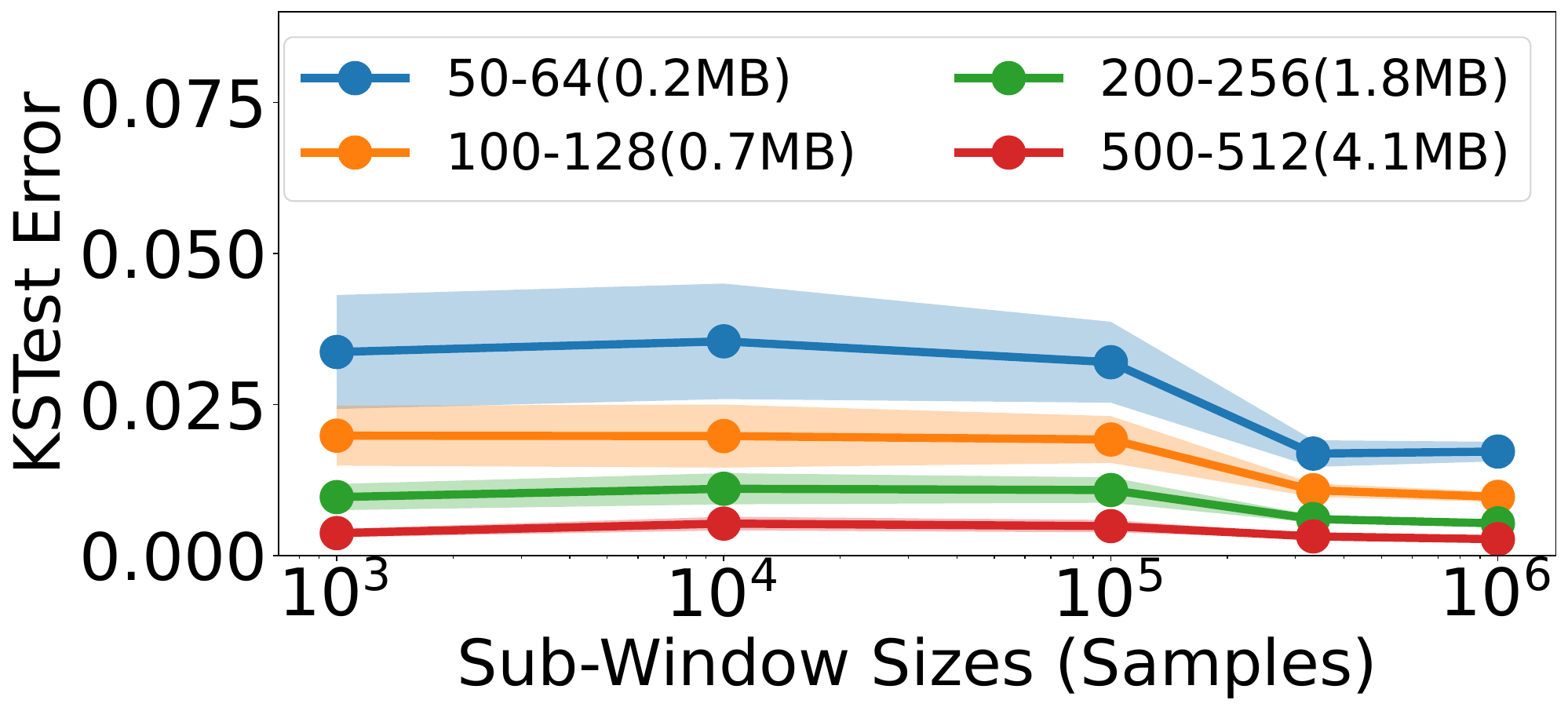}
    	\centering
    	\vspace{-.2in}
        
    	{\footnotesize(b) Multi Sub-windows, Google}
    \end{minipage}
    \vspace{-3mm}
\mycaption{fig:quantile_error}{EHKLL's mean quantile estimation KSTest error for (a) memory-error in 300K-600K intervals, and (b) various sub-window sizes}{\footnotesize \change{A legend label \texttt{x-y(z MB)} denotes an EHKLL configuration with $k_{EH}=x$, $k_{KLL}=y$, and $z$ MB total memory.}}
\end{figure}

\begin{figure}[t]
\centering
\begin{minipage}[t]{.5\linewidth}
     \includegraphics[width=1\linewidth]{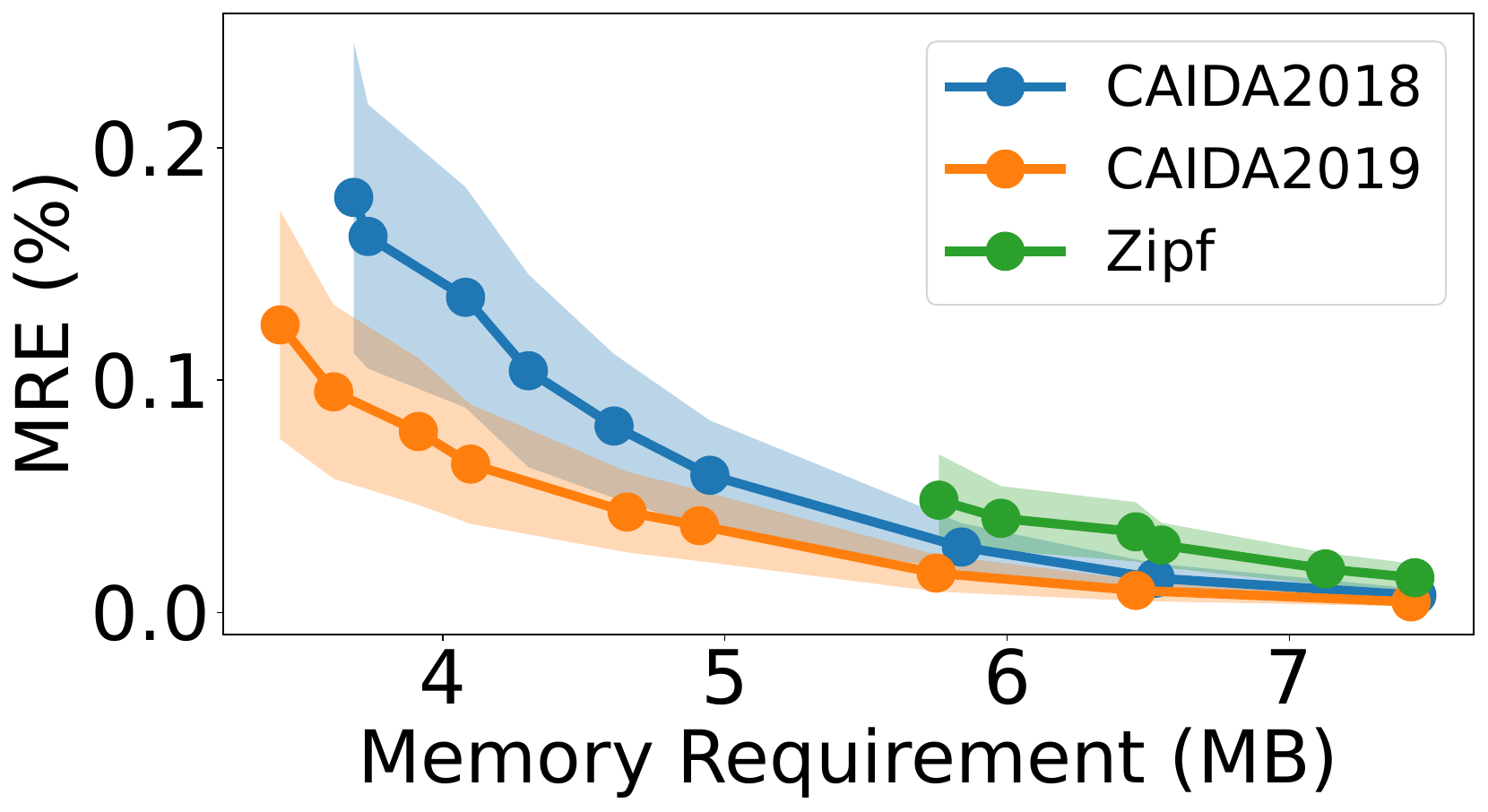}
    	\centering
    	\vspace{-.2in}
        
    	{\footnotesize(a) Entropy, 300-600K Interval}
    \end{minipage}
    \begin{minipage}[t]{.48\linewidth}
     \includegraphics[width=1\linewidth]{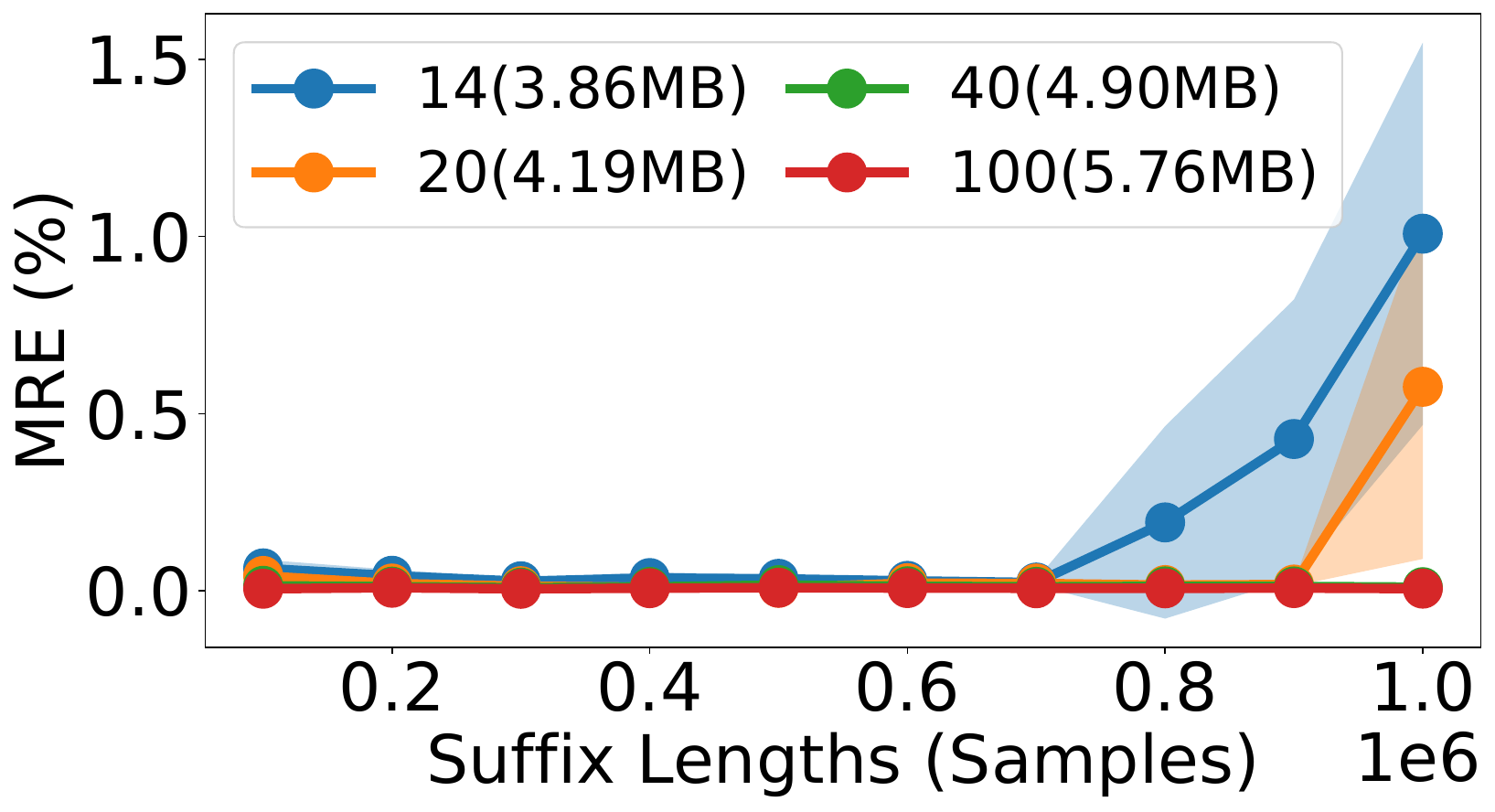}
    	\centering
    	\vspace{-.2in}
        
    	{\footnotesize(b) Entropy, Multi Suffix}
    \end{minipage}
    
    \begin{minipage}[t]{.49\linewidth}
     \includegraphics[width=1\linewidth]{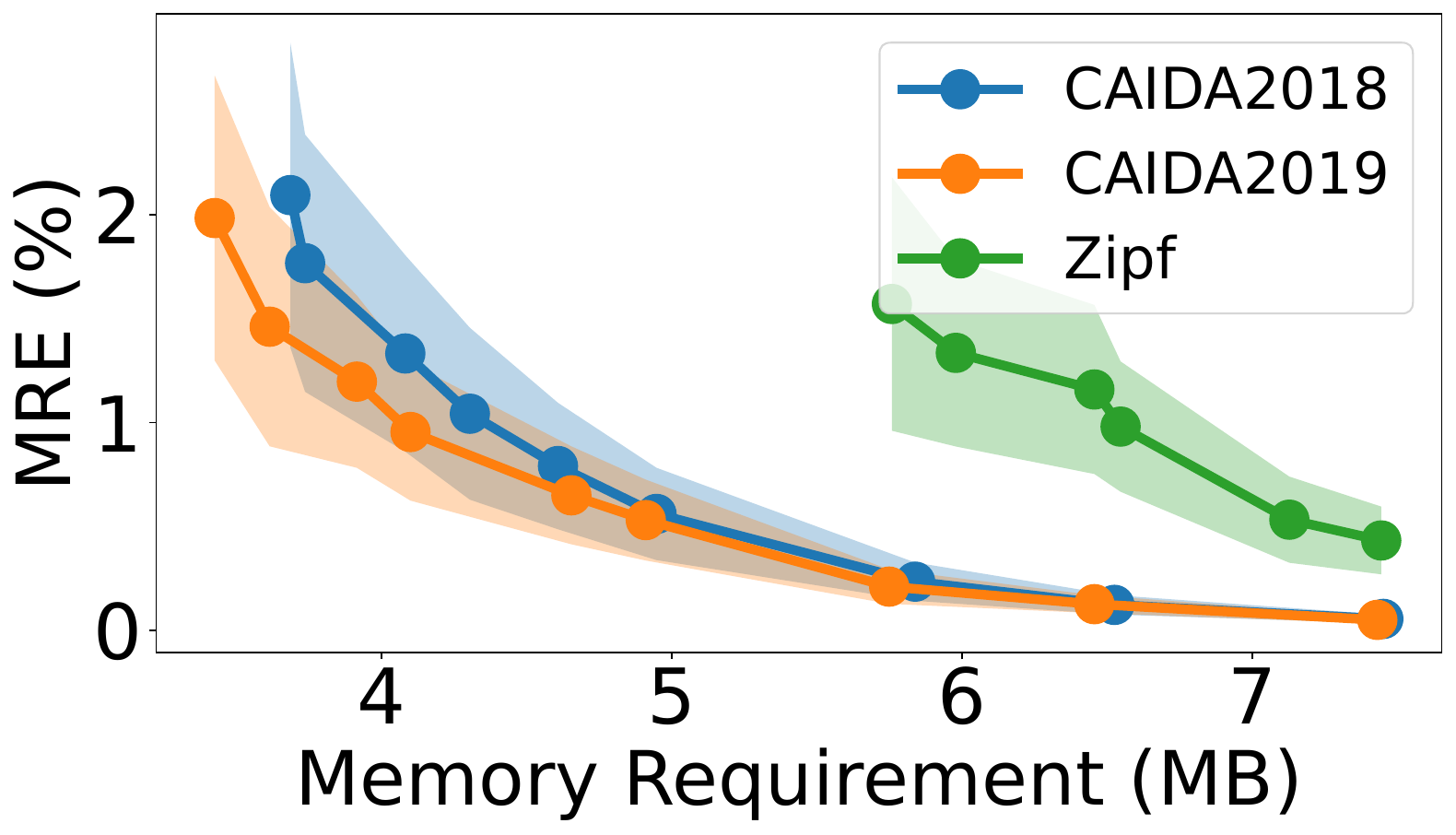}
    	\centering
    	\vspace{-.2in}
        
    	{\footnotesize(c) Distinct counting, 300-600K Interval}
    \end{minipage}
    \begin{minipage}[t]{.49\linewidth}
     \includegraphics[width=1\linewidth]{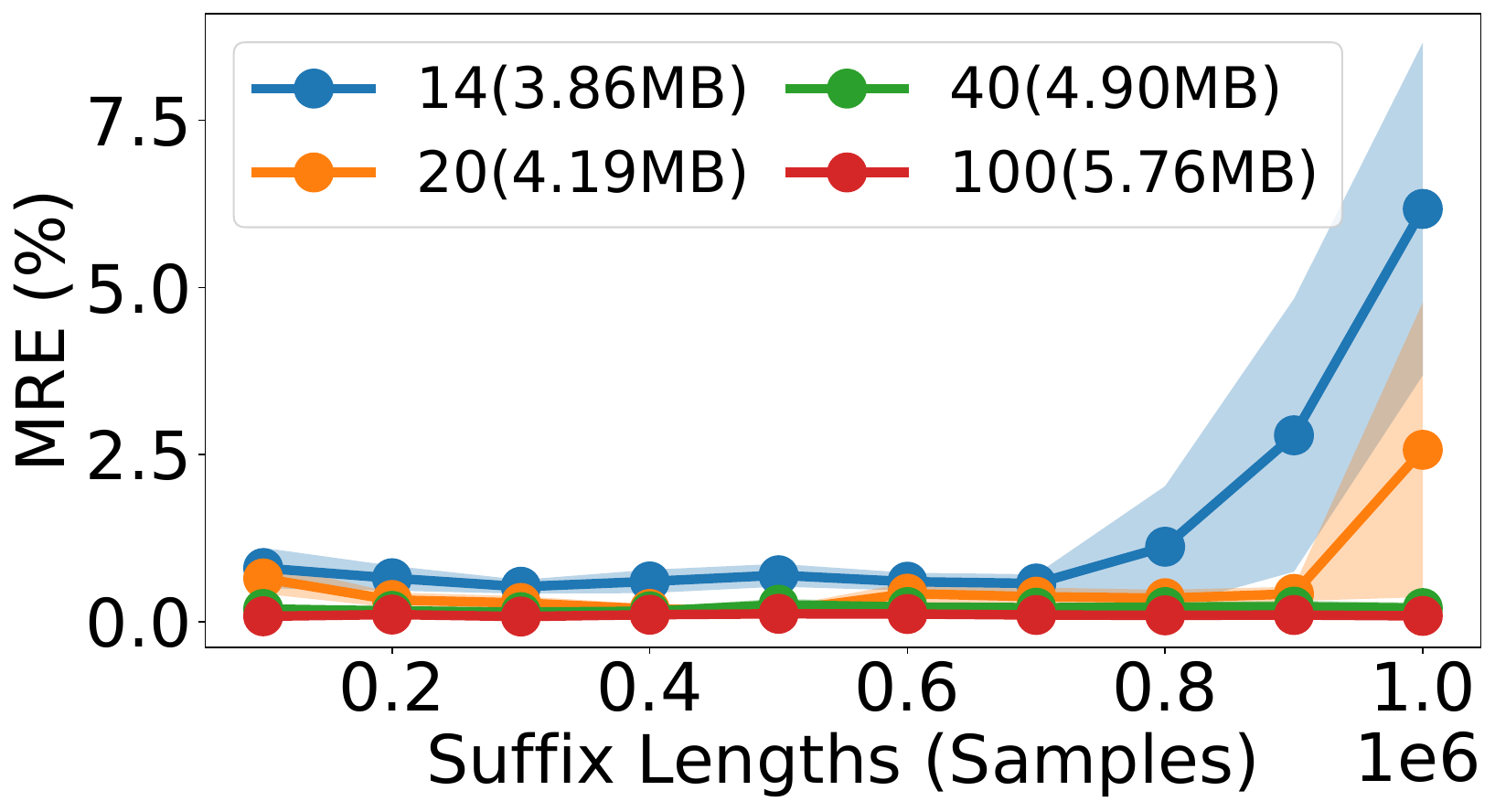}
    	\centering
    	\vspace{-.2in}	
        
    	{\footnotesize(d) Distinct counting, Multi Suffix}
    \end{minipage}
    
    \begin{minipage}[t]{.49\linewidth}
     \includegraphics[width=1\linewidth]{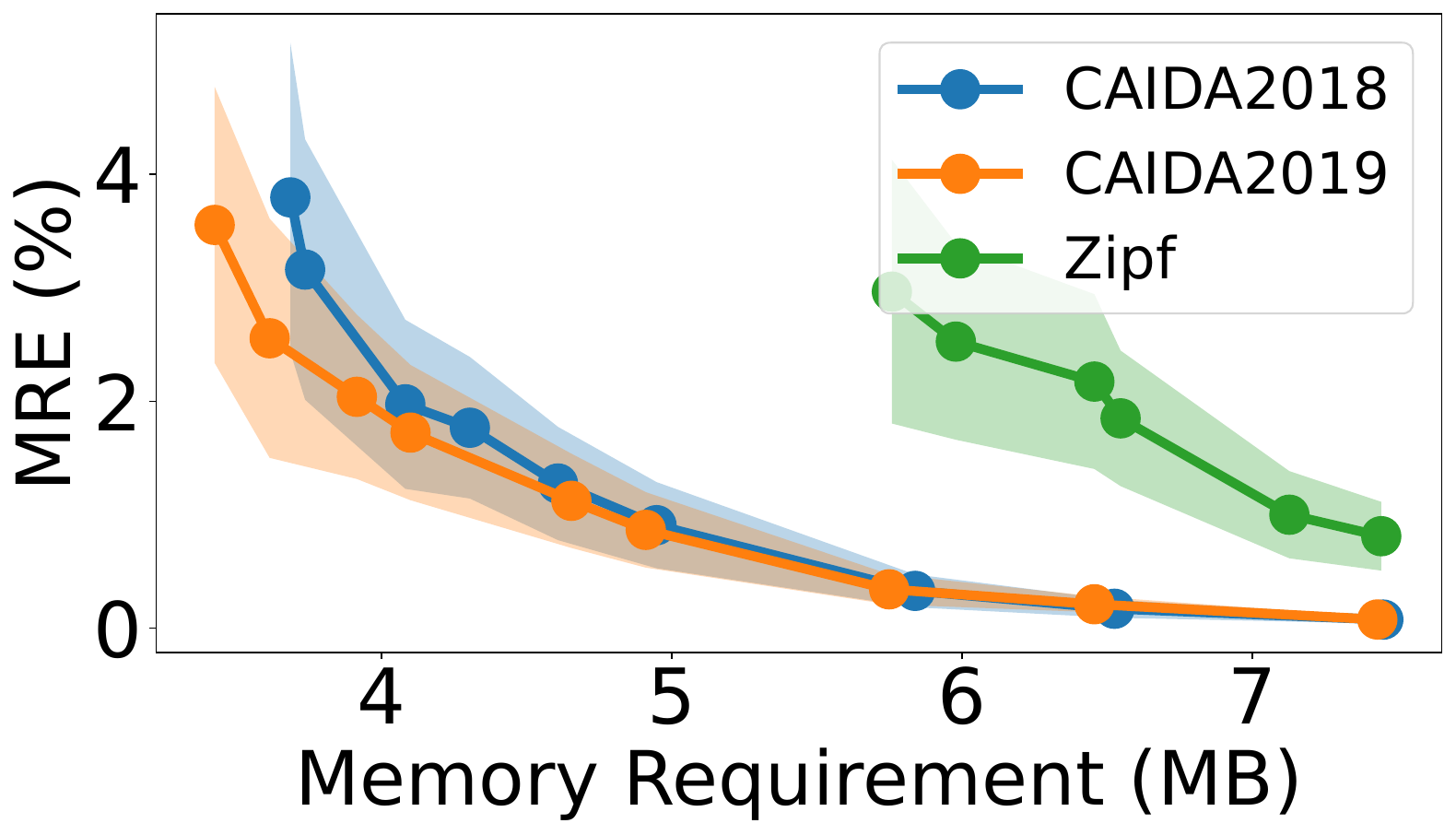}
    	\centering
    	\vspace{-.2in}
        
    	{\footnotesize(e) $L_2$ norm, 300-600K Interval}
    \end{minipage}
    \begin{minipage}[t]{.49\linewidth}
     \includegraphics[width=1\linewidth]{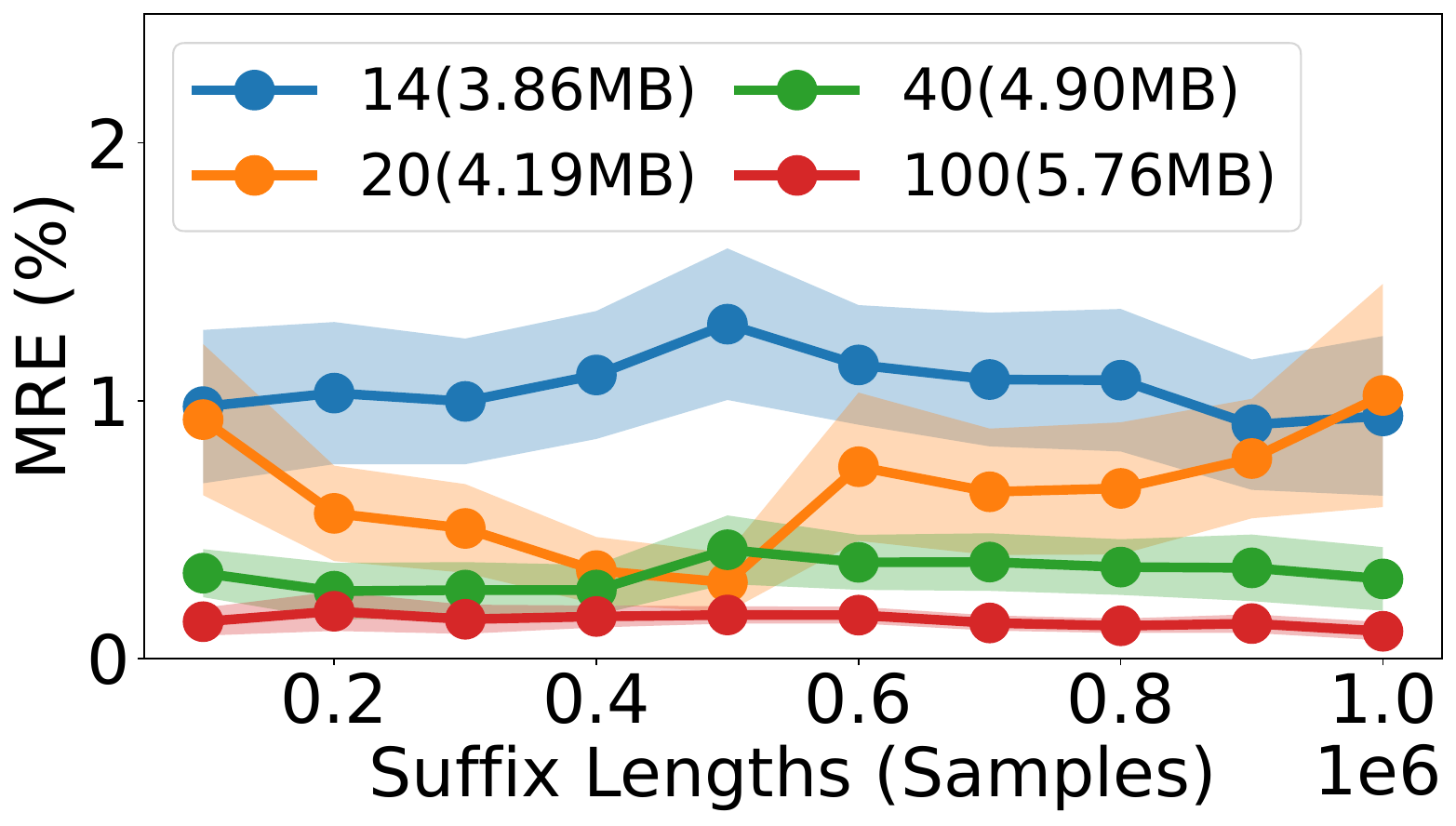}
    	\centering
    	\vspace{-.2in}
        
    	{\footnotesize(f) $L_2$ norm, Multi Suffix}
    \end{minipage}
    \vspace{-3mm}
\mycaption{fig:entropy_error}{EHUniv's mean relative errors of entropy, distinct counting, and $L_2$ norm estimation for (a) memory-error in 300K-600K intervals, and (b) various  suffix lengths on CAIDA2018}{\footnotesize \change{A legend label \texttt{x(z MB)} denotes an EHUniv configuration with $k_{EH}=x$, a 16-layer universal sketch (8 layers with 3-rows, 2048-column CS, and 8 layers with 3-row, 512-column CS), and $z$ MB total memory, where CS referes to Count Sketch.}}
\end{figure}

\subsection{\change{\sysname{} Sensitivity Analysis}}  \label{sec:sensitivity_analysis}
\subsubsection{Accuracy with different memory consumption and sub-window sizes.} \label{sec:eval_error} 
To evaluate memory consumption and empirical errors, we set the sliding window size to 1M samples and query sub-windows ranging from 300K to 600K samples and different suffix lengths on real-world traces. Differences in datasets primarily stem from the skewness of workload distribution.  
Fig.~\ref{fig:avg_error} and Fig.~\ref{fig:quantile_error} present the average statistic and quantile statistic results. Both Fig.~\ref{fig:avg_error}(a) and Fig.~\ref{fig:quantile_error}(a) show that  increased memory improves estimation accuracy. Fig.~\ref{fig:avg_error}(b) and Fig.~\ref{fig:quantile_error}(b) show that larger sub-windows also yield smaller estimation errors compared to smaller ones, and more memory enhances accuracy across all sub-window sizes. 
We repeat the above evaluation for entropy, distinct counting (cardinality), and $L_2$ norm statistics, as shown in Fig.~\ref{fig:entropy_error}.
As memory increases, the relative errors for these estimates decrease. With 4MB memory, EHUniv can achieve 2\% MRE for $L_2$ norm, 1\% MRE for entropy, and 2\% MRE for distinct counting on both CAIDA datasets for the $\frac{1}{3}$ sub-window in a 1M sliding window. Additionally, as suffix lengths increase, estimation errors for entropy and distinct counting rise due to the fixed memory space accommodating more data samples. Errors for $L_2$ norm may vary depending on how well suffix lengths align with EH bucket boundaries.  EHUniv approaches near-zero errors when its memory approaches that of the exact algorithm.
\change{Fig.~\ref{fig:tune_sketch_parameter} shows the impact of different configurations to EHKLL and EHUniv. Given a KLL or Universal sketch configuration, larger $k_{EH}$ reduces window alignment error and thus has smaller estimation errors. Given a $k_{EH}$, smaller KLL and Universal sketch memory configuration has larger errors. EHUniv shows smaller memory gap when $k_{EH}$ is larger than 20 because each EH bucket is smaller and thus uses exact map, with errors solely due to window alignment.}

\begin{figure}[t]
\centering
\begin{minipage}[t]{.49\linewidth}
     \includegraphics[width=1\linewidth]{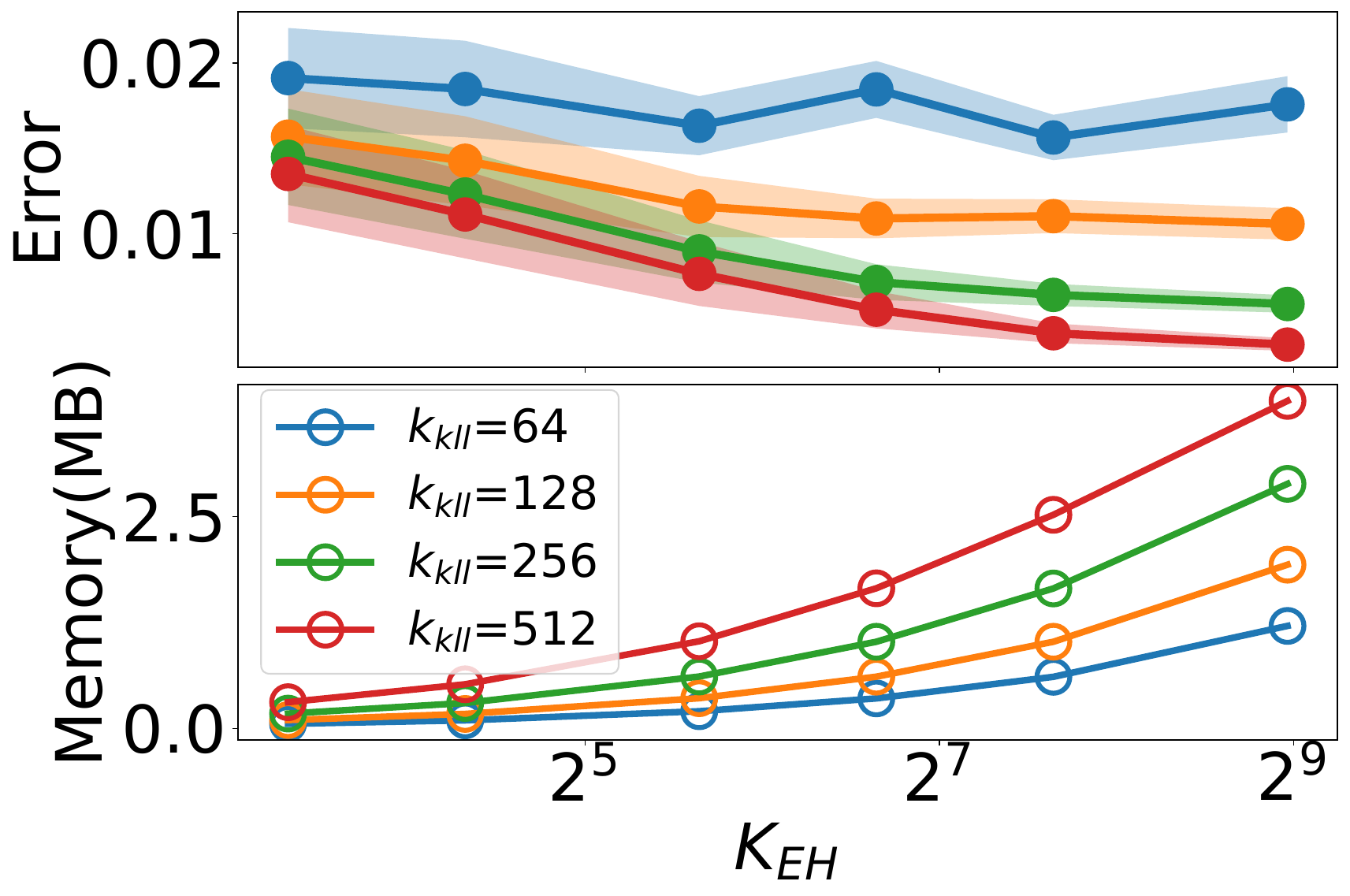}
    	\centering
    	\vspace{-.2in}
    	
    	{\footnotesize(a) 300-600K Interval, Google}
    \end{minipage}
    \begin{minipage}[t]{.455\linewidth}
     \includegraphics[width=1\linewidth]{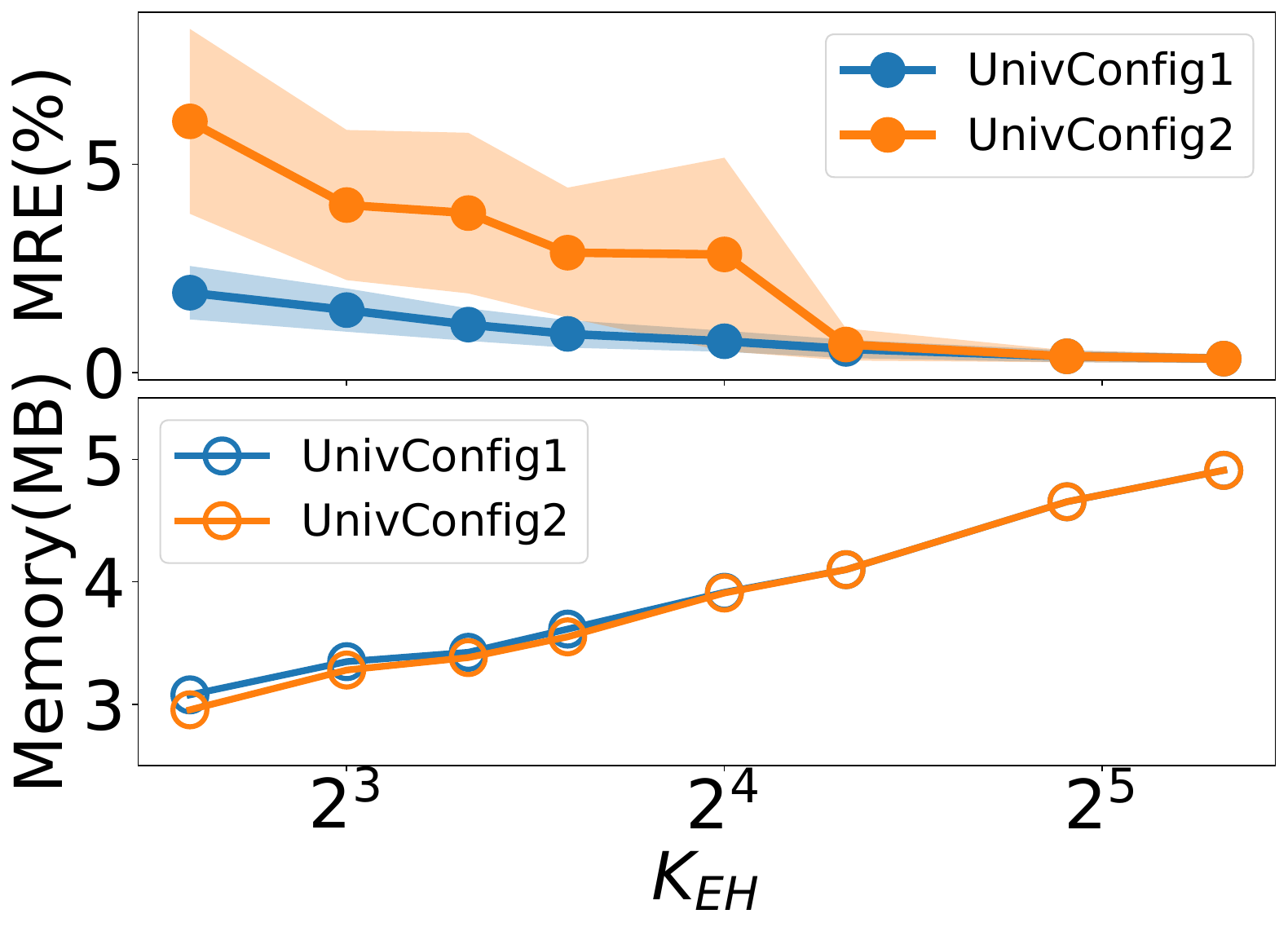}
    	\centering
    	\vspace{-.2in}

    	{\footnotesize(b) 1M-Window, CAIDA2019}
    \end{minipage}
    \vspace{-3mm}
\mycaption{fig:tune_sketch_parameter}{\change{Tuning parameters for EHKLL and EHUniv}}{\footnotesize \change{We show KSTest error for EHKLL. UnivConfig1 is a 16-layer universal sketch (8 layers with 3-row, 1024-column CS, and 8 layers with 3-row, 128-column CS). UnivConfig2 is a 14-layer universal sketch (8 layers with 3-row, 256-column CS, and 6 layers with 3-row, 64-column CS). }}
\end{figure}

\begin{figure}[t]
\centering
    \begin{minipage}[t]{.49\linewidth}
     \includegraphics[width=1\linewidth]{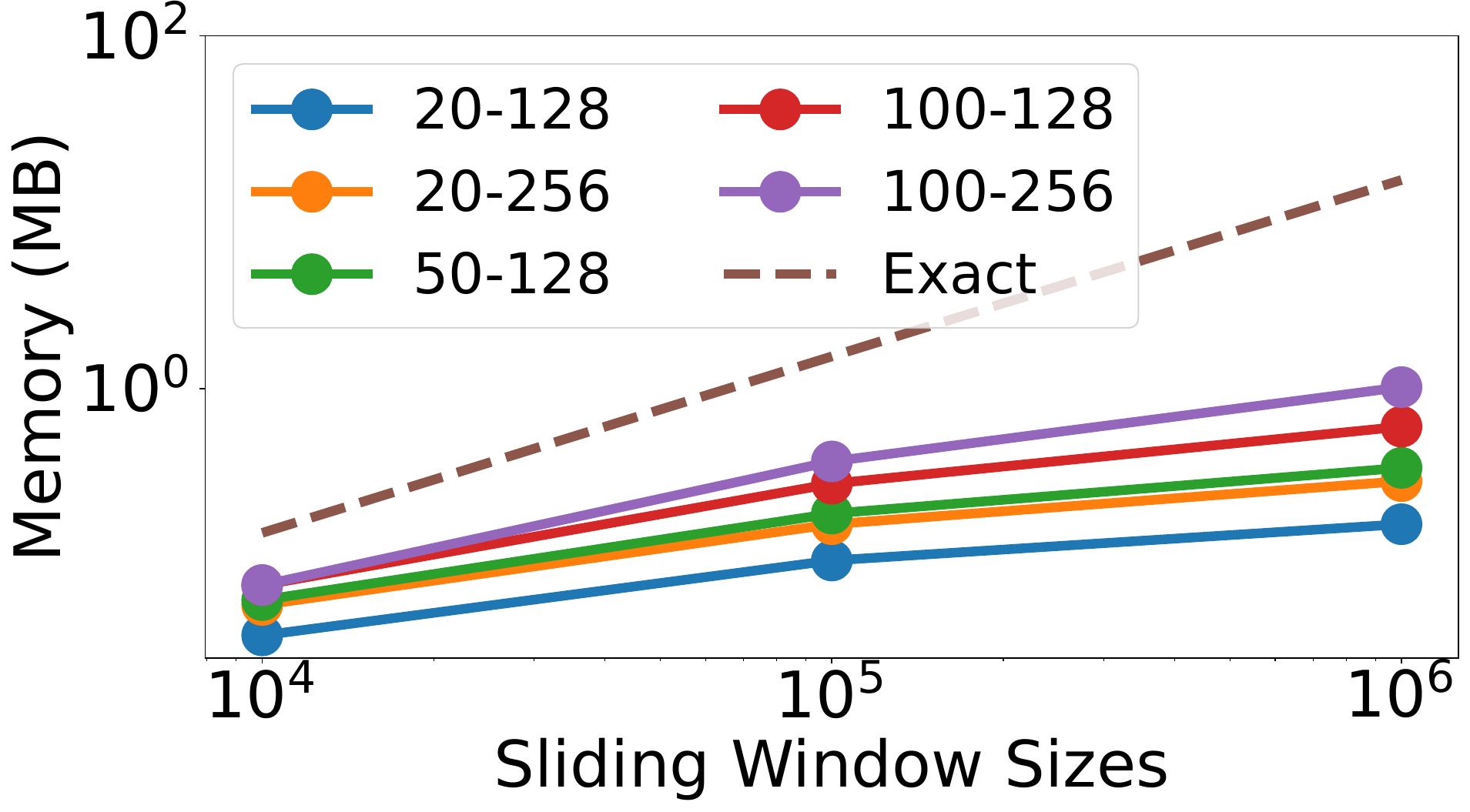}
    	\centering
    	\vspace{-.2in}
    	
    	{\footnotesize(a) EHKLL, Google}
    \end{minipage}
    \begin{minipage}[t]{.49\linewidth}
     \includegraphics[width=1\linewidth]{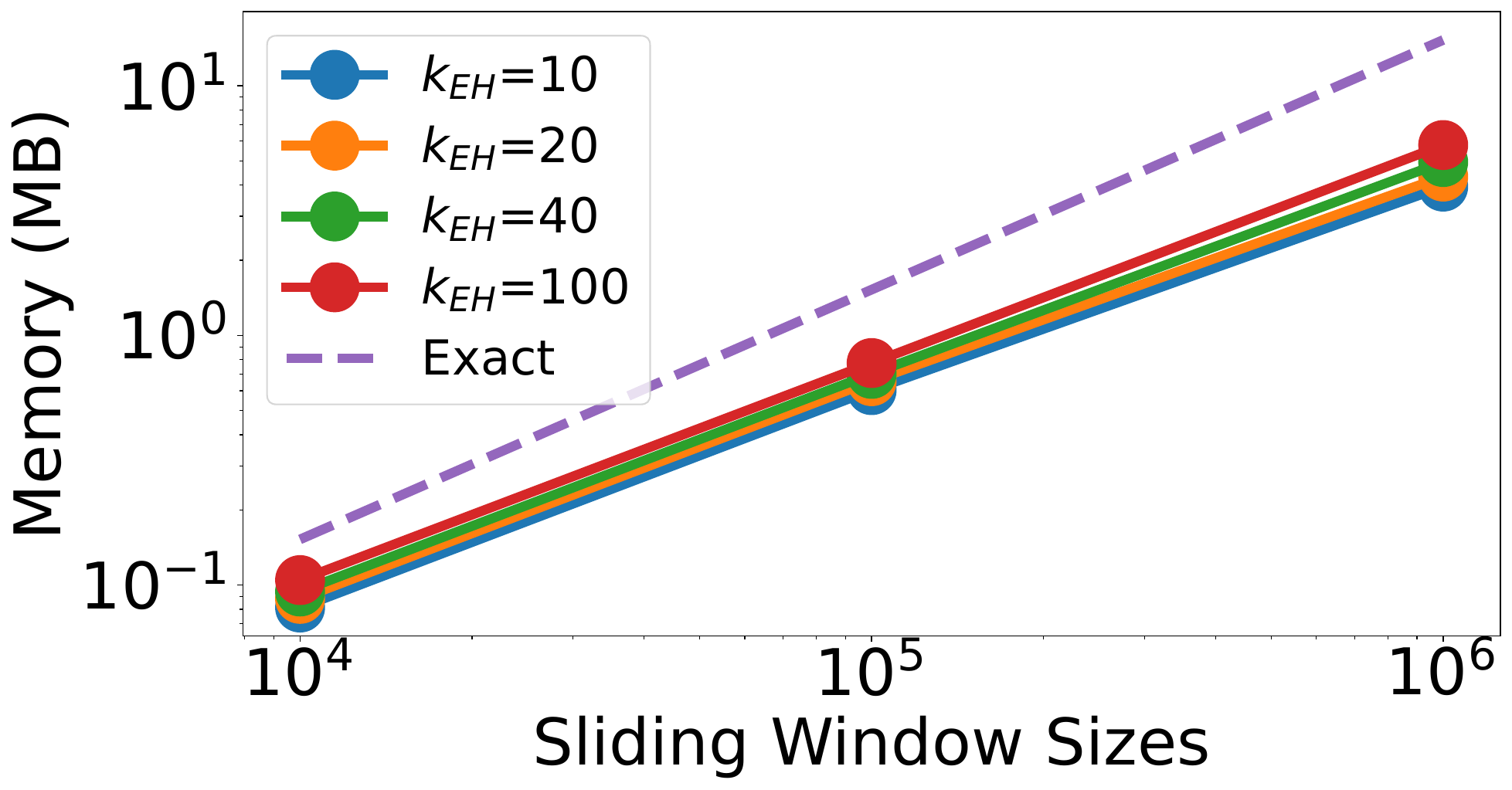}
    	\centering
    	\vspace{-.2in}
    	
    	{\footnotesize(b) EHUniv, CAIDA2018}
    \end{minipage}
    \vspace{-3mm}
\mycaption{fig:memory_timeseries}{\change{Memory usage with sliding window sizes}}{\footnotesize \change{In (a), a legend label \texttt{x-y} denotes an EHKLL configuration with $k_{EH}=x$, $k_{KLL}=y$. The exact baseline stores every point with its associated timestamp.}} 
\end{figure}

\subsubsection{\change{Memory with different timeseries numbers and sliding window sizes. }} \change{We evaluate \sysname{} memory consumption under 5\% error target and 1M-sample sliding window for each timeseries.  Fig.~\ref{fig:insertion_throughput}(e) shows the memory increases linearly as the timeseries number increases, because we allocate one \sysname{} instance per series. Since Dynamic dataset contains uniformly distributed data, EHUniv's memory usage is higher compared to the more skewed CAIDA2019 dataset.
Fig.~\ref{fig:memory_timeseries}(a) and (b) shows the memory consumption of EHKLL and EHUniv with different sliding window sizes and parameter configurations for each timeseries. For each configuration of both EHKLL and EHUniv, the memory usage increases sublinearly as sliding window sizes grows. Larger $k_{EH}$ and $k_{KLL}$ in EHKLL and $k_{EH}$ in EHUniv use more memory. Given a window size, a larger  $k_{EH}$ of EHUniv uses more memory because smaller EH buckets use hash maps for exact computation. Conversely, a smaller  $k_{EH}$  reduces memory usage since larger buckets leverage sketches for compression.}

\subsubsection{\change{Comparing with sliding window algorithm.}}\label{sec:fixed_sliding_compare} 
\change{We compare \sysname{} with fixed sliding window algorithm, e.g., MicroscopeSketch~\cite{wu2023microscopesketch}, on concurrent window queries, evaluating insertion throughput and estimation error across all query windows. We query TopK-frequent item finding over time of EHUniv, and compare MRE and average recall rate over 10 sub-windows (ranging from 100K- to 1M-sample sub-windows in a 1M-sample sliding window) against MicroscopeSketch with HeavyGuardian~\cite{yang2018heavyguardian} and SpaceSaving~\cite{mitzenmacher2012hierarchical} on CAIDA2019. Fig.~\ref{fig:microscope_compare} shows between 1.7MB and 3.3MB memory, \sysname{} has up to 8$\times$ higher insertion throughputs, smaller errors, and higher recall rates than MicroscopeSketch, because of EH's ability to support multiple windows simultaneously. }

%% file: RelatedWork.tex
\section{Related Work}\label{sec:related_work}

\begin{figure}[t]
\centering
    \begin{minipage}[t]{.31\linewidth}
     \includegraphics[width=1\linewidth]{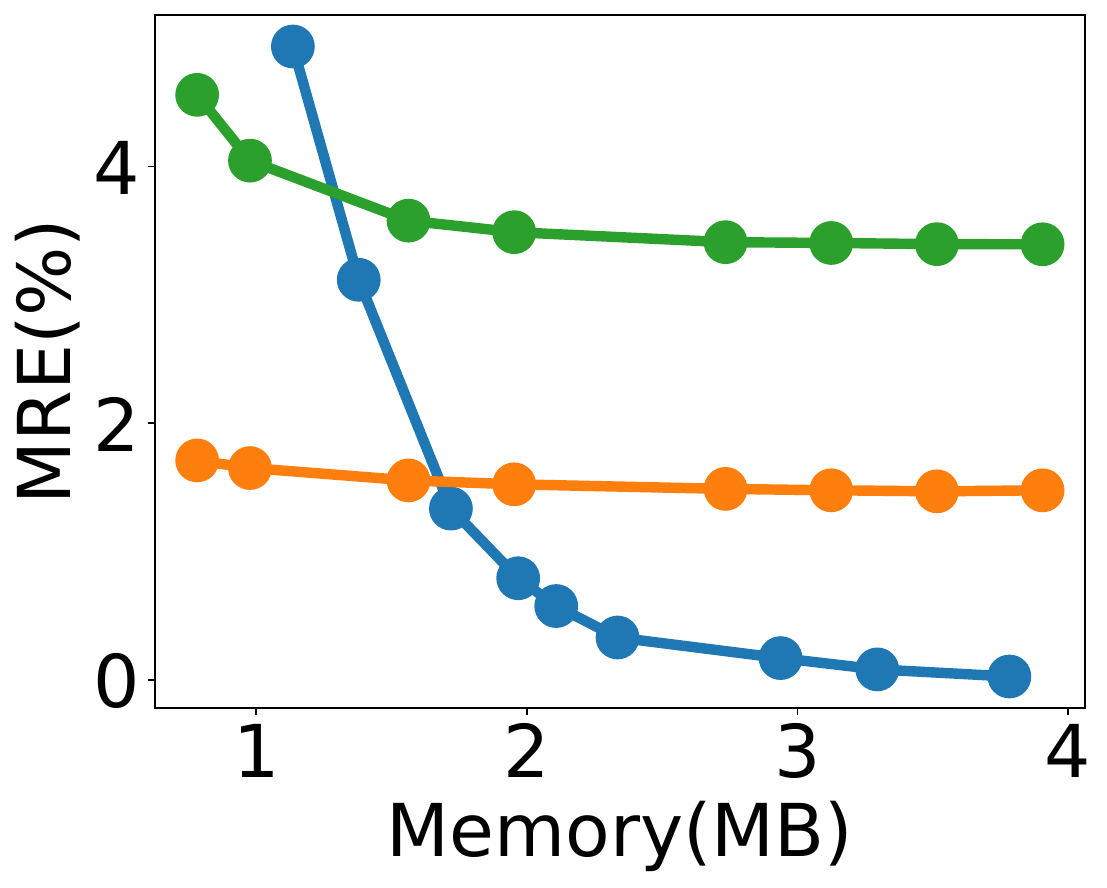}
    	\centering
    	\vspace{-.2in}
    	
    	{\footnotesize(a) Mean Relative Error}
    \end{minipage}
    \begin{minipage}[t]{.34\linewidth}
     \includegraphics[width=1\linewidth]{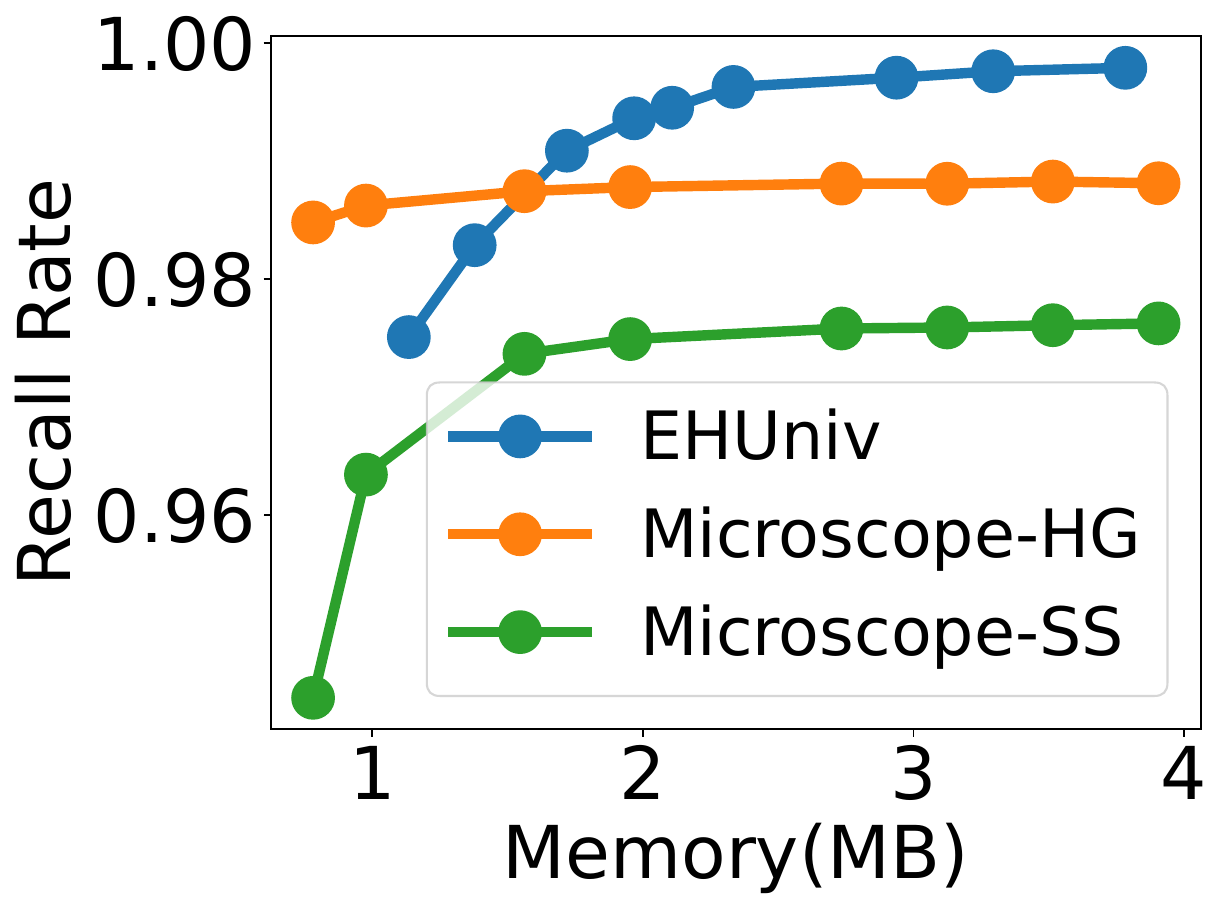}
    	\centering
    	\vspace{-.2in}
    	
    	{\footnotesize(b) Recall Rate}
    \end{minipage}
    \begin{minipage}[t]{.32\linewidth}
     \includegraphics[width=1\linewidth]{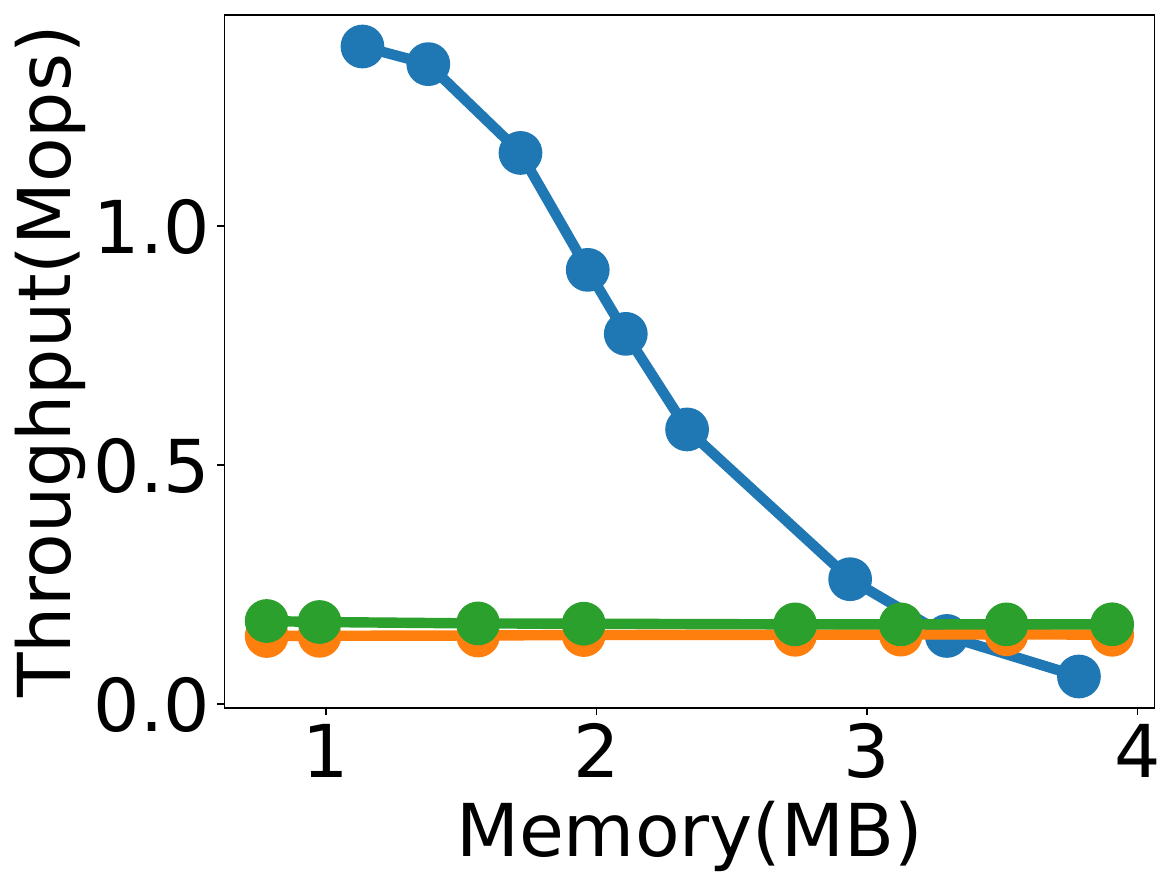}
    	\centering
    	\vspace{-.2in}
    	
    	{\footnotesize(c) Insertion Throughput}
    \end{minipage}
    \vspace{-3mm}
\mycaption{fig:microscope_compare}{\change{TopK-frequent item estimation comparing to MicroscopeSketch with HeavyGuardian (HG) and SpaceSaving (SS)}}{\footnotesize \change{}} 
\end{figure}

\smallskip\noindent\textbf{Window-based summaries.} \change{While various sliding window methods exist, most do not support arbitrary sub-window queries, incurring per-window maintenance efforts. } \cite{arasu2004approximate} addresses approximate frequency counting and quantiles. SlidingSketches~\cite{gou2020sliding} optimizes hash-based sketches like Bloom filter~\cite{bloom1970space} and CountSketch~\cite{charikar2002finding} but lacks support for quantile sketches like KLL. ECM-sketch~\cite{papapetrou2012sketch} enhances CountMin Sketch~\cite{cormode2005improved} by replacing counters with Exponential Histogram for frequency estimation, whereas \sysname{} uses sketches as EH buckets. WCSS~\cite{ben2016heavy} \change{and MicroscopeSketch~\cite{wu2023microscopesketch} support frequency estimation and TopK-frequent item finding, but they lack sub-window querying support and MicroscopeSketch is limited to counter-based sketches. CoopStore~\cite{gan2020coopstore} focuses on offline precomputation of quantiles and frequencies with fixed window sizes and luxury memory during aggregation to reduce errors. SummaryStore~\cite{agrawal2017low} designs approximate timeseries storage with PowerLaw-based sub-window queries, while \sysname{} functions as an in-memory cache and integrates to Prometheus-like monitoring systems with rule query support.}

\change{
\smallskip\noindent\textbf{Approximate timeseries visualization. }
M4~\cite{jugel2014m4} and MinMaxCache~\cite{maroulis2024visualization} solve timeseries data point visualization problem with approximate pixel positions in a canvas, by querying min and max values of data points of a time range. They are orthogonal to our work and don't address the window query bottlenecks.}

\smallskip\noindent\textbf{Approximate query processing (AQP) systems.} Another type of approximate query system focuses on label dimensional queries.  PASS~\cite{liang2021combining, https://doi.org/10.48550/arxiv.2103.15994} and AQP++~\cite{peng2018aqp++} combine precomputed-aggregation and sampling to support SQL queries, including sum, count, min, max, and variance statistics. VerdictDB~\cite{park2018verdictdb} acts as a sampling middlebox between the user interface and the backend database, enabling approximate SQL queries without requiring backend modifications. \change{DHS~\cite{zhao2021dhs} offers streaming data estimation of network traffic frequencies with dynamic sketch memory layout but does not target the time window queries.}

%% file: Conclustions.tex
\section{Conclusions}
We present \sysname{} an approximation-first query caching solution that enhances scalability of queries in timeseries monitoring systems by reducing query latency and operational costs. \sysname{} leverages approximate (sub)window-based query frameworks and sketches for efficient in-memory intermediate query result caching. Our evaluation shows that \sysname{} reduces query latency by up to \change{two orders of magnitude} compared to Prometheus and VictoriaMetrics, and reduces operational costs by two orders of magnitude compared to the state-of-the-art, with only 5\% errors across statistics.

%% file: main.bbl

%% file: main.bbl
\begin{thebibliography}{105}


\ifx \showCODEN    \undefined \def \showCODEN     #1{\unskip}     \fi
\ifx \showDOI      \undefined \def \showDOI       #1{#1}\fi
\ifx \showISBNx    \undefined \def \showISBNx     #1{\unskip}     \fi
\ifx \showISBNxiii \undefined \def \showISBNxiii  #1{\unskip}     \fi
\ifx \showISSN     \undefined \def \showISSN      #1{\unskip}     \fi
\ifx \showLCCN     \undefined \def \showLCCN      #1{\unskip}     \fi
\ifx \shownote     \undefined \def \shownote      #1{#1}          \fi
\ifx \showarticletitle \undefined \def \showarticletitle #1{#1}   \fi
\ifx \showURL      \undefined \def \showURL       {\relax}        \fi
\providecommand\bibfield[2]{#2}
\providecommand\bibinfo[2]{#2}
\providecommand\natexlab[1]{#1}
\providecommand\showeprint[2][]{arXiv:#2}

\bibitem[\protect\citeauthoryear{??}{pro}{2015}]%
        {promethues-soundcloud}
 \bibinfo{year}{2015}\natexlab{}.
\newblock \bibinfo{title}{Prometheus: Monitoring at SoundCloud}.
\newblock
\newblock
\urldef\tempurl%
\url{https://github.com/pingcap/docs/blob/master/tidb-monitoring-framework.md}
\showURL{%
\tempurl}


\bibitem[\protect\citeauthoryear{??}{ec2}{2019}]%
        {ec2-pps}
 \bibinfo{year}{2019}\natexlab{}.
\newblock \bibinfo{title}{Packets-per-second limits in EC2}.
\newblock
\newblock
\urldef\tempurl%
\url{https://stressgrid.com/blog/pps_limits_in_ec2/}
\showURL{%
Retrieved 2024 from \tempurl}


\bibitem[\protect\citeauthoryear{??}{goo}{2020}]%
        {google-cluster-trace}
 \bibinfo{year}{2020}\natexlab{}.
\newblock \bibinfo{title}{Google ClusterData 2019 traces}.
\newblock
\newblock
\urldef\tempurl%
\url{https://github.com/google/cluster-data/blob/master/ClusterData2019.md}
\showURL{%
Retrieved 2024 from \tempurl}


\bibitem[\protect\citeauthoryear{??}{ec2}{2022}]%
        {ec2-billing}
 \bibinfo{year}{2022}\natexlab{}.
\newblock \bibinfo{title}{Amazon EC2 On-Demand Pricing}.
\newblock
\newblock
\urldef\tempurl%
\url{https://aws.amazon.com/ec2/pricing/on-demand/}
\showURL{%
Retrieved 2024 from \tempurl}


\bibitem[\protect\citeauthoryear{??}{kub}{2022}]%
        {kubernetes}
 \bibinfo{year}{2022}\natexlab{}.
\newblock \bibinfo{title}{Kubernetes}.
\newblock
\newblock
\urldef\tempurl%
\url{https://kubernetes.io/}
\showURL{%
\tempurl}


\bibitem[\protect\citeauthoryear{??}{pro}{2022}]%
        {prometheus-ood}
 \bibinfo{year}{2022}\natexlab{}.
\newblock \bibinfo{title}{Prometheus handling out-of-order samples}.
\newblock
\newblock
\urldef\tempurl%
\url{https://promlabs.com/blog/2022/12/15/understanding-duplicate-samples-and-out-of-order-timestamp-errors-in-prometheus/}
\showURL{%
Retrieved 2024 from \tempurl}


\bibitem[\protect\citeauthoryear{??}{Vic}{2022}]%
        {VictoriaMetrics}
 \bibinfo{year}{2022}\natexlab{}.
\newblock \bibinfo{title}{VictoriaMetrics}.
\newblock
\newblock
\urldef\tempurl%
\url{https://victoriametrics.com}
\showURL{%
\tempurl}


\bibitem[\protect\citeauthoryear{??}{ins}{2023}]%
        {instant-query-grafana}
 \bibinfo{year}{2023}\natexlab{}.
\newblock \bibinfo{title}{PromCon 2023 - Yet Another Streaming PromQL Engine}.
\newblock
\newblock
\urldef\tempurl%
\url{https://www.youtube.com/watch?v=3kM2Asj6hcg}
\showURL{%
Retrieved 2024 from \tempurl}


\bibitem[\protect\citeauthoryear{??}{pro}{2023}]%
        {prom-autoscale}
 \bibinfo{year}{2023}\natexlab{}.
\newblock \bibinfo{title}{Prometheus Metrics based autoscaling in Kubernetes}.
\newblock
\newblock
\urldef\tempurl%
\url{https://gcore.com/learning/prometheus-metrics-based-autoscaling-in-kubernetes/}
\showURL{%
Retrieved 2024 from \tempurl}


\bibitem[\protect\citeauthoryear{??}{aws}{2024}]%
        {aws-price}
 \bibinfo{year}{2024}\natexlab{}.
\newblock \bibinfo{title}{Amazon Managed Service for Prometheus pricing}.
\newblock
\newblock
\urldef\tempurl%
\url{https://aws.amazon.com/prometheus/pricing/}
\showURL{%
Retrieved 2024 from \tempurl}


\bibitem[\protect\citeauthoryear{??}{awe}{2024}]%
        {awesomealerts}
 \bibinfo{year}{2024}\natexlab{}.
\newblock \bibinfo{title}{Awesome Prometheus alerts}.
\newblock
\newblock
\urldef\tempurl%
\url{https://samber.github.io/awesome-prometheus-alerts/}
\showURL{%
\tempurl}


\bibitem[\protect\citeauthoryear{??}{clo}{2024}]%
        {cloudflare}
 \bibinfo{year}{2024}\natexlab{}.
\newblock \bibinfo{title}{Cloudflare Blog - Monitoring our monitoring: how we
  validate our Prometheus alert rules}.
\newblock
\newblock
\urldef\tempurl%
\url{https://blog.cloudflare.com/monitoring-our-monitoring/}
\showURL{%
\tempurl}


\bibitem[\protect\citeauthoryear{??}{clu}{2024}]%
        {cluster-vm}
 \bibinfo{year}{2024}\natexlab{}.
\newblock \bibinfo{title}{Cluster VictoriaMetrics}.
\newblock
\newblock
\urldef\tempurl%
\url{https://docs.victoriametrics.com/cluster-victoriametrics/}
\showURL{%
Retrieved 2024 from \tempurl}


\bibitem[\protect\citeauthoryear{??}{dat}{2024}]%
        {datadog-anomaly-detection}
 \bibinfo{year}{2024}\natexlab{}.
\newblock \bibinfo{title}{{DataDog Anomaly Detection}}.
\newblock
  \bibinfo{howpublished}{\url{https://docs.datadoghq.com/monitors/types/anomaly/}}.
\newblock


\bibitem[\protect\citeauthoryear{??}{fas}{2024}]%
        {fastcache}
 \bibinfo{year}{2024}\natexlab{}.
\newblock \bibinfo{title}{{Fastcache used by VictoriaMetrics}}.
\newblock
\newblock
\urldef\tempurl%
\url{https://github.com/VictoriaMetrics/fastcache}
\showURL{%
\tempurl}


\bibitem[\protect\citeauthoryear{??}{goo}{2024}]%
        {google-scaling}
 \bibinfo{year}{2024}\natexlab{}.
\newblock \bibinfo{title}{{Google Cloud scales based on Monitoring metrics}}.
\newblock
\newblock
\urldef\tempurl%
\url{https://cloud.google.com/compute/docs/autoscaler/scaling-cloud-monitoring-metrics}
\showURL{%
\tempurl}


\bibitem[\protect\citeauthoryear{??}{gke}{2024}]%
        {gke-quotas}
 \bibinfo{year}{2024}\natexlab{}.
\newblock \bibinfo{title}{{Google Kubernetes Engine Quotas and Limits}}.
\newblock
\newblock
\urldef\tempurl%
\url{https://cloud.google.com/kubernetes-engine/quotas}
\showURL{%
\tempurl}


\bibitem[\protect\citeauthoryear{??}{gra}{2024}]%
        {grafana-dashboards}
 \bibinfo{year}{2024}\natexlab{}.
\newblock \bibinfo{title}{{Grafana Dashboards}}.
\newblock
  \bibinfo{howpublished}{\url{https://grafana.com/grafana/dashboards/}}.
\newblock


\bibitem[\protect\citeauthoryear{??}{mim}{2024a}]%
        {mimir}
 \bibinfo{year}{2024}\natexlab{a}.
\newblock \bibinfo{title}{Grafana Mimir}.
\newblock
\newblock
\urldef\tempurl%
\url{https://grafana.com/docs/mimir/latest/}
\showURL{%
\tempurl}


\bibitem[\protect\citeauthoryear{??}{mim}{2024b}]%
        {mimir-redis}
 \bibinfo{year}{2024}\natexlab{b}.
\newblock \bibinfo{title}{{Grafana Mimir uses Redis or Memcached as
  chunks-cache, index-cache, results-cache and metadata-cache}}.
\newblock
\newblock
\urldef\tempurl%
\url{https://grafana.com/docs/helm-charts/mimir-distributed/latest/configure/configure-redis-cache/}
\showURL{%
\tempurl}


\bibitem[\protect\citeauthoryear{??}{Inf}{2024}]%
        {InfluxDB}
 \bibinfo{year}{2024}\natexlab{}.
\newblock \bibinfo{title}{InfluxDB}.
\newblock
\newblock
\urldef\tempurl%
\url{https://www.influxdata.com/}
\showURL{%
\tempurl}


\bibitem[\protect\citeauthoryear{??}{kub}{2024}]%
        {kubernetesmetric}
 \bibinfo{year}{2024}\natexlab{}.
\newblock \bibinfo{title}{Kubernetes monitoring with Prometheus}.
\newblock
\newblock
\urldef\tempurl%
\url{https://prometheus.io/docs/prometheus/latest/configuration/configuration/#kubernetes_sd_config}
\showURL{%
Retrieved 2024 from \tempurl}


\bibitem[\protect\citeauthoryear{??}{mem}{2024}]%
        {memcached}
 \bibinfo{year}{2024}\natexlab{}.
\newblock \bibinfo{title}{{Memcached}}.
\newblock \bibinfo{howpublished}{\url{https://memcached.org}}.
\newblock


\bibitem[\protect\citeauthoryear{??}{jun}{2024}]%
        {junos}
 \bibinfo{year}{2024}\natexlab{}.
\newblock \bibinfo{title}{Monitoring Juniper Networks with Prometheus}.
\newblock
\newblock
\urldef\tempurl%
\url{https://github.com/czerwonk/junos_exporter}
\showURL{%
Retrieved 2024 from \tempurl}


\bibitem[\protect\citeauthoryear{??}{5gm}{2024}]%
        {5gmetric}
 \bibinfo{year}{2024}\natexlab{}.
\newblock \bibinfo{title}{Open5GS Metrics with Prometheus}.
\newblock
\newblock
\urldef\tempurl%
\url{https://open5gs.org/open5gs/docs/tutorial/04-metrics-prometheus/}
\showURL{%
Retrieved 2024 from \tempurl}


\bibitem[\protect\citeauthoryear{??}{pro}{2024a}]%
        {prometheus-rules}
 \bibinfo{year}{2024}\natexlab{a}.
\newblock \bibinfo{title}{Prometheus Configurations}.
\newblock
\newblock
\urldef\tempurl%
\url{https://prometheus.io/docs/prometheus/latest/configuration/configuration/}
\showURL{%
Retrieved 2024 from \tempurl}


\bibitem[\protect\citeauthoryear{??}{pro}{2024b}]%
        {promethuesfunction}
 \bibinfo{year}{2024}\natexlab{b}.
\newblock \bibinfo{title}{Prometheus functions}.
\newblock
\newblock
\urldef\tempurl%
\url{https://prometheus.io/docs/prometheus/latest/querying/functions/}
\showURL{%
Retrieved 2024 from \tempurl}


\bibitem[\protect\citeauthoryear{??}{pro}{2024c}]%
        {promql}
 \bibinfo{year}{2024}\natexlab{c}.
\newblock \bibinfo{title}{Prometheus Query Language.}
\newblock
\newblock
\urldef\tempurl%
\url{https://prometheus.io/docs/prometheus/latest/querying/basics/}
\showURL{%
Retrieved 2024 from \tempurl}


\bibitem[\protect\citeauthoryear{??}{snm}{2024}]%
        {snmp}
 \bibinfo{year}{2024}\natexlab{}.
\newblock \bibinfo{title}{Prometheus SNMP exporter}.
\newblock
\newblock
\urldef\tempurl%
\url{https://github.com/prometheus/snmp_exporter}
\showURL{%
Retrieved 2024 from \tempurl}


\bibitem[\protect\citeauthoryear{??}{red}{2024}]%
        {redis.io}
 \bibinfo{year}{2024}\natexlab{}.
\newblock \bibinfo{title}{{Redis}}.
\newblock \bibinfo{howpublished}{\url{https://redis.io}}.
\newblock


\bibitem[\protect\citeauthoryear{??}{tha}{2024a}]%
        {thanos}
 \bibinfo{year}{2024}\natexlab{a}.
\newblock \bibinfo{title}{Thanos}.
\newblock
\newblock
\urldef\tempurl%
\url{https://thanos.io/}
\showURL{%
\tempurl}


\bibitem[\protect\citeauthoryear{??}{tha}{2024b}]%
        {thanos-downsampling}
 \bibinfo{year}{2024}\natexlab{b}.
\newblock \bibinfo{title}{Thanos Downsampling, Resolution and Retention}.
\newblock
\newblock
\urldef\tempurl%
\url{https://thanos.io/v0.8/components/compact/}
\showURL{%
Retrieved 2024 from \tempurl}


\bibitem[\protect\citeauthoryear{??}{cai}{2024}]%
        {caida-dataset}
 \bibinfo{year}{2024}\natexlab{}.
\newblock \bibinfo{title}{{The CAIDA UCSD Anonymized Internet Traces}}.
\newblock
  \bibinfo{howpublished}{\url{https://www.caida.org/catalog/datasets/passive_dataset/}}.
\newblock
\newblock
\shownote{Online.}


\bibitem[\protect\citeauthoryear{??}{vma}{2024}]%
        {vmanomaly}
 \bibinfo{year}{2024}\natexlab{}.
\newblock \bibinfo{title}{VictoriaMetrics Anomaly Detection.}
\newblock
\newblock
\urldef\tempurl%
\url{https://victoriametrics.com/blog/victoriametrics-anomaly-detection-handbook-chapter-2/index.html}
\showURL{%
Retrieved 2024 from \tempurl}


\bibitem[\protect\citeauthoryear{??}{vm-}{2024a}]%
        {vm-backfilling}
 \bibinfo{year}{2024}\natexlab{a}.
\newblock \bibinfo{title}{VictoriaMetrics backfilling support for out-of-order
  samples}.
\newblock
\newblock
\urldef\tempurl%
\url{https://docs.victoriametrics.com/#backfilling}
\showURL{%
Retrieved 2025 from \tempurl}


\bibitem[\protect\citeauthoryear{??}{vm-}{2024b}]%
        {vm-dedup}
 \bibinfo{year}{2024}\natexlab{b}.
\newblock \bibinfo{title}{VictoriaMetrics Deduplication}.
\newblock
\newblock
\urldef\tempurl%
\url{https://docs.victoriametrics.com/#deduplication}
\showURL{%
Retrieved 2024 from \tempurl}


\bibitem[\protect\citeauthoryear{??}{vm-}{2024c}]%
        {vm-parallel-query}
 \bibinfo{year}{2024}\natexlab{c}.
\newblock \bibinfo{title}{{VictoriaMetrics parallel query in vm-select}}.
\newblock
  \bibinfo{howpublished}{\url{https://github.com/VictoriaMetrics/VictoriaMetrics/issues/2886
  }}.
\newblock


\bibitem[\protect\citeauthoryear{??}{vm-}{2024d}]%
        {vm-price}
 \bibinfo{year}{2024}\natexlab{d}.
\newblock \bibinfo{title}{VictoriaMetrics Pricing compared to Prometheus.}
\newblock
\newblock
\urldef\tempurl%
\url{https://victoriametrics.com/blog/managed-prometheus-pricing/}
\showURL{%
Retrieved 2024 from \tempurl}


\bibitem[\protect\citeauthoryear{??}{vic}{2024}]%
        {victoriarollup}
 \bibinfo{year}{2024}\natexlab{}.
\newblock \bibinfo{title}{VictoriaMetrics rollup functions}.
\newblock
\newblock
\urldef\tempurl%
\url{https://docs.victoriametrics.com/metricsql/#rollup-functions}
\showURL{%
Retrieved 2024 from \tempurl}


\bibitem[\protect\citeauthoryear{??}{vm-}{2024e}]%
        {vm-single-version}
 \bibinfo{year}{2024}\natexlab{e}.
\newblock \bibinfo{title}{VictoriaMetrics Single Version.}
\newblock
\newblock
\urldef\tempurl%
\url{https://docs.victoriametrics.com/single-server-victoriametrics/}
\showURL{%
Retrieved 2024 from \tempurl}


\bibitem[\protect\citeauthoryear{Abraham, Allen, Barykin, Borkar, Chopra,
  Gerea, Merl, Metzler, Reiss, Subramanian, et~al\mbox{.}}{Abraham
  et~al\mbox{.}}{2013}]%
        {abraham2013scuba}
\bibfield{author}{\bibinfo{person}{Lior Abraham}, \bibinfo{person}{John Allen},
  \bibinfo{person}{Oleksandr Barykin}, \bibinfo{person}{Vinayak Borkar},
  \bibinfo{person}{Bhuwan Chopra}, \bibinfo{person}{Ciprian Gerea},
  \bibinfo{person}{Daniel Merl}, \bibinfo{person}{Josh Metzler},
  \bibinfo{person}{David Reiss}, \bibinfo{person}{Subbu Subramanian},
  {et~al\mbox{.}}} \bibinfo{year}{2013}\natexlab{}.
\newblock \showarticletitle{Scuba: Diving into data at facebook}.
\newblock \bibinfo{journal}{\emph{Proceedings of the VLDB Endowment}}
  \bibinfo{volume}{6}, \bibinfo{number}{11} (\bibinfo{year}{2013}),
  \bibinfo{pages}{1057--1067}.
\newblock


\bibitem[\protect\citeauthoryear{Acharya, Gibbons, Poosala, and
  Ramaswamy}{Acharya et~al\mbox{.}}{1999}]%
        {acharya1999aqua}
\bibfield{author}{\bibinfo{person}{Swarup Acharya}, \bibinfo{person}{Phillip~B
  Gibbons}, \bibinfo{person}{Viswanath Poosala}, {and} \bibinfo{person}{Sridhar
  Ramaswamy}.} \bibinfo{year}{1999}\natexlab{}.
\newblock \showarticletitle{The aqua approximate query answering system}. In
  \bibinfo{booktitle}{\emph{Proceedings of the 1999 ACM SIGMOD international
  conference on Management of data}}. \bibinfo{pages}{574--576}.
\newblock


\bibitem[\protect\citeauthoryear{Agarwal, Mozafari, Panda, Milner, Madden, and
  Stoica}{Agarwal et~al\mbox{.}}{2013}]%
        {agarwal2013blinkdb}
\bibfield{author}{\bibinfo{person}{Sameer Agarwal}, \bibinfo{person}{Barzan
  Mozafari}, \bibinfo{person}{Aurojit Panda}, \bibinfo{person}{Henry Milner},
  \bibinfo{person}{Samuel Madden}, {and} \bibinfo{person}{Ion Stoica}.}
  \bibinfo{year}{2013}\natexlab{}.
\newblock \showarticletitle{BlinkDB: queries with bounded errors and bounded
  response times on very large data}. In \bibinfo{booktitle}{\emph{Proceedings
  of the 8th ACM European conference on computer systems}}.
  \bibinfo{pages}{29--42}.
\newblock


\bibitem[\protect\citeauthoryear{Agrawal and Vulimiri}{Agrawal and
  Vulimiri}{2017}]%
        {agrawal2017low}
\bibfield{author}{\bibinfo{person}{Nitin Agrawal} {and} \bibinfo{person}{Ashish
  Vulimiri}.} \bibinfo{year}{2017}\natexlab{}.
\newblock \showarticletitle{Low-latency analytics on colossal data streams with
  summarystore}. In \bibinfo{booktitle}{\emph{Proceedings of the 26th Symposium
  on Operating Systems Principles}}. \bibinfo{pages}{647--664}.
\newblock


\bibitem[\protect\citeauthoryear{Antonakakis, April, Bailey, Bernhard,
  Bursztein, Cochran, Durumeric, Halderman, Invernizzi, Kallitsis,
  et~al\mbox{.}}{Antonakakis et~al\mbox{.}}{2017}]%
        {antonakakis2017understanding}
\bibfield{author}{\bibinfo{person}{Manos Antonakakis}, \bibinfo{person}{Tim
  April}, \bibinfo{person}{Michael Bailey}, \bibinfo{person}{Matt Bernhard},
  \bibinfo{person}{Elie Bursztein}, \bibinfo{person}{Jaime Cochran},
  \bibinfo{person}{Zakir Durumeric}, \bibinfo{person}{J~Alex Halderman},
  \bibinfo{person}{Luca Invernizzi}, \bibinfo{person}{Michalis Kallitsis},
  {et~al\mbox{.}}} \bibinfo{year}{2017}\natexlab{}.
\newblock \showarticletitle{Understanding the mirai botnet}. In
  \bibinfo{booktitle}{\emph{26th USENIX security symposium (USENIX Security
  17)}}. \bibinfo{pages}{1093--1110}.
\newblock


\bibitem[\protect\citeauthoryear{Arasu and Manku}{Arasu and Manku}{2004}]%
        {arasu2004approximate}
\bibfield{author}{\bibinfo{person}{Arvind Arasu} {and}
  \bibinfo{person}{Gurmeet~Singh Manku}.} \bibinfo{year}{2004}\natexlab{}.
\newblock \showarticletitle{Approximate counts and quantiles over sliding
  windows}. In \bibinfo{booktitle}{\emph{Proceedings of the twenty-third ACM
  SIGMOD-SIGACT-SIGART symposium on Principles of database systems}}.
  \bibinfo{pages}{286--296}.
\newblock


\bibitem[\protect\citeauthoryear{Basat, Einziger, Keslassy, Orda, Vargaftik,
  and Waisbard}{Basat et~al\mbox{.}}{2018}]%
        {basat2018memento}
\bibfield{author}{\bibinfo{person}{Ran~Ben Basat}, \bibinfo{person}{Gil
  Einziger}, \bibinfo{person}{Isaac Keslassy}, \bibinfo{person}{Ariel Orda},
  \bibinfo{person}{Shay Vargaftik}, {and} \bibinfo{person}{Erez Waisbard}.}
  \bibinfo{year}{2018}\natexlab{}.
\newblock \showarticletitle{Memento: Making sliding windows efficient for heavy
  hitters}. In \bibinfo{booktitle}{\emph{Proceedings of the 14th International
  Conference on Emerging Networking EXperiments and Technologies}}.
  \bibinfo{pages}{254--266}.
\newblock


\bibitem[\protect\citeauthoryear{Ben-Basat, Einziger, Friedman, and
  Kassner}{Ben-Basat et~al\mbox{.}}{2016}]%
        {ben2016heavy}
\bibfield{author}{\bibinfo{person}{Ran Ben-Basat}, \bibinfo{person}{Gil
  Einziger}, \bibinfo{person}{Roy Friedman}, {and} \bibinfo{person}{Yaron
  Kassner}.} \bibinfo{year}{2016}\natexlab{}.
\newblock \showarticletitle{Heavy hitters in streams and sliding windows}. In
  \bibinfo{booktitle}{\emph{IEEE INFOCOM 2016-The 35th Annual IEEE
  International Conference on Computer Communications}}. IEEE,
  \bibinfo{pages}{1--9}.
\newblock


\bibitem[\protect\citeauthoryear{Ben~Basat, Einziger, Friedman, and
  Kassner}{Ben~Basat et~al\mbox{.}}{2019}]%
        {ben2019succinct}
\bibfield{author}{\bibinfo{person}{Ran Ben~Basat}, \bibinfo{person}{Gil
  Einziger}, \bibinfo{person}{Roy Friedman}, {and} \bibinfo{person}{Yaron
  Kassner}.} \bibinfo{year}{2019}\natexlab{}.
\newblock \showarticletitle{Succinct summing over sliding windows}.
\newblock \bibinfo{journal}{\emph{Algorithmica}}  \bibinfo{volume}{81}
  (\bibinfo{year}{2019}), \bibinfo{pages}{2072--2091}.
\newblock


\bibitem[\protect\citeauthoryear{Berger and Zhou}{Berger and Zhou}{2014}]%
        {berger2014kolmogorov}
\bibfield{author}{\bibinfo{person}{Vance~W Berger} {and}
  \bibinfo{person}{YanYan Zhou}.} \bibinfo{year}{2014}\natexlab{}.
\newblock \showarticletitle{Kolmogorov--smirnov test: Overview}.
\newblock \bibinfo{journal}{\emph{Wiley statsref: Statistics reference online}}
  (\bibinfo{year}{2014}).
\newblock


\bibitem[\protect\citeauthoryear{Bloom}{Bloom}{1970}]%
        {bloom1970space}
\bibfield{author}{\bibinfo{person}{Burton~H Bloom}.}
  \bibinfo{year}{1970}\natexlab{}.
\newblock \showarticletitle{Space/time trade-offs in hash coding with allowable
  errors}.
\newblock \bibinfo{journal}{\emph{Commun. ACM}} \bibinfo{volume}{13},
  \bibinfo{number}{7} (\bibinfo{year}{1970}), \bibinfo{pages}{422--426}.
\newblock


\bibitem[\protect\citeauthoryear{Braverman, Chestnut, Woodruff, and
  Yang}{Braverman et~al\mbox{.}}{2016}]%
        {braverman2016streaming}
\bibfield{author}{\bibinfo{person}{Vladimir Braverman},
  \bibinfo{person}{Stephen~R Chestnut}, \bibinfo{person}{David~P Woodruff},
  {and} \bibinfo{person}{Lin~F Yang}.} \bibinfo{year}{2016}\natexlab{}.
\newblock \showarticletitle{Streaming space complexity of nearly all functions
  of one variable on frequency vectors}. In
  \bibinfo{booktitle}{\emph{Proceedings of the 35th ACM SIGMOD-SIGACT-SIGAI
  Symposium on Principles of Database Systems}}. \bibinfo{pages}{261--276}.
\newblock


\bibitem[\protect\citeauthoryear{Braverman and Ostrovsky}{Braverman and
  Ostrovsky}{2007}]%
        {braverman2007smooth}
\bibfield{author}{\bibinfo{person}{Vladimir Braverman} {and}
  \bibinfo{person}{Rafail Ostrovsky}.} \bibinfo{year}{2007}\natexlab{}.
\newblock \showarticletitle{Smooth histograms for sliding windows}. In
  \bibinfo{booktitle}{\emph{48th Annual IEEE Symposium on Foundations of
  Computer Science (FOCS'07)}}. IEEE, \bibinfo{pages}{283--293}.
\newblock


\bibitem[\protect\citeauthoryear{Braverman and Ostrovsky}{Braverman and
  Ostrovsky}{2010}]%
        {braverman2010zero}
\bibfield{author}{\bibinfo{person}{Vladimir Braverman} {and}
  \bibinfo{person}{Rafail Ostrovsky}.} \bibinfo{year}{2010}\natexlab{}.
\newblock \showarticletitle{Zero-one frequency laws}. In
  \bibinfo{booktitle}{\emph{Proceedings of the forty-second ACM symposium on
  Theory of computing}}. \bibinfo{pages}{281--290}.
\newblock


\bibitem[\protect\citeauthoryear{Chabchoub and He{\'e}brail}{Chabchoub and
  He{\'e}brail}{2010}]%
        {chabchoub2010sliding}
\bibfield{author}{\bibinfo{person}{Yousra Chabchoub} {and}
  \bibinfo{person}{Georges He{\'e}brail}.} \bibinfo{year}{2010}\natexlab{}.
\newblock \showarticletitle{Sliding hyperloglog: Estimating cardinality in a
  data stream over a sliding window}. In \bibinfo{booktitle}{\emph{2010 IEEE
  International Conference on Data Mining Workshops}}. IEEE,
  \bibinfo{pages}{1297--1303}.
\newblock


\bibitem[\protect\citeauthoryear{Chang, Zhang, Tang, Yin, Chang,
  Hasegawa-Johnson, and Huang}{Chang et~al\mbox{.}}{2017}]%
        {chang2017streaming}
\bibfield{author}{\bibinfo{person}{Shiyu Chang}, \bibinfo{person}{Yang Zhang},
  \bibinfo{person}{Jiliang Tang}, \bibinfo{person}{Dawei Yin},
  \bibinfo{person}{Yi Chang}, \bibinfo{person}{Mark~A Hasegawa-Johnson}, {and}
  \bibinfo{person}{Thomas~S Huang}.} \bibinfo{year}{2017}\natexlab{}.
\newblock \showarticletitle{Streaming recommender systems}. In
  \bibinfo{booktitle}{\emph{Proceedings of the 26th international conference on
  world wide web}}. \bibinfo{pages}{381--389}.
\newblock


\bibitem[\protect\citeauthoryear{Charikar, Chen, and Farach-Colton}{Charikar
  et~al\mbox{.}}{2002}]%
        {charikar2002finding}
\bibfield{author}{\bibinfo{person}{Moses Charikar}, \bibinfo{person}{Kevin
  Chen}, {and} \bibinfo{person}{Martin Farach-Colton}.}
  \bibinfo{year}{2002}\natexlab{}.
\newblock \showarticletitle{Finding frequent items in data streams}. In
  \bibinfo{booktitle}{\emph{International Colloquium on Automata, Languages,
  and Programming}}. Springer, \bibinfo{pages}{693--703}.
\newblock


\bibitem[\protect\citeauthoryear{Cormode and Muthukrishnan}{Cormode and
  Muthukrishnan}{2005}]%
        {cormode2005improved}
\bibfield{author}{\bibinfo{person}{Graham Cormode} {and} \bibinfo{person}{Shan
  Muthukrishnan}.} \bibinfo{year}{2005}\natexlab{}.
\newblock \showarticletitle{An improved data stream summary: the count-min
  sketch and its applications}.
\newblock \bibinfo{journal}{\emph{Journal of Algorithms}} \bibinfo{volume}{55},
  \bibinfo{number}{1} (\bibinfo{year}{2005}), \bibinfo{pages}{58--75}.
\newblock


\bibitem[\protect\citeauthoryear{Cranor, Johnson, Spataschek, and
  Shkapenyuk}{Cranor et~al\mbox{.}}{2003}]%
        {cranor2003gigascope}
\bibfield{author}{\bibinfo{person}{Chuck Cranor}, \bibinfo{person}{Theodore
  Johnson}, \bibinfo{person}{Oliver Spataschek}, {and}
  \bibinfo{person}{Vladislav Shkapenyuk}.} \bibinfo{year}{2003}\natexlab{}.
\newblock \showarticletitle{Gigascope: A stream database for network
  applications}. In \bibinfo{booktitle}{\emph{Proceedings of the 2003 ACM
  SIGMOD international conference on Management of data}}.
  \bibinfo{pages}{647--651}.
\newblock


\bibitem[\protect\citeauthoryear{Datar, Gionis, Indyk, and Motwani}{Datar
  et~al\mbox{.}}{2002}]%
        {datar2002maintaining}
\bibfield{author}{\bibinfo{person}{Mayur Datar}, \bibinfo{person}{Aristides
  Gionis}, \bibinfo{person}{Piotr Indyk}, {and} \bibinfo{person}{Rajeev
  Motwani}.} \bibinfo{year}{2002}\natexlab{}.
\newblock \showarticletitle{Maintaining stream statistics over sliding
  windows}.
\newblock \bibinfo{journal}{\emph{SIAM journal on computing}}
  \bibinfo{volume}{31}, \bibinfo{number}{6} (\bibinfo{year}{2002}),
  \bibinfo{pages}{1794--1813}.
\newblock


\bibitem[\protect\citeauthoryear{David and Barr}{David and Barr}{2021}]%
        {david2021kubernetes}
\bibfield{author}{\bibinfo{person}{Ronen~Ben David} {and}
  \bibinfo{person}{Anat~Bremler Barr}.} \bibinfo{year}{2021}\natexlab{}.
\newblock \showarticletitle{Kubernetes autoscaling: Yoyo attack vulnerability
  and mitigation}.
\newblock \bibinfo{journal}{\emph{arXiv preprint arXiv:2105.00542}}
  (\bibinfo{year}{2021}).
\newblock


\bibitem[\protect\citeauthoryear{Estan and Varghese}{Estan and
  Varghese}{2002}]%
        {estan2002new}
\bibfield{author}{\bibinfo{person}{Cristian Estan} {and}
  \bibinfo{person}{George Varghese}.} \bibinfo{year}{2002}\natexlab{}.
\newblock \showarticletitle{New directions in traffic measurement and
  accounting}. In \bibinfo{booktitle}{\emph{Proceedings of the 2002 conference
  on Applications, technologies, architectures, and protocols for computer
  communications}}. \bibinfo{pages}{323--336}.
\newblock


\bibitem[\protect\citeauthoryear{Gan, Bailis, and Charikar}{Gan
  et~al\mbox{.}}{2020}]%
        {gan2020coopstore}
\bibfield{author}{\bibinfo{person}{Edward Gan}, \bibinfo{person}{Peter Bailis},
  {and} \bibinfo{person}{Moses Charikar}.} \bibinfo{year}{2020}\natexlab{}.
\newblock \showarticletitle{Coopstore: Optimizing precomputed summaries for
  aggregation}.
\newblock \bibinfo{journal}{\emph{Proceedings of the VLDB Endowment}}
  \bibinfo{volume}{13}, \bibinfo{number}{12} (\bibinfo{year}{2020}),
  \bibinfo{pages}{2174--2187}.
\newblock


\bibitem[\protect\citeauthoryear{Gou, He, Zhang, Wang, Liu, Yang, Wang, and
  Cui}{Gou et~al\mbox{.}}{2020}]%
        {gou2020sliding}
\bibfield{author}{\bibinfo{person}{Xiangyang Gou}, \bibinfo{person}{Long He},
  \bibinfo{person}{Yinda Zhang}, \bibinfo{person}{Ke Wang},
  \bibinfo{person}{Xilai Liu}, \bibinfo{person}{Tong Yang}, \bibinfo{person}{Yi
  Wang}, {and} \bibinfo{person}{Bin Cui}.} \bibinfo{year}{2020}\natexlab{}.
\newblock \showarticletitle{Sliding sketches: A framework using time zones for
  data stream processing in sliding windows}. In
  \bibinfo{booktitle}{\emph{Proceedings of the 26th ACM SIGKDD International
  Conference on Knowledge Discovery \& Data Mining}}.
  \bibinfo{pages}{1015--1025}.
\newblock


\bibitem[\protect\citeauthoryear{Greenwald and Khanna}{Greenwald and
  Khanna}{2001}]%
        {greenwald2001space}
\bibfield{author}{\bibinfo{person}{Michael Greenwald} {and}
  \bibinfo{person}{Sanjeev Khanna}.} \bibinfo{year}{2001}\natexlab{}.
\newblock \showarticletitle{Space-efficient online computation of quantile
  summaries}.
\newblock \bibinfo{journal}{\emph{ACM SIGMOD Record}} \bibinfo{volume}{30},
  \bibinfo{number}{2} (\bibinfo{year}{2001}), \bibinfo{pages}{58--66}.
\newblock


\bibitem[\protect\citeauthoryear{Hebrail and Berard}{Hebrail and
  Berard}{2012}]%
        {misc_individual_household_electric_power_consumption_235}
\bibfield{author}{\bibinfo{person}{Georges Hebrail} {and}
  \bibinfo{person}{Alice Berard}.} \bibinfo{year}{2012}\natexlab{}.
\newblock \bibinfo{title}{{Individual Household Electric Power Consumption}}.
\newblock \bibinfo{howpublished}{UCI Machine Learning Repository}.
\newblock
\newblock
\shownote{{DOI}: https://doi.org/10.24432/C58K54.}


\bibitem[\protect\citeauthoryear{Ho, Agrawal, Megiddo, and Srikant}{Ho
  et~al\mbox{.}}{1997}]%
        {ho1997range}
\bibfield{author}{\bibinfo{person}{Ching-Tien Ho}, \bibinfo{person}{Rakesh
  Agrawal}, \bibinfo{person}{Nimrod Megiddo}, {and}
  \bibinfo{person}{Ramakrishnan Srikant}.} \bibinfo{year}{1997}\natexlab{}.
\newblock \showarticletitle{Range queries in OLAP data cubes}.
\newblock \bibinfo{journal}{\emph{ACM SIGMOD Record}} \bibinfo{volume}{26},
  \bibinfo{number}{2} (\bibinfo{year}{1997}), \bibinfo{pages}{73--88}.
\newblock


\bibitem[\protect\citeauthoryear{Huang, Guo, Zhou, Lorch, Dang, Chintalapati,
  and Yao}{Huang et~al\mbox{.}}{2017}]%
        {huang2017gray}
\bibfield{author}{\bibinfo{person}{Peng Huang}, \bibinfo{person}{Chuanxiong
  Guo}, \bibinfo{person}{Lidong Zhou}, \bibinfo{person}{Jacob~R Lorch},
  \bibinfo{person}{Yingnong Dang}, \bibinfo{person}{Murali Chintalapati}, {and}
  \bibinfo{person}{Randolph Yao}.} \bibinfo{year}{2017}\natexlab{}.
\newblock \showarticletitle{Gray failure: The achilles' heel of cloud-scale
  systems}. In \bibinfo{booktitle}{\emph{Proceedings of the 16th Workshop on
  Hot Topics in Operating Systems}}. \bibinfo{pages}{150--155}.
\newblock


\bibitem[\protect\citeauthoryear{Ivkin, Basat, Liu, Einziger, Friedman, and
  Braverman}{Ivkin et~al\mbox{.}}{2019}]%
        {ivkin2019know}
\bibfield{author}{\bibinfo{person}{Nikita Ivkin}, \bibinfo{person}{Ran~Ben
  Basat}, \bibinfo{person}{Zaoxing Liu}, \bibinfo{person}{Gil Einziger},
  \bibinfo{person}{Roy Friedman}, {and} \bibinfo{person}{Vladimir Braverman}.}
  \bibinfo{year}{2019}\natexlab{}.
\newblock \showarticletitle{I know what you did last summer: Network monitoring
  using interval queries}.
\newblock \bibinfo{journal}{\emph{Proceedings of the ACM on Measurement and
  Analysis of Computing Systems}} \bibinfo{volume}{3}, \bibinfo{number}{3}
  (\bibinfo{year}{2019}), \bibinfo{pages}{1--28}.
\newblock


\bibitem[\protect\citeauthoryear{Jugel, Jerzak, Hackenbroich, and Markl}{Jugel
  et~al\mbox{.}}{2014}]%
        {jugel2014m4}
\bibfield{author}{\bibinfo{person}{Uwe Jugel}, \bibinfo{person}{Zbigniew
  Jerzak}, \bibinfo{person}{Gregor Hackenbroich}, {and} \bibinfo{person}{Volker
  Markl}.} \bibinfo{year}{2014}\natexlab{}.
\newblock \showarticletitle{M4: a visualization-oriented time series data
  aggregation}.
\newblock \bibinfo{journal}{\emph{Proceedings of the VLDB Endowment}}
  \bibinfo{volume}{7}, \bibinfo{number}{10} (\bibinfo{year}{2014}),
  \bibinfo{pages}{797--808}.
\newblock


\bibitem[\protect\citeauthoryear{Karger, Lehman, Leighton, Panigrahy, Levine,
  and Lewin}{Karger et~al\mbox{.}}{1997}]%
        {karger1997consistent}
\bibfield{author}{\bibinfo{person}{David Karger}, \bibinfo{person}{Eric
  Lehman}, \bibinfo{person}{Tom Leighton}, \bibinfo{person}{Rina Panigrahy},
  \bibinfo{person}{Matthew Levine}, {and} \bibinfo{person}{Daniel Lewin}.}
  \bibinfo{year}{1997}\natexlab{}.
\newblock \showarticletitle{Consistent hashing and random trees: Distributed
  caching protocols for relieving hot spots on the world wide web}. In
  \bibinfo{booktitle}{\emph{Proceedings of the twenty-ninth annual ACM
  symposium on Theory of computing}}. \bibinfo{pages}{654--663}.
\newblock


\bibitem[\protect\citeauthoryear{Karnin, Lang, and Liberty}{Karnin
  et~al\mbox{.}}{2016}]%
        {karnin2016optimal}
\bibfield{author}{\bibinfo{person}{Zohar Karnin}, \bibinfo{person}{Kevin Lang},
  {and} \bibinfo{person}{Edo Liberty}.} \bibinfo{year}{2016}\natexlab{}.
\newblock \showarticletitle{Optimal quantile approximation in streams}. In
  \bibinfo{booktitle}{\emph{2016 ieee 57th annual symposium on foundations of
  computer science (focs)}}. IEEE, \bibinfo{pages}{71--78}.
\newblock


\bibitem[\protect\citeauthoryear{Koren}{Koren}{2009}]%
        {koren2009collaborative}
\bibfield{author}{\bibinfo{person}{Yehuda Koren}.}
  \bibinfo{year}{2009}\natexlab{}.
\newblock \showarticletitle{Collaborative filtering with temporal dynamics}. In
  \bibinfo{booktitle}{\emph{Proceedings of the 15th ACM SIGKDD international
  conference on Knowledge discovery and data mining}}.
  \bibinfo{pages}{447--456}.
\newblock


\bibitem[\protect\citeauthoryear{Lall, Sekar, Ogihara, Xu, and Zhang}{Lall
  et~al\mbox{.}}{2006}]%
        {lall2006data}
\bibfield{author}{\bibinfo{person}{Ashwin Lall}, \bibinfo{person}{Vyas Sekar},
  \bibinfo{person}{Mitsunori Ogihara}, \bibinfo{person}{Jun Xu}, {and}
  \bibinfo{person}{Hui Zhang}.} \bibinfo{year}{2006}\natexlab{}.
\newblock \showarticletitle{Data streaming algorithms for estimating entropy of
  network traffic}.
\newblock \bibinfo{journal}{\emph{ACM SIGMETRICS Performance Evaluation
  Review}} \bibinfo{volume}{34}, \bibinfo{number}{1} (\bibinfo{year}{2006}),
  \bibinfo{pages}{145--156}.
\newblock


\bibitem[\protect\citeauthoryear{Liang, Sintos, Shang, and Krishnan}{Liang
  et~al\mbox{.}}{2021a}]%
        {liang2021combining}
\bibfield{author}{\bibinfo{person}{Xi Liang}, \bibinfo{person}{Stavros Sintos},
  \bibinfo{person}{Zechao Shang}, {and} \bibinfo{person}{Sanjay Krishnan}.}
  \bibinfo{year}{2021}\natexlab{a}.
\newblock \showarticletitle{Combining aggregation and sampling (nearly)
  optimally for approximate query processing}. In
  \bibinfo{booktitle}{\emph{Proceedings of the 2021 International Conference on
  Management of Data}}. \bibinfo{pages}{1129--1141}.
\newblock


\bibitem[\protect\citeauthoryear{Liang, Sintos, Shang, and Krishnan}{Liang
  et~al\mbox{.}}{2021b}]%
        {https://doi.org/10.48550/arxiv.2103.15994}
\bibfield{author}{\bibinfo{person}{Xi Liang}, \bibinfo{person}{Stavros Sintos},
  \bibinfo{person}{Zechao Shang}, {and} \bibinfo{person}{Sanjay Krishnan}.}
  \bibinfo{year}{2021}\natexlab{b}.
\newblock \bibinfo{title}{Combining Aggregation and Sampling (Nearly) Optimally
  for Approximate Query Processing}.
\newblock
\newblock
\urldef\tempurl%
\url{https://doi.org/10.48550/ARXIV.2103.15994}
\showDOI{\tempurl}


\bibitem[\protect\citeauthoryear{Lim, Hassan, Jin, Volos, and Jeon}{Lim
  et~al\mbox{.}}{2020}]%
        {lim2020approximate}
\bibfield{author}{\bibinfo{person}{Gangmuk Lim}, \bibinfo{person}{Mohamed~S
  Hassan}, \bibinfo{person}{Ze Jin}, \bibinfo{person}{Stavros Volos}, {and}
  \bibinfo{person}{Myeongjae Jeon}.} \bibinfo{year}{2020}\natexlab{}.
\newblock \showarticletitle{Approximate quantiles for datacenter telemetry
  monitoring}. In \bibinfo{booktitle}{\emph{2020 IEEE 36th International
  Conference on Data Engineering (ICDE)}}. IEEE, \bibinfo{pages}{1914--1917}.
\newblock


\bibitem[\protect\citeauthoryear{Liu, Manousis, Vorsanger, Sekar, and
  Braverman}{Liu et~al\mbox{.}}{2016}]%
        {liu2016one}
\bibfield{author}{\bibinfo{person}{Zaoxing Liu}, \bibinfo{person}{Antonis
  Manousis}, \bibinfo{person}{Gregory Vorsanger}, \bibinfo{person}{Vyas Sekar},
  {and} \bibinfo{person}{Vladimir Braverman}.} \bibinfo{year}{2016}\natexlab{}.
\newblock \showarticletitle{One sketch to rule them all: Rethinking network
  flow monitoring with univmon}. In \bibinfo{booktitle}{\emph{Proceedings of
  the 2016 ACM SIGCOMM Conference}}. \bibinfo{pages}{101--114}.
\newblock


\bibitem[\protect\citeauthoryear{Liu, Namkung, Nikolaidis, Lee, Kim, Jin,
  Braverman, Yu, and Sekar}{Liu et~al\mbox{.}}{2021}]%
        {liu2021jaqen}
\bibfield{author}{\bibinfo{person}{Zaoxing Liu}, \bibinfo{person}{Hun Namkung},
  \bibinfo{person}{Georgios Nikolaidis}, \bibinfo{person}{Jeongkeun Lee},
  \bibinfo{person}{Changhoon Kim}, \bibinfo{person}{Xin Jin},
  \bibinfo{person}{Vladimir Braverman}, \bibinfo{person}{Minlan Yu}, {and}
  \bibinfo{person}{Vyas Sekar}.} \bibinfo{year}{2021}\natexlab{}.
\newblock \showarticletitle{Jaqen: A High-Performance Switch-Native Approach
  for Detecting and Mitigating Volumetric DDoS Attacks with Programmable
  Switches}. In \bibinfo{booktitle}{\emph{30th USENIX Security Symposium
  (USENIX Security 21)}}. \bibinfo{pages}{3829--3846}.
\newblock


\bibitem[\protect\citeauthoryear{Lu, Tok, Raissi, and Bressan}{Lu
  et~al\mbox{.}}{2010}]%
        {lu2010simple}
\bibfield{author}{\bibinfo{person}{Xuesong Lu}, \bibinfo{person}{Wee~Hyong
  Tok}, \bibinfo{person}{Chedy Raissi}, {and} \bibinfo{person}{St{\'e}phane
  Bressan}.} \bibinfo{year}{2010}\natexlab{}.
\newblock \showarticletitle{A simple, yet effective and efficient, sliding
  window sampling algorithm}. In \bibinfo{booktitle}{\emph{Database Systems for
  Advanced Applications: 15th International Conference, DASFAA 2010, Tsukuba,
  Japan, April 1-4, 2010, Proceedings, Part I 15}}. Springer,
  \bibinfo{pages}{337--351}.
\newblock


\bibitem[\protect\citeauthoryear{Madden, Franklin, Hellerstein, and
  Hong}{Madden et~al\mbox{.}}{2003}]%
        {madden2003design}
\bibfield{author}{\bibinfo{person}{Samuel Madden}, \bibinfo{person}{Michael~J
  Franklin}, \bibinfo{person}{Joseph~M Hellerstein}, {and} \bibinfo{person}{Wei
  Hong}.} \bibinfo{year}{2003}\natexlab{}.
\newblock \showarticletitle{The design of an acquisitional query processor for
  sensor networks}. In \bibinfo{booktitle}{\emph{Proceedings of the 2003 ACM
  SIGMOD international conference on Management of data}}.
  \bibinfo{pages}{491--502}.
\newblock


\bibitem[\protect\citeauthoryear{Manousis, Cheng, Basat, Liu, and
  Sekar}{Manousis et~al\mbox{.}}{2022}]%
        {manousis2022enabling}
\bibfield{author}{\bibinfo{person}{Antonis Manousis}, \bibinfo{person}{Zhuo
  Cheng}, \bibinfo{person}{Ran~Ben Basat}, \bibinfo{person}{Zaoxing Liu}, {and}
  \bibinfo{person}{Vyas Sekar}.} \bibinfo{year}{2022}\natexlab{}.
\newblock \showarticletitle{Enabling efficient and general subpopulation
  analytics in multidimensional data streams}.
\newblock \bibinfo{journal}{\emph{arXiv preprint arXiv:2208.04927}}
  (\bibinfo{year}{2022}).
\newblock


\bibitem[\protect\citeauthoryear{Maroulis, Stamatopoulos, Papastefanatos, and
  Terrovitis}{Maroulis et~al\mbox{.}}{2024}]%
        {maroulis2024visualization}
\bibfield{author}{\bibinfo{person}{Stavros Maroulis}, \bibinfo{person}{Vassilis
  Stamatopoulos}, \bibinfo{person}{George Papastefanatos}, {and}
  \bibinfo{person}{Manolis Terrovitis}.} \bibinfo{year}{2024}\natexlab{}.
\newblock \showarticletitle{Visualization-aware Time Series Min-Max Caching
  with Error Bound Guarantees}.
\newblock \bibinfo{journal}{\emph{Proceedings of the VLDB Endowment}}
  \bibinfo{volume}{17}, \bibinfo{number}{8} (\bibinfo{year}{2024}),
  \bibinfo{pages}{2091--2103}.
\newblock


\bibitem[\protect\citeauthoryear{Mart, Negru, Pop, and Castiglione}{Mart
  et~al\mbox{.}}{2020}]%
        {mart2020observability}
\bibfield{author}{\bibinfo{person}{Octavian Mart}, \bibinfo{person}{Catalin
  Negru}, \bibinfo{person}{Florin Pop}, {and} \bibinfo{person}{Aniello
  Castiglione}.} \bibinfo{year}{2020}\natexlab{}.
\newblock \showarticletitle{Observability in kubernetes cluster: Automatic
  anomalies detection using prometheus}. In \bibinfo{booktitle}{\emph{2020 IEEE
  22nd International Conference on High Performance Computing and
  Communications; IEEE 18th International Conference on Smart City; IEEE 6th
  International Conference on Data Science and Systems (HPCC/SmartCity/DSS)}}.
  IEEE, \bibinfo{pages}{565--570}.
\newblock


\bibitem[\protect\citeauthoryear{Marzano, Alexander, Fonseca, Fazzion, Hoepers,
  Steding-Jessen, Chaves, Cunha, Guedes, and Meira}{Marzano
  et~al\mbox{.}}{2018}]%
        {marzano2018evolution}
\bibfield{author}{\bibinfo{person}{Artur Marzano}, \bibinfo{person}{David
  Alexander}, \bibinfo{person}{Osvaldo Fonseca}, \bibinfo{person}{Elverton
  Fazzion}, \bibinfo{person}{Cristine Hoepers}, \bibinfo{person}{Klaus
  Steding-Jessen}, \bibinfo{person}{Marcelo~HPC Chaves},
  \bibinfo{person}{{\'I}talo Cunha}, \bibinfo{person}{Dorgival Guedes}, {and}
  \bibinfo{person}{Wagner Meira}.} \bibinfo{year}{2018}\natexlab{}.
\newblock \showarticletitle{The evolution of bashlite and mirai iot botnets}.
  In \bibinfo{booktitle}{\emph{2018 IEEE Symposium on Computers and
  Communications (ISCC)}}. IEEE, \bibinfo{pages}{00813--00818}.
\newblock


\bibitem[\protect\citeauthoryear{Masson, Rim, and Lee}{Masson
  et~al\mbox{.}}{2019}]%
        {masson2019ddsketch}
\bibfield{author}{\bibinfo{person}{Charles Masson}, \bibinfo{person}{Jee~E
  Rim}, {and} \bibinfo{person}{Homin~K Lee}.} \bibinfo{year}{2019}\natexlab{}.
\newblock \showarticletitle{DDSketch: A fast and fully-mergeable quantile
  sketch with relative-error guarantees}.
\newblock \bibinfo{journal}{\emph{arXiv preprint arXiv:1908.10693}}
  (\bibinfo{year}{2019}).
\newblock


\bibitem[\protect\citeauthoryear{Mitzenmacher, Steinke, and
  Thaler}{Mitzenmacher et~al\mbox{.}}{2012}]%
        {mitzenmacher2012hierarchical}
\bibfield{author}{\bibinfo{person}{Michael Mitzenmacher},
  \bibinfo{person}{Thomas Steinke}, {and} \bibinfo{person}{Justin Thaler}.}
  \bibinfo{year}{2012}\natexlab{}.
\newblock \showarticletitle{Hierarchical heavy hitters with the space saving
  algorithm}. In \bibinfo{booktitle}{\emph{2012 Proceedings of the Fourteenth
  Workshop on Algorithm Engineering and Experiments (ALENEX)}}. SIAM,
  \bibinfo{pages}{160--174}.
\newblock


\bibitem[\protect\citeauthoryear{Moosa, Vangujar, and Mahajan}{Moosa
  et~al\mbox{.}}{2023}]%
        {moosa2023detection}
\bibfield{author}{\bibinfo{person}{Muhammad~Aashiq Moosa},
  \bibinfo{person}{Apurva~K Vangujar}, {and} \bibinfo{person}{Dnyanesh~Pramod
  Mahajan}.} \bibinfo{year}{2023}\natexlab{}.
\newblock \showarticletitle{Detection and Analysis of DDoS Attack Using a
  Collaborative Network Monitoring Stack}. In \bibinfo{booktitle}{\emph{2023
  16th International Conference on Security of Information and Networks
  (SIN)}}. IEEE, \bibinfo{pages}{1--9}.
\newblock


\bibitem[\protect\citeauthoryear{Ohsita, Ata, and Murata}{Ohsita
  et~al\mbox{.}}{2004}]%
        {1378371}
\bibfield{author}{\bibinfo{person}{Y. Ohsita}, \bibinfo{person}{S. Ata}, {and}
  \bibinfo{person}{M. Murata}.} \bibinfo{year}{2004}\natexlab{}.
\newblock \showarticletitle{Detecting distributed denial-of-service attacks by
  analyzing TCP SYN packets statistically}. In \bibinfo{booktitle}{\emph{IEEE
  Global Telecommunications Conference, 2004. GLOBECOM '04.}},
  Vol.~\bibinfo{volume}{4}. \bibinfo{pages}{2043--2049 Vol.4}.
\newblock
\urldef\tempurl%
\url{https://doi.org/10.1109/GLOCOM.2004.1378371}
\showDOI{\tempurl}


\bibitem[\protect\citeauthoryear{Papadopoulos, Ali-Eldin, {\AA}rz{\'e}n,
  Tordsson, and Elmroth}{Papadopoulos et~al\mbox{.}}{2016}]%
        {papadopoulos2016peas}
\bibfield{author}{\bibinfo{person}{Alessandro~Vittorio Papadopoulos},
  \bibinfo{person}{Ahmed Ali-Eldin}, \bibinfo{person}{Karl-Erik {\AA}rz{\'e}n},
  \bibinfo{person}{Johan Tordsson}, {and} \bibinfo{person}{Erik Elmroth}.}
  \bibinfo{year}{2016}\natexlab{}.
\newblock \showarticletitle{PEAS: A performance evaluation framework for
  auto-scaling strategies in cloud applications}.
\newblock \bibinfo{journal}{\emph{ACM Transactions on Modeling and Performance
  Evaluation of Computing Systems (TOMPECS)}} \bibinfo{volume}{1},
  \bibinfo{number}{4} (\bibinfo{year}{2016}), \bibinfo{pages}{1--31}.
\newblock


\bibitem[\protect\citeauthoryear{Papapetrou, Garofalakis, and
  Deligiannakis}{Papapetrou et~al\mbox{.}}{2012}]%
        {papapetrou2012sketch}
\bibfield{author}{\bibinfo{person}{Odysseas Papapetrou}, \bibinfo{person}{Minos
  Garofalakis}, {and} \bibinfo{person}{Antonios Deligiannakis}.}
  \bibinfo{year}{2012}\natexlab{}.
\newblock \showarticletitle{Sketch-based querying of distributed sliding-window
  data streams}.
\newblock \bibinfo{journal}{\emph{arXiv preprint arXiv:1207.0139}}
  (\bibinfo{year}{2012}).
\newblock


\bibitem[\protect\citeauthoryear{Park, Mozafari, Sorenson, and Wang}{Park
  et~al\mbox{.}}{2018}]%
        {park2018verdictdb}
\bibfield{author}{\bibinfo{person}{Yongjoo Park}, \bibinfo{person}{Barzan
  Mozafari}, \bibinfo{person}{Joseph Sorenson}, {and} \bibinfo{person}{Junhao
  Wang}.} \bibinfo{year}{2018}\natexlab{}.
\newblock \showarticletitle{Verdictdb: Universalizing approximate query
  processing}. In \bibinfo{booktitle}{\emph{Proceedings of the 2018
  International Conference on Management of Data}}.
  \bibinfo{pages}{1461--1476}.
\newblock


\bibitem[\protect\citeauthoryear{Pelkonen, Franklin, Teller, Cavallaro, Huang,
  Meza, and Veeraraghavan}{Pelkonen et~al\mbox{.}}{2015}]%
        {pelkonen2015gorilla}
\bibfield{author}{\bibinfo{person}{Tuomas Pelkonen}, \bibinfo{person}{Scott
  Franklin}, \bibinfo{person}{Justin Teller}, \bibinfo{person}{Paul Cavallaro},
  \bibinfo{person}{Qi Huang}, \bibinfo{person}{Justin Meza}, {and}
  \bibinfo{person}{Kaushik Veeraraghavan}.} \bibinfo{year}{2015}\natexlab{}.
\newblock \showarticletitle{Gorilla: A fast, scalable, in-memory time series
  database}.
\newblock \bibinfo{journal}{\emph{Proceedings of the VLDB Endowment}}
  \bibinfo{volume}{8}, \bibinfo{number}{12} (\bibinfo{year}{2015}),
  \bibinfo{pages}{1816--1827}.
\newblock


\bibitem[\protect\citeauthoryear{Peng, Zhang, Wang, and Pei}{Peng
  et~al\mbox{.}}{2018}]%
        {peng2018aqp++}
\bibfield{author}{\bibinfo{person}{Jinglin Peng}, \bibinfo{person}{Dongxiang
  Zhang}, \bibinfo{person}{Jiannan Wang}, {and} \bibinfo{person}{Jian Pei}.}
  \bibinfo{year}{2018}\natexlab{}.
\newblock \showarticletitle{Aqp++ connecting approximate query processing with
  aggregate precomputation for interactive analytics}. In
  \bibinfo{booktitle}{\emph{Proceedings of the 2018 International Conference on
  Management of Data}}. \bibinfo{pages}{1477--1492}.
\newblock


\bibitem[\protect\citeauthoryear{Pope, Raimondo, Kumar, McConville, Piechocki,
  Oikonomou, Pasquier, Luo, Howarth, Mavromatis, et~al\mbox{.}}{Pope
  et~al\mbox{.}}{2021}]%
        {pope2021container}
\bibfield{author}{\bibinfo{person}{James Pope}, \bibinfo{person}{Francesco
  Raimondo}, \bibinfo{person}{Vijay Kumar}, \bibinfo{person}{Ryan McConville},
  \bibinfo{person}{Rob Piechocki}, \bibinfo{person}{George Oikonomou},
  \bibinfo{person}{Thomas Pasquier}, \bibinfo{person}{Bo Luo},
  \bibinfo{person}{Dan Howarth}, \bibinfo{person}{Ioannis Mavromatis},
  {et~al\mbox{.}}} \bibinfo{year}{2021}\natexlab{}.
\newblock \showarticletitle{Container escape detection for edge devices}. In
  \bibinfo{booktitle}{\emph{Proceedings of the 19th ACM Conference on Embedded
  Networked Sensor Systems}}. \bibinfo{pages}{532--536}.
\newblock


\bibitem[\protect\citeauthoryear{Priovolos, Lioprasitis, Gardikis, and
  Costicoglou}{Priovolos et~al\mbox{.}}{2021}]%
        {priovolos2021using}
\bibfield{author}{\bibinfo{person}{Athanasios Priovolos},
  \bibinfo{person}{Dimitris Lioprasitis}, \bibinfo{person}{Georgios Gardikis},
  {and} \bibinfo{person}{Socrates Costicoglou}.}
  \bibinfo{year}{2021}\natexlab{}.
\newblock \showarticletitle{Using anomaly detection techniques for securing 5G
  infrastructure and applications}. In \bibinfo{booktitle}{\emph{2021 IEEE
  International Mediterranean Conference on Communications and Networking
  (MeditCom)}}. IEEE, \bibinfo{pages}{519--524}.
\newblock


\bibitem[\protect\citeauthoryear{Shen, Ouyang, Li, Liu, Zhu, Zou, Su, Yu, Yi,
  Hu, et~al\mbox{.}}{Shen et~al\mbox{.}}{2023}]%
        {shen2023lindorm}
\bibfield{author}{\bibinfo{person}{Chunhui Shen}, \bibinfo{person}{Qianyu
  Ouyang}, \bibinfo{person}{Feibo Li}, \bibinfo{person}{Zhipeng Liu},
  \bibinfo{person}{Longcheng Zhu}, \bibinfo{person}{Yujie Zou},
  \bibinfo{person}{Qing Su}, \bibinfo{person}{Tianhuan Yu}, \bibinfo{person}{Yi
  Yi}, \bibinfo{person}{Jianhong Hu}, {et~al\mbox{.}}}
  \bibinfo{year}{2023}\natexlab{}.
\newblock \showarticletitle{Lindorm TSDB: A Cloud-Native Time-Series Database
  for Large-Scale Monitoring Systems}.
\newblock \bibinfo{journal}{\emph{Proceedings of the VLDB Endowment}}
  \bibinfo{volume}{16}, \bibinfo{number}{12} (\bibinfo{year}{2023}),
  \bibinfo{pages}{3715--3727}.
\newblock


\bibitem[\protect\citeauthoryear{Sides, Bremler-Barr, and Rosensweig}{Sides
  et~al\mbox{.}}{2015}]%
        {sides2015yo}
\bibfield{author}{\bibinfo{person}{Mor Sides}, \bibinfo{person}{Anat
  Bremler-Barr}, {and} \bibinfo{person}{Elisha Rosensweig}.}
  \bibinfo{year}{2015}\natexlab{}.
\newblock \showarticletitle{Yo-Yo Attack: vulnerability in auto-scaling
  mechanism}. In \bibinfo{booktitle}{\emph{Proceedings of the 2015 ACM
  Conference on Special Interest Group on Data Communication}}.
  \bibinfo{pages}{103--104}.
\newblock


\bibitem[\protect\citeauthoryear{Turnbull}{Turnbull}{2018}]%
        {turnbull2018monitoring}
\bibfield{author}{\bibinfo{person}{James Turnbull}.}
  \bibinfo{year}{2018}\natexlab{}.
\newblock \bibinfo{booktitle}{\emph{Monitoring with Prometheus}}.
\newblock \bibinfo{publisher}{Turnbull Press}.
\newblock


\bibitem[\protect\citeauthoryear{Wang, Xue, and Shao}{Wang
  et~al\mbox{.}}{2021}]%
        {wang2021heracles}
\bibfield{author}{\bibinfo{person}{Zhiqi Wang}, \bibinfo{person}{Jin Xue},
  {and} \bibinfo{person}{Zili Shao}.} \bibinfo{year}{2021}\natexlab{}.
\newblock \showarticletitle{Heracles: an efficient storage model and data
  flushing for performance monitoring timeseries}.
\newblock \bibinfo{journal}{\emph{Proceedings of the VLDB Endowment}}
  \bibinfo{volume}{14}, \bibinfo{number}{6} (\bibinfo{year}{2021}),
  \bibinfo{pages}{1080--1092}.
\newblock


\bibitem[\protect\citeauthoryear{Wilkinson}{Wilkinson}{2016}]%
        {google-alerting}
\bibfield{author}{\bibinfo{person}{Jamie Wilkinson}.}
  \bibinfo{year}{2016}\natexlab{}.
\newblock \bibinfo{title}{Google Promethues: A practical guide to alerting at
  scale}.
\newblock
\newblock
\urldef\tempurl%
\url{https://docs.google.com/presentation/d/1X1rKozAUuF2MVc1YXElFWq9wkcWv3Axdldl8LOH9Vik/edit#slide=id.g598ef96a6_0_341}
\showURL{%
Retrieved 2024 from \tempurl}


\bibitem[\protect\citeauthoryear{Wu, Jiang, Dong, Zhong, Chen, Hu, Yang, Uhlig,
  and Cui}{Wu et~al\mbox{.}}{2023}]%
        {wu2023microscopesketch}
\bibfield{author}{\bibinfo{person}{Yuhan Wu}, \bibinfo{person}{Shiqi Jiang},
  \bibinfo{person}{Siyuan Dong}, \bibinfo{person}{Zheng Zhong},
  \bibinfo{person}{Jiale Chen}, \bibinfo{person}{Yutong Hu},
  \bibinfo{person}{Tong Yang}, \bibinfo{person}{Steve Uhlig}, {and}
  \bibinfo{person}{Bin Cui}.} \bibinfo{year}{2023}\natexlab{}.
\newblock \showarticletitle{MicroscopeSketch: Accurate Sliding Estimation Using
  Adaptive Zooming}. In \bibinfo{booktitle}{\emph{Proceedings of the 29th ACM
  SIGKDD Conference on Knowledge Discovery and Data Mining}}.
  \bibinfo{pages}{2660--2671}.
\newblock


\bibitem[\protect\citeauthoryear{Yang, Zhang, Gadre, Liu, Kumar, and
  Sekar}{Yang et~al\mbox{.}}{2020}]%
        {yang2020joltik}
\bibfield{author}{\bibinfo{person}{Mingran Yang}, \bibinfo{person}{Junbo
  Zhang}, \bibinfo{person}{Akshay Gadre}, \bibinfo{person}{Zaoxing Liu},
  \bibinfo{person}{Swarun Kumar}, {and} \bibinfo{person}{Vyas Sekar}.}
  \bibinfo{year}{2020}\natexlab{}.
\newblock \showarticletitle{Joltik: enabling energy-efficient" future-proof"
  analytics on low-power wide-area networks}. In
  \bibinfo{booktitle}{\emph{Proceedings of the 26th Annual International
  Conference on Mobile Computing and Networking}}. \bibinfo{pages}{1--14}.
\newblock


\bibitem[\protect\citeauthoryear{Yang, Gong, Zhang, Zou, Shi, and Li}{Yang
  et~al\mbox{.}}{2018}]%
        {yang2018heavyguardian}
\bibfield{author}{\bibinfo{person}{Tong Yang}, \bibinfo{person}{Junzhi Gong},
  \bibinfo{person}{Haowei Zhang}, \bibinfo{person}{Lei Zou},
  \bibinfo{person}{Lei Shi}, {and} \bibinfo{person}{Xiaoming Li}.}
  \bibinfo{year}{2018}\natexlab{}.
\newblock \showarticletitle{Heavyguardian: Separate and guard hot items in data
  streams}. In \bibinfo{booktitle}{\emph{Proceedings of the 24th ACM SIGKDD
  International Conference on Knowledge Discovery \& Data Mining}}.
  \bibinfo{pages}{2584--2593}.
\newblock


\bibitem[\protect\citeauthoryear{Zhao, Li, Tian, Mei, and Wu}{Zhao
  et~al\mbox{.}}{2021}]%
        {zhao2021dhs}
\bibfield{author}{\bibinfo{person}{Bohan Zhao}, \bibinfo{person}{Xiang Li},
  \bibinfo{person}{Boyu Tian}, \bibinfo{person}{Zhiyu Mei}, {and}
  \bibinfo{person}{Wenfei Wu}.} \bibinfo{year}{2021}\natexlab{}.
\newblock \showarticletitle{Dhs: Adaptive memory layout organization of sketch
  slots for fast and accurate data stream processing}. In
  \bibinfo{booktitle}{\emph{Proceedings of ACM SIGKDD}}.
  \bibinfo{pages}{2285--2293}.
\newblock


\end{thebibliography}
